\documentclass[a4paper,11pt]{article}
\pdfoutput=1 

\usepackage{jcappub} 

\usepackage[T1]{fontenc} 
\usepackage{tensor}
\allowdisplaybreaks
\usepackage{tabularx} 
\usepackage{array}
\usepackage{multirow}
\usepackage{subfig}
\providecommand{\tabularnewline}{\\}

\usepackage{makecell}

\title{\boldmath Quintessence in the Weyl-Gauss-Bonnet Model}

\author[a]{Jos\'e Jaime Terente D\'iaz,}
\author[b]{Konstantinos Dimopoulos,}
\author[a,c]{Mindaugas Kar\v{c}iauskas,}
\author[d]{Antonio Racioppi}

\affiliation[a]{Departamento de F\'isica Te\'orica, Universidad Complutense de Madrid,\\E-28040 Madrid, Spain}
\affiliation[b]{Consortium for Fundamental Physics, Physics Department, Lancaster University,\\Lancaster LA1 4YB, UK}
\affiliation[c]{Center for Physical Sciences and Technology,\\Saul\.{e}tekio av. 3, 10257 Vilnius, Lithuania}
\affiliation[d]{National Institute of Chemical Physics and Biophysics,\\R\"avala 10 10143 Tallinn, Estonia}

\abstract{Quintessence models have been widely examined in the context of scalar-Gauss-Bonnet gravity, a subclass of Horndeski's theory, and were proposed as viable candidates for Dark Energy. However, the relatively recent observational constraints on the speed of gravitational waves $c_{\textrm{GW}}$ have resulted in many of those models being ruled out because they predict $c_{\textrm{GW}} \neq c$ generally. While these were formulated in the metric formalism of gravity, 
we put forward a new quintessence model with the scalar-Gauss-Bonnet action but in Weyl geometry, where the connection is not metric compatible. We find the fixed points of the dynamical system under some assumptions and determine their stability via linear analysis. The past evolution of the Universe can be reproduced correctly, but the late Universe constraints on $c_{\textrm{GW}}$ are grossly violated. Moreover, at these later stages tensor modes suffer from the gradient instabilities. We also consider the implications of imposing an additional constraint $c_{\textrm{GW}} = c$, but this does not lead to evolution that is consistent with cosmological observations.}

\begin{document}
\maketitle
\flushbottom

\section{Introduction}
\label{sec:intro}
Late Dark Energy (DE) models have become quite popular following the observational discovery of the current accelerated expansion of the Universe from Type Ia supernovae \cite{SupernovaSearchTeam:1998fmf,SupernovaCosmologyProject:1998vns}. \emph{Planck} Collaboration in particular finds that the equation of state (EoS) parameter of this DE is consistent with a cosmological constant (CC) \cite{Planck:2018vyg}, arguably the simplest candidate for DE. This CC is interpreted as vacuum energy in particle physics \cite{Carroll:2000fy,Martin:2012bt,Joyce:2014kja}, but no convincing scenario has been found where the actual observed energy scale of DE can be naturally accounted for by vacuum energy ascribed to a particle physics model \cite{Weinberg:1988cp,Weinberg:2000yb}.\  

\emph{Planck} measurements also leave the door open to a dynamical scalar field (named `quintessence' \cite{Caldwell:1997ii,Zlatev:1998tr}), with changing DE EoS parameter. This latter possibility allows for a more sensible mechanism to explain DE \cite{Copeland:2006wr,Tsujikawa:2010sc,Tsujikawa:2013fta} and paves the way for a unified description (called `quintessential inflation' \cite{Peebles:1998qn,Konstantinos:2001MQI,deHaro:2016ftq,Dimopoulos:2021xld}) of the late accelerated expansion era and early inflation. The inflationary phase in the early evolution of the Universe serves to explain the primordial origin of the Large Scale Structures in the Universe while resolving the horizon and flatness problems of the Hot Big Bang model \cite{Starobinsky:1980te,Guth:1980zm,Linde:1981mu}. \ 

Many of those quintessence DE models rely on modifications of gravity at large and fall within the scalar-tensor theories chiefly \cite{Capozziello:2002rd,vandeBruck:2017voa,deHaro:2021swo,Nojiri:2017ncd,Carroll:2003wy,Koivisto:2006xf}. Consequently, they have been (and still are) under close examination after the detection of gravitational waves (GWs) made by the Advanced LIGO and Virgo detectors \cite{LIGOScientific:2016aoc,LIGOScientific:2016sjg,LIGOScientific:2017vwq} (see Refs.~\cite{Sakstein:2017xjx,Ezquiaga:2017ekz,Baker:2017hug,Arai:2017hxj,Creminelli:2017sry,Langlois:2017dyl}). The multi-messenger observation of a binary neutron star system's merger \cite{LIGOScientific:2017ync} placed stringent bounds on the relative difference between the speed of propagation of GWs, $c_{\textrm{GW}}$, and the speed of light in vacuum, $c$ \cite{LIGOScientific:2017zic}. Defining the parameter \begin{equation} \alpha_T \equiv c_{\textrm{GW}}^2-1~,\end{equation} in units where $c=1$, those bounds yield \begin{equation}\label{bounds-on-cgw2} |\alpha_T| < 10^{-15},\end{equation} given the delay between arrival times of the GWs and $\gamma$-rays emitted by the system's merger.\

Horndeski's theory \cite{Horndeski:1974wa} is an example of theory of gravity that has been strongly constrained by Eq.~\eqref{bounds-on-cgw2}.\footnote{Regarding this bound, its consistency across the whole spectrum of GW frequencies has been discussed within the context of Horndeski's theory in Ref.~\cite{deRham:2018red}.} This theory is the most general one in four dimensions leading to second-order field equations of the metric tensor $g_{\mu\nu}(x)$ and a scalar field $\phi(x)$. It can be shown that, in order for GWs to propagate at the speed of light ($\alpha_T = 0$), only the standard, field-dependent conformal coupling to gravity $G(\phi)R$, where $R$ is the Ricci scalar, is allowed. This reduces significantly the number of subclasses of Horndeski's theory conducive to modelling quintessence \cite{Kobayashi:2011nu,Ezquiaga:2017ekz}.\

Among those subclasses affected by the constraint \eqref{bounds-on-cgw2}, the scalar-Gauss-Bonnet (SGB) model, the scalar-tensor generalisation of Einstein-Gauss-Bonnet (EGB) gravity \cite{Fernandes:2022zrq}, is a well-known case \cite{Kobayashi:2011nu} in cosmology. This model consists of a field-dependent coupling $\xi(\phi)$ to the Gauss-Bonnet (GB) term \begin{equation}\label{eq:GB-term} \mathcal{G} \equiv R^2-4R^{\mu\nu}R_{\mu\nu} + R^{\alpha\beta\mu\nu} R_{\alpha\beta\mu\nu}~,\end{equation} where $R_{\mu\nu}$ and $\tensor{R}{^{\alpha}_{\beta\mu\nu}}$ are the Ricci and Riemann tensors, respectively, in addition to the Einstein-Hilbert action of General Relativity (GR). The upper bound on $|\alpha_T|$ is satisfied if $\xi$ is constant. This is not surprising as the GB term is a topological term in four dimensions and does not contribute to the field equations of the metric \cite{Fernandes:2022zrq}. Hence, one is left with Einstein's theory of GR, which predicts that GWs propagate at the speed of light. Another possibility is to impose the following condition: \begin{equation}\label{eq:cgweqc-metric}\ddot\xi = H\dot \xi~,\end{equation} which restricts the functional form of the coupling $\xi(t)$, such that $\dot \xi(t) \propto a(t)$. $H(t) \equiv \dot a(t)/a(t)$ is the Hubble parameter and $a(t)$ the scale factor. Overdots denote time derivatives and a homogeneous, isotropic and spatially-flat spacetime, described by the flat Friedmann-Lema\^itre-Robertson-Walker (FLRW) metric $g_{\mu\nu} = \textrm{diag} \left[-1,a^2(t),a^2(t),a^2(t)\right]$, was assumed. The consequences of the scaling of $\xi$ after Eq.~\eqref{eq:cgweqc-metric} have been studied in the literature in the context of cosmic inflation \cite{Odintsov:2019clh,Odintsov:2020zkl,Odintsov:2020sqy}, despite the fact that the neutron star system's merger is a low-redshift event, significantly lower than the estimated redshift value at the onset of DE domination \cite{Hjorth:2017yza} and at a time much later than inflation. For this latter reason, we considered Gauss-Bonnet Dark Energy (GBDE) models with $\alpha_T = 0$ in Ref.~\cite{TerenteDiaz:2023iqk}. Unfortunately, we found that one cannot make $c_{\textrm{GW}} = 1$ if the density parameters of matter and DE are such that $\Omega_{\textrm{M}}\sim \Omega_{\textrm{DE}}$.\ 

Leaving those two possibilities aside, there are works in which the effects of the constraint on $|\alpha_T|$ are examined and illustrated \cite{Gong:2017kim,Granda:2018tzi}, the first reference including a case of kinetic coupling to curvature besides the coupling to the GB term. In Ref.~\cite{TerenteDiaz:2023iqk}, those effects were explored in the GBDE models mentioned above. We came to the realisation that the bound does not seem very constraining for those models. Contrary to this intuition however, it was found that the constraint is easily violated at present time.\footnote{Modifications to the SGB model including a coupling between the matter sector and the quintessence field or a non-minimal coupling of the scalar field to gravity seem to satisfy the constraint \eqref{bounds-on-cgw2} for very specific initial conditions set in radiation domination and $\xi(\phi) \propto \phi^2$. See results reported in Refs.~\cite{MohseniSadjadi:2023amn,MohseniSadjadi:2023cjd}.}\ 

Additionally, there has been an intensive research on Horndeski models in the framework of \emph{the Palatini formalism} \cite{Helpin:2019kcq,Helpin:2019vrv,Kubota:2020ehu,Dong:2021jtd,Dong:2022cvf}.\footnote{Recent quintessential inflation models have been proposed in this formalism in $f(R)$ and $f(\phi,R)$ theories \cite{Dimopoulos:2020pas,Dimopoulos:2022rdp}. In the case of $f(\phi)R$ gravity, an example of quintessence scalar field couplings with metric and torsion can be found in Ref.~\cite{Sharma:2021fou}, where the growth of linear matter perturbations is studied.} A striking result of that investigation has been the realisation that Horndeski models in this formalism, with a conformal coupling depending on the kinetic term $X\equiv -\frac{1}{2} g^{\mu\nu} \partial_{\mu} \phi \partial_{\nu}\phi$ (that is, $G(\phi,X)R$), do allow GWs propagating at the speed of light \cite{Kubota:2020ehu}. This is in contrast to what the metric one prescribes. The reason lies in the fact that no second order derivatives of the metric are present in the action in the Palatini formalism, because the metric and the connection are assumed to be independent variables. Therefore, no `counter terms' are added to the action in order to avoid the Ostrogradski ghost \cite{Joyce:2014kja,deRham:2016wji}. These `counter terms' are precisely those which cause $c_{\textrm{GW}} \neq 1$ in the metric formalism \cite{Kobayashi:2011nu}. Similar attempts were made to address this question for more complicated Horndeski models \cite{Dong:2021jtd,Dong:2022cvf}, but the connection field equations turn out to be \emph{differential} rather than algebraic, so the connection $\Gamma^{\alpha}_{\mu\nu}$ becomes a propagating degree of freedom of the theory. This entails relevant new aspects in the physical content of the theory \cite{Helpin:2019kcq}. Even the fact of whether the SGB model remains a subclass of Horndeski theory in the Palatini formalism is an unexplored issue in the literature to the best of our knowledge.\

Motivated by the negative results regarding the SGB model and the referenced works involving the speed of GWs in the metric formalism 
we propose a quintessence model with the SGB action assuming Weyl geometry, and determine the behaviour of the $\alpha_T$ parameter. A previous analysis of EGB gravity with a Weyl connection in higher dimensions was carried out in Ref.~\cite{BeltranJimenez:2014iie}, but the present work considers a field-dependent coupling function $\xi(\phi)$ instead. On top of this, we find that the very same homogeneous and tensor perturbation equations derived here apply to another connection which, as opposed to the Weyl connection, has non-vanishing torsion and is metric compatible. The role the projective transformations play to achieve this is discussed in Sec.~\ref{sec:projective_trans_and_torsion} of the \textbf{Appendix}.\  

Once the equations are derived, we perform a dynamical systems analysis assuming an exponential potential, in line with Ref.~\cite{TerenteDiaz:2023iqk}. This potential is very common in quintessence scenarios \cite{Copeland:1997et,Ferreira:1997hj,Dimopoulos:2022rdp,Barreiro:1999zs,Copeland:2006wr,Tsujikawa:2010sc}. For an exponential coupling function $\xi(\phi)$, the fixed points are calculated, those being relevant from the cosmological standpoint. We analyse their stability and compute the $\alpha_T$ parameter along trajectories that reproduce the observationally constrained values of the density parameter of matter and the effective EoS parameter of the DE fluid at the present time. While in the SGB model, Eq.~\eqref{eq:cgweqc-metric} must be satisfied in order for $\alpha_T$ to vanish, here a different equation is obtained and its stability around the de Sitter and scaling regimes is examined.\

Natural units for which $M_{\textrm{Pl}} \equiv (8\pi G)^{-1/2}$, $M_{\textrm{Pl}}=2.43\times 10^{18}\,$GeV being the reduced Planck mass, have been assumed in this work. $G$ is Newton's gravitational constant.

\section{Theoretical Framework}
\label{sec:theoretical_framework_ref}
\subsection{Quadratic Gravity in the Palatini Formalism}
In the metric formalism of gravity, the connection is set to be
\begin{equation}\label{eq:LC-conn}\mathring{\Gamma}^{\alpha}_{\mu\nu} \equiv \frac{1}{2} g^{\alpha\beta} \left(\partial_{\mu} g_{\beta\nu} + \partial_{\nu} g_{\beta\mu} - \partial_{\beta} g_{\mu\nu} \right),\end{equation} which is the well-known `Levi-Civita' (LC) connection \cite{Carroll:1997ar}.\footnote{Rings (overcircles) are used to denote geometric quantities defined with respect to the LC connection and its first derivatives. We shall adopt this convention hereon, unless otherwise stated.}\ 
It is torsion free and metric compatible. Consequently \begin{align}\label{eqqq:eqqq2.2}&\mathring{\Gamma}^{\alpha}_{\mu\nu} = \mathring{\Gamma}^{\alpha}_{\nu\mu}~,\\
\label{eqqq:eqqq2.3}&\mathring{\nabla}_{\alpha} g_{\mu\nu} = 0~,\end{align} respectively, where $\mathring{\nabla}_{\mu}$ is the covariant derivative induced by the LC connection. 

In the Palatini formalism of gravity however, the connection $\Gamma^{\alpha}_{\mu\nu}(x)$ does not depend on the metric tensor $g_{\mu\nu}(x)$. The dynamics of $\Gamma^{\alpha}_{\mu\nu}$ is governed by appropriate field equations. While $\Gamma^{\alpha}_{\mu\nu}$ does not necessarily fulfill Eqs.~\eqref{eqqq:eqqq2.2} and \eqref{eqqq:eqqq2.3}, one can relate it to the LC connection by \begin{equation}\label{eq:conn-generic}\tensor{\kappa}{_{\mu\nu}^{\alpha}}\equiv \Gamma^{\alpha}_{\mu\nu} - \mathring{\Gamma}^{\alpha}_{\mu\nu}~.\end{equation} Since $\tensor{\kappa}{_{\mu\nu}^{\alpha}}$ is the difference of two connections, it transforms as a tensor under coordinate transformations. This tensor is called the `distortion tensor' \cite{Bahamonde:2021gfp}. By defining the torsion and non-metricity tensors as \begin{align}&\tensor{T}{_{\mu\nu}^{\alpha}} \equiv \Gamma^{\alpha}_{\left[\mu\nu\right]} = \Gamma^{\alpha}_{\mu\nu} - \Gamma^{\alpha}_{\nu\mu}~,\\
&Q_{\alpha\mu\nu} \equiv \nabla_{\alpha} g_{\mu\nu}~,\end{align} respectively, we find the following equations: \begin{align}\label{eq:T-rel-kappa}&\tensor{T}{_{\mu\nu}^{\alpha}} = \tensor{\kappa}{_{\mu\nu}^{\alpha}} - \tensor{\kappa}{_{\nu\mu}^{\alpha}},\\
\label{eq:Q-rel-kappa}&Q_{\alpha\mu\nu} = -\kappa_{\alpha\mu\nu} - \kappa_{\alpha\nu\mu}~,\end{align} where indices are raised and lowered with respect to the metric tensor $g_{\mu\nu}$. We see that $\tensor{T}{_{\mu\nu}^{\alpha}}$ is antisymmetric in its first two indices, while $Q_{\alpha\mu\nu}$ is symmetric in the last two.\ 

In this work, we consider $\tensor{\kappa}{_{\mu\nu}^{\alpha}}$ of the form \begin{equation}\label{weyl-connection-consideration} \tensor{\kappa}{_{\mu\nu}^{\alpha}} = \delta_{\mu}^{\alpha} A_{\nu} + \delta_{\nu}^{\alpha} A_{\mu} - g_{\mu\nu} A^{\alpha}.\end{equation} $A_{\mu}$ in the above expression is a vector field that we call `Weyl vector' and $\Gamma^{\alpha}_{\mu\nu}$ with that distortion tensor is known as the `Weyl connection' \cite{BeltranJimenez:2014iie}. Then, the corresponding torsion and non-metricity tensors, using Eqs.~\eqref{eq:T-rel-kappa} and \eqref{eq:Q-rel-kappa}, are respectively, $\tensor{T}{^{\alpha}_{\mu\nu}} = 0$ and \begin{equation}\label{eqqq:__QQ2.10} Q_{\alpha\mu\nu} = -2g_{\mu\nu} A_{\alpha}~.\end{equation} Weyl connection then has non-vanishing non-metricity tensor. A more general connection is regarded in App.~\ref{app:I}.\

Given the GB term in the metric formalism $\mathring{\mathcal{G}}$ \begin{equation}\label{eq:GB-term-again} \mathring{\mathcal{G}} \equiv \mathring{R}^2-4\mathring{R}^{\mu\nu} \mathring{R}_{\mu\nu} + \mathring{R}^{\alpha\beta\mu\nu} \mathring{R}_{\alpha\beta\mu\nu}~,\end{equation} one might regard a similar collection of quadratic curvature scalars as the GB term in the Palatini formalism. However, given the lack of symmetries in the latter formalism (see App.~\ref{app:I}), one must take into account that the most general combination of quadratic terms is more complicated. In fact, such a combination can be written as \cite{Borunda:2008kf} \begin{align}\nonumber&\mathcal{G} = \alpha R^2 +R^{\mu\nu}\left(\beta_1 R_{\mu\nu} +\beta_2 R_{\nu\mu}\right) +\tilde{R}^{\mu\nu}\left( \beta_3 R_{\mu\nu} +\beta_4 R_{\nu\mu} +\beta_5 \tilde{R}_{\mu\nu} +\beta_6\tilde{R}_{\nu\mu}\right) +\bar{R}^{\mu\nu}\left(\beta_7 R_{\mu\nu} +\right.\\
\nonumber&\left.+\beta_8 \tilde{R}_{\mu\nu} +\beta_9 \bar{R}_{\mu\nu}\right)+R^{\alpha\beta\mu\nu}\left(\gamma_1R_{\alpha\beta\mu\nu} + \gamma_2 R_{\beta\alpha\mu\nu} +\gamma_3 R_{\mu\nu\alpha\beta}+\gamma_4 R_{\alpha\mu\nu\beta} +\gamma_5 R_{\mu\alpha\nu\beta} +\right.\\
\label{eq:G-Palatini}&\left.+\gamma_6 R_{\nu\beta\mu\alpha} + \gamma_7 R_{\beta\nu\mu\alpha}\right),\end{align} 
where $\tilde R_{\mu\nu}$ and $\bar R_{\mu\nu}$ are the co-Ricci tensor and the homothetic curvature tensor,
defined in Eqs.~\eqref{eq:coRicci} and \eqref{eq:homothetic} respectively.\footnote{We refer the reader to Ref.~\cite{BeltranJimenez:2014iie}, where `co-Ricci' and `homothetic' are used to name the respective tensors as well.}  In the above, the constant coefficients $\alpha$, $\left\{\beta_i\right\}_{i=1,...,6}$ and $\left\{\gamma_i\right\}_{i=1,...,7}$ are arbitrary. In order to recover Eq.~\eqref{eq:GB-term-again} when $\tensor{\kappa}{_{\mu\nu}^{\alpha}} = 0$ (i.e. $\Gamma^{\alpha}_{\mu\nu} = \mathring{\Gamma}^{\alpha}_{\mu\nu}$), we set $(\alpha,\beta,\gamma) = (1,-4,1)$, where \begin{align}&\beta \equiv \sum^{6}_{i=1} \beta_i~,\\
&\gamma \equiv \gamma_1 - \gamma_2 + \gamma_3~.\end{align} The rest of $\gamma$ coefficients are chosen such that $\gamma_4 = \gamma_5 = \gamma_6 = \gamma_7$ \cite{Borunda:2008kf}. Since $\mathring{\bar{R}} = 0$, there is no need to fix $\beta_7$, $\beta_8$ and $\beta_9$.\ 

With that parametrisation, $\mathcal{G}$ becomes the Lagrangian of quadratic terms that we use in the action of the SGB model in the Palatini formalism, which reads \begin{equation}\label{eq:action-GB-metric}S = \int \textrm{d}^4x \sqrt{-g} \left[\frac{M^2_{\textrm{Pl}}}{2}R-\frac{1}{2}\xi(\phi)\mathcal{G}-\frac{1}{2} g^{\mu\nu} \partial_{\mu} \phi \partial_{\nu} \phi -V(\phi)\right].\end{equation} Also, $R = \mathring{R} + \kappa$ (see Eq.~\eqref{eq:eqqq-def-kappa-scalar} in App.~\ref{app:I}). $V(\phi)$ is the scalar field potential and $\xi(\phi)$ the GB coupling function. 

\subsection{Weyl-Gauss-Bonnet Action}\label{WGM}

Using the tensors defined in Eqs.~\eqref{eqqq:eqqq2.13}-\eqref{eqqq:eqqq2.18}, their symmetry properties that were shown in App.~\ref{app:I}, and the scalar quantity $\kappa$ defined in Eq.~\eqref{eq:eqqq-def-kappa-scalar}, we can write $\mathcal{G}$ as \begin{align}\label{eqqq:eqqqq2.32}\mathcal{G} = \mathring{\mathcal{G}} + 2\mathring{R} \kappa - 8 \mathring{R}^{\mu\nu}\tilde{\kappa}_{\mu\nu} + 2\mathring{R}^{\alpha\beta\mu\nu} \bar{\kappa}_{\alpha\beta\mu\nu} + \kappa^2 - 4\tilde{\kappa}^{\mu\nu} \tilde{\kappa}_{\mu\nu} + \bar{\kappa}^{\alpha\beta\mu\nu} \bar{\kappa}_{\alpha\beta\mu\nu} -\frac{1}{4} \Upsilon \bar{\kappa}^{\mu\nu} \bar{\kappa}_{\mu\nu}~,\end{align} where $\mathring{\mathcal{G}}$ is defined in Eq.~\eqref{eq:GB-term-again} and $\Upsilon$ is given by \begin{align}&\Upsilon \equiv 1-\beta_1 +\beta_2 -2\beta_7 -4\beta_9-2\gamma_1 +\gamma_4~.\end{align} As one can see, $\Upsilon$ depends on the arbitrary coefficients, whereas the rest of the terms in Eq.~\eqref{eqqq:eqqqq2.32} are oblivious to them.\ 

In terms of the Weyl vector and its derivatives, $\mathcal{G}$ reads \begin{align}\nonumber&\mathcal{G}  = \mathring{\mathcal{G}} +8\left[\left(\mathring{G}^{\mu\nu}-\mathring{\nabla}^{\nu}A^{\mu} \right)\mathring{\nabla}_{\mu} A_{\nu} -\left(\mathring{R}^{\mu\nu} -2\mathring{\nabla}^{\mu} A^{\nu} \right) A_{\mu} A_{\nu}+\left(\mathring{\nabla}_{\sigma} A^{\sigma} + A_{\sigma} A^{\sigma} \right) \mathring{\nabla}_{\rho} A^{\rho}-\right.\\
\label{eq:G-Weyl}&\left.-\Upsilon \mathring{\nabla}^{\mu} A^{\nu} \left(\mathring{\nabla}_{\mu} A_{\nu} - \mathring{\nabla}_{\nu} A_{\mu} \right) \right].\end{align} Plugging this $\mathcal{G}$ into Eq.~\eqref{eq:action-GB-metric}, we obtain the action of what we call the `Weyl-Gauss-Bonnet' (WGB) model, \begin{align}\nonumber&S_{\textrm{WGB}}= \frac{1}{2} \int \textrm{d}^4x \sqrt{-g} \left[M^2_{\textrm{Pl}}\mathring{R} - \xi(\phi) \mathring{\mathcal{G}} \right] -4\int\textrm{d}^4x \sqrt{-g}\left\{\frac{3}{4}M^2_{\textrm{Pl}}A_{\sigma} A^{\sigma}-\left(\mathring{G}^{\mu\nu}-\mathring{\nabla}^{\mu} A^{\nu}\right) A_{\mu}\partial_{\nu} \xi -\right.\\
\nonumber&\left.-\left(\mathring{\nabla}_{\sigma} A^{\sigma} + A_{\sigma} A^{\sigma} \right) \partial_{\rho} \xi A^{\rho}-\xi(\phi) \Upsilon\mathring{\nabla}^{\mu} A^{\nu}\left(\partial_{\mu} A_{\nu} - \partial_{\nu} A_{\mu} \right)\right\}- \int \textrm{d}^4x \sqrt{-g} \left[\frac{1}{2}g^{\mu\nu} \partial_{\mu} \phi \partial_{\nu} \phi +V(\phi) \right]+\\
\label{eq:full-actionv2}&+\int \textrm{d}^4x \sqrt{-g} \mathcal{L}_{\textrm{M}}(g_{\mu\nu},\Psi)~,\end{align} where we have integrated by parts and used some well-known properties of the curvature tensors in the metric formalism \cite{Carroll:1997ar}. We drop the rings from now on as it is clear that every geometric quantity is defined with respect to the LC connection and its derivatives once the variables depending on the distortion tensor have been rewritten in terms of the Weyl vector and its covariant derivative. Notice also that we have included the matter action $S_{\textrm{M}}$, \begin{equation}\label{eqqq:eqqqq2.35}S_{\textrm{M}}[g_{\mu\nu}, \Psi] = \int \textrm{d}^4x \sqrt{-g} \mathcal{L}_{\textrm{M}}(g_{\mu\nu},\Psi)~,\end{equation} given our interest in the late Universe. `$\Psi$' is used to denote collectively the matter fields and $\mathcal{L}_{\textrm{M}}$ is the corresponding Lagrangian. As we see, the matter fields are minimally coupled to gravity and the action $S_{\textrm{M}}$ is \emph{not} a functional of the connection, only of the metric tensor. This choice is made for simplicity, although one might explore the possibility of matter fields coupled to the connection. In that case, besides the energy-momentum tensor, one obtains the so-called `hypermomentum tensor' from the variation of the action with respect to the connection \cite{Clifton:2011jh}.

\subsection{Homogeneous Field Equations}
We now present the homogeneous equations of the WGB model. The field equation of the vector field $A_{\mu}$, derived from Eq.~\eqref{eq:full-actionv2}, is (remember that rings are dropped for simplicity) \begin{align}\label{eq:A-full}&\frac{3}{2} \left(M^2_{\textrm{Pl}} -\frac{4}{3} \partial_{\nu} \xi A^{\nu} \right) A_{\mu} = \left(\tensor{G}{_{\mu}^{\nu}}-2\nabla_{\mu} A^{\nu} \right) \partial_{\nu} \xi +\left(2\nabla_{\sigma} A^{\sigma} + A_{\sigma} A^{\sigma} \right) \partial_{\mu} \xi  +\frac{1}{2} \Upsilon \nabla_{\nu} \left(\xi \tensor{\bar{\kappa}}{_{\mu}^{\nu}}\right).\end{align} A vector field of the form $A_{\mu}(t) \equiv (-\Phi(t),\mathbf{0})$ is considered. That is, we take the spatial components of the vector field to vanish at the background level. This is required to keep the expansion of the Universe isotropic (see App.~\ref{app:II}). Such a choice implies $\bar{\kappa}_{\mu\nu}=0$ in the background. Consequently, $\Upsilon$ does not play any role in the dynamics of $\Phi$. In particular, the homogeneous part of Eq.~\eqref{eq:A-full} reads \begin{equation}\label{eq:eq-Phi} \Phi = 2\frac{\dot \xi}{M^2_{\textrm{Pl}}} \left(H-\Phi\right)^2.\end{equation} For $\dot \xi \neq 0$, one can write this equation as \begin{equation}\label{eq:eq-Phi-v1} \Phi^2 -2\Phi \left(H+ \frac{M^2_{\textrm{Pl}}}{4\dot \xi} \right) + H^2 = 0~.\end{equation}

It can be noticed from Eq.~\eqref{eq:eq-Phi} that $\Phi = 0$ for $\dot \xi = 0$ (constant $\xi$), and the LC connection is recovered at the background level. Another possibility that leads to $\Phi = 0$ is $H=\Phi$ (see Eq.~\eqref{eq:eq-Phi}), but this implies $H=0$, which is not a valid solution when we regard a time dependent scale factor.\

Eq.~\eqref{eq:eq-Phi-v1} is an algebraic equation that can be readily solved for $\dot \xi \neq 0$ \begin{equation}\label{eq:Phi-sol}\Phi = H\left[1+ \frac{1}{4 \mu} \left(1\pm \sqrt{1+8\mu}\right)\right],\end{equation} where $\mu$ is defined as \begin{equation}\label{eq:def-mu-dimensionless}\mu \equiv \frac{\dot \xi H}{M_{\textrm{Pl}}^2}~.\end{equation} Notice that $\mu$ cannot be smaller than $-1/8$, otherwise $\Phi$ becomes complex. Since we shall consider expanding universes ($H>0$), this bound has to be taken into account if $\dot \xi$ turns negative.\

The \emph{Weyl parameter} $W\equiv H-\Phi$ is introduced, such that \begin{equation}\label{eq:ratio-W-H} \frac{W}{H} = -\frac{1}{4\mu} \left(1-\sqrt{1+8\mu} \right),\end{equation} where the positive root in Eq.~\eqref{eq:Phi-sol} has been discarded. The reason for this is that if one takes $\dot \xi = 0$ in Eq.~\eqref{eq:eq-Phi}, then $\Phi = 0$. But only the negative root in Eq.~\eqref{eq:Phi-sol} yields $\Phi\simeq 0$ (i.e. $W\simeq H$) when $|\mu| \ll 1$. On the other hand, $|\mu| \gg 1$ in Eq.~\eqref{eq:ratio-W-H} leads to \begin{equation} \frac{W}{H} \simeq \frac{1}{\sqrt{2\mu}}~,\end{equation} and hence $W\ll H$. For $W$ to be real, $\mu>0$ when $|\mu| \gg 1$ as was pointed out above in relation to $\Phi$.\ 

Later we will make use of the inverse relation to Eq.~\eqref{eq:ratio-W-H}. Using Eq.~\eqref{eq:eq-Phi}, we can write the $H(W,\dot \xi)$ function as \begin{equation}\label{eq:H-func-of-W} H = W+\Phi = W\left(1+2\frac{\dot \xi W}{M^2_{\textrm{Pl}}}\right).\end{equation}    

The homogeneous field equations of the metric and the scalar field are given by (see App.~\ref{app:II}) \begin{align}\label{eq:00-v2}&H^2 = \frac{1}{3M^2_{\textrm{Pl}}}\left[\frac{1}{2} \dot \phi^2+V(\phi)+\rho_{\textrm{M}}+ 12\dot \xi W^3\right] + \Phi^2,\\
\label{eq:ij-00-v2}&\dot H = -\frac{1}{2M^2_{\textrm{Pl}}} \left[\dot \phi^2+\rho_{\textrm{M}}+P_{\textrm{M}}+4\dot \xi W^3 -4\left(\dot \xi W^2\right)^{\bullet}-12\dot \xi \Phi W^2\right]-3\Phi^2,\\
\label{eq:sf-eq}&\ddot \phi + 3H \dot \phi + V_{,\phi} = - 12\xi_{,\phi}W^2 \left(\dot W+ H W\right).\end{align} $\rho_{\textrm{M}}$ and $P_{\textrm{M}}$ are the homogeneous energy density and isotropic pressure of the matter fields, respectively. The energy density satisfies the following continuity equation: \begin{equation}\label{eq:3.26eq} \dot \rho_{\textrm{M}} = -3H\rho_{\textrm{M}} \left(1+w_{\textrm{M}}\right),\end{equation} where $w_{\textrm{M}}\equiv P_{\textrm{M}}/\rho_{\textrm{M}}$ is the EoS parameter of the background fluid under consideration, which we will assume to lie in the range $0\leq w_{\textrm{M}}<1$. This range includes the values corresponding to pressureless matter ($w_{\textrm{M}}=w_m = 0$) and radiation ($w_{\textrm{M}}=w_r = 1/3$).\ 

If $\Phi = 0$, Eqs.~\eqref{eq:00-v2}-\eqref{eq:sf-eq} are the same as those of the SGB model \cite{Odintsov:2020zkl}\begin{align}\label{eq:Friedmann-GB-metric}&H^2 = \frac{1}{3M^2_{\textrm{Pl}}} \left[\frac{1}{2} \dot \phi^2+V(\phi)+\rho_{\textrm{M}}+ 12\dot \xi H^3\right],\\
\label{eq:dot-H-eq-GB-metric}&\dot H = -\frac{1}{2M^2_{\textrm{Pl}}}\left[\dot \phi^2+\rho_{\textrm{M}}+P_{\textrm{M}}+4\dot \xi H^3 -4\left(\dot \xi H^2 \right)^{\bullet} \right],\\
\label{eq:eq-sf-metric}&\ddot \phi + 3H \dot \phi + V_{,\phi} = -12\xi_{,\phi} H^2 \left(\dot H + H^2 \right).\end{align} This is reasonable as we have not made use of the background equation of $\Phi$, Eq.~\eqref{eq:eq-Phi}, yet. Now, using Eq.~\eqref{eq:eq-Phi}, Eqs.~\eqref{eq:00-v2}-\eqref{eq:sf-eq} take on the form \begin{align}\label{eq:eq-W2-matter}&W^2 = \frac{1}{3M^2_{\textrm{Pl}}}\left[\frac{1}{2} \dot \phi^2+V(\phi)+\rho_{\textrm{M}}\right],\\
\label{eq:eq-dotW-matter}&\dot W + \Phi W = -\frac{1}{2M^2_{\textrm{Pl}}}\left(\dot \phi^2+\rho_{\textrm{M}}+P_{\textrm{M}} \right),\\
\label{eqqq:eqqqq3.18}&\ddot \phi + 3W\dot \phi + V_{,\phi} = -12\xi_{,\phi} W^2 \left(W^2 -\rho_{\textrm{M}}\frac{1+w_{\textrm{M}}}{2M^2_{\textrm{Pl}}}\right).\end{align} For $\Phi = 0$, these equations reduce to  \begin{align}\label{eq:limit-GR1}&H^2 = \frac{1}{3M^2_{\textrm{Pl}}}\left[\frac{1}{2} \dot \phi^2+V(\phi)+\rho_{\textrm{M}}\right],\\
\label{eq:limit-GR2}&\dot H = -\frac{1}{2M^2_{\textrm{Pl}}}\left(\dot \phi^2+\rho_{\textrm{M}}+P_{\textrm{M}}\right),\\
&\ddot \phi +3H \dot \phi + V_{,\phi} = 0~.\end{align} We recover the equations of GR above because $\Phi = 0$ implies $\dot \xi = 0$ (unless $H=0$) by virtue of Eq.~\eqref{eq:eq-Phi}.

\subsection{Equations for Tensor Perturbations}
\label{sec:speed-GWs}
We derived the homogeneous equations and solved the only non-trivial equation \eqref{eq:eq-Phi} that governs the evolution of $A_{\mu}$. Now, we perturb about the flat FLRW metric and consider tensor perturbations \begin{equation}\textrm{d}s^2 =-\textrm{d}t^2 +a^2(t)\left[\delta_{ij} + h_{ij}(t,\mathbf{x}) \right] \textrm{d}x^{i} \textrm{d}x^{j},\end{equation} where $h_{ij}(t,\mathbf{x})$ are the traceless-transverse tensor components, such that $\tensor{h}{^{i}_{i}} = 0 = \partial_{i} \tensor{h}{^{i}_{j}}$. Perturbing the action \eqref{eq:full-actionv2} up to quadratic terms, we derive the equation of tensor perturbations\footnote{We assume there is no contribution from the matter action in the form of an anisotropic stress tensor.} \begin{equation}\label{eqqqq:eqqqqq3.23}\left(M^2_{\textrm{Pl}}-4\dot \xi W\right) \ddot{h_{ij}} +\left[3H\left(M^2_{\textrm{Pl}} -4\dot \xi  W\right)-4\left(\ddot \xi W +\dot \xi \dot W\right)\right]\dot{h_{ij}}-\left[M^2_{\textrm{Pl}}-4\left(\ddot \xi +\dot \xi \Phi \right) \right] a^{-2} \partial_k \partial^{k} h_{ij} = 0~,\end{equation} where $\partial_k \partial^{k} \equiv \delta^{kl} \partial_k \partial_l$. \emph{Assuming that} $4\dot \xi W \neq M^2_{\textrm{Pl}}$, the equation may be written as \begin{equation}\ddot{h_{ij}} + \left(3+\alpha_{M} \right) H \dot{h_{ij}}-\left(1+\alpha_T\right) a^{-2} \partial_k \partial^{k} h_{ij} = 0~,\end{equation} where $\alpha_M$ and $\alpha_T$ are given by \begin{align}&\alpha_M \equiv -\frac{4}{H} \frac{\ddot \xi W+ \dot \xi \dot W}{M^2_{\textrm{Pl}}-4\dot \xi W}~,\\
\label{eq:alphaT-wgb-eq}&\alpha_T \equiv -4\frac{\ddot \xi - \left(H-2\Phi \right)\dot \xi}{M^2_{\textrm{Pl}}-4\dot \xi W}~.\end{align} The speed of GWs predicted by the WGB model is then given by $c_{\textrm{GW}}^2 = 1+ \alpha_T$, with $\alpha_T$ of Eq.~\eqref{eq:alphaT-wgb-eq}. Using Eq.~\eqref{eq:eq-Phi}, we may write the $\alpha_T$ parameter entirely in terms of $W$, $\dot \xi$ and $\ddot \xi$ \begin{equation}\label{eq:alphaT-wgb-eqv2} \alpha_T = -\frac{4}{M^2_{\textrm{Pl}}}\frac{M^2_{\textrm{Pl}}\left(\ddot \xi -\dot \xi W\right) +2\dot \xi^2 W^2}{M^2_{\textrm{Pl}}-4\dot \xi W}~.\end{equation} 

\section{Dynamical Systems Analysis}
\label{sec:dynamical-systems-analysis-full}
\subsection{Dynamical System}
Given the homogeneous field equations \eqref{eq:eq-W2-matter}-\eqref{eqqq:eqqqq3.18}, we analyse the generic behaviour of the dynamical system. We find the corresponding fixed points and determine their stability keeping in mind the known history of the Universe. To that end, the following dimensionless variables are defined: \begin{align}\label{eqqq:eqqq4.3}&x \equiv \frac{\dot \phi}{\sqrt{6} WM_{\textrm{Pl}}}~,~\ ~y \equiv \frac{1}{WM_{\textrm{Pl}}}\sqrt{\frac{V}{3}}~,~\ ~z \equiv \frac{1}{WM_{\textrm{Pl}}}\sqrt{\frac{\rho_{\mathrm{M}}}{3}}~,\end{align} such that Eq.~\eqref{eq:eq-W2-matter}, which is the analogue of the Friedmann equation in GR, can be written as \begin{equation}\label{eq:constraint-eq-v1-xyz}1=x^2+y^2+z^2.\end{equation} It is noticed that, in contrast to the definitions of $x$, $y$ and $z$ in Ref.~\cite{TerenteDiaz:2023iqk}, these are given in terms of the Weyl parameter, $W$, not the Hubble one. $W$ and $H$ are related by Eq.~\eqref{eq:H-func-of-W}. Using the dimensionless variables, this relation can be written as \begin{equation}\label{eqqq:eqqqq4.4}H = W\left(1+ux\right),\end{equation} where another dimensionless variable, \begin{equation}\label{eq:def-u-wgb-model} u \equiv 2\sqrt{6} \frac{\xi_{,\phi} W^2}{M_{\textrm{Pl}}}~,\end{equation} was defined.\ 

A second difference between those definitions and the ones discussed in Ref.~\cite{TerenteDiaz:2023iqk} is that $u$ is not constrained by Eq.~\eqref{eq:constraint-eq-v1-xyz}. The reason for this independence is related to the fact that the modifications of gravity are included in the definitions of $x$, $y$ and $z$ in Eq.~\eqref{eqqq:eqqq4.3} via the Weyl parameter. In Ref.~\cite{TerenteDiaz:2023iqk}, on the other hand, the effects of modified gravity are accounted for solely by the $u$ variable (see Eq.~(17) in that paper).\

With the definitions of the dimensionless variables above, we can write the density parameter of the background fluid $\Omega_{\textrm{M}}$ as \begin{equation}\label{eqqq:eqqqq4.86} \Omega_{\textrm{M}}\equiv \frac{\rho_{\textrm{M}}}{3H^2 M^2_{\textrm{Pl}}} = \frac{1-x^2-y^2}{(1+ux)^2}~,\end{equation} where Eqs.~\eqref{eq:constraint-eq-v1-xyz} and \eqref{eqqq:eqqqq4.4} were used.\

Using those very same equations, we define the density parameter of the scalar field $\phi$ too and write it in terms of the dimensionless variables \eqref{eqqq:eqqq4.3} and \eqref{eq:def-u-wgb-model} \begin{equation}\label{eqqq:eqqqq4.55}\Omega_{\phi} \equiv \frac{\frac{1}{2} \dot \phi^2 + V}{3H^2 M_{\textrm{Pl}}^2} = \frac{x^2+y^2}{(1+ux)^2}~.\end{equation} Finally,  the `effective' density parameter associated with $\xi$ is \begin{equation}\label{eq:constraint-eq-Friedmannv2} \Omega_{\xi} \equiv \frac{(2+ux)ux}{(1+ux)^2}~,\end{equation} such that $1=\Omega_{\phi} + \Omega_{\textrm{M}} + \Omega_{\xi}$.\ 

Assuming an exponential potential \begin{equation}\label{eq:Vphi}
V(\phi) = V_0 e^{-\lambda\phi/M_{\textrm{Pl}}},\end{equation} such that $V_{,\phi} = -\frac{\lambda}{M_{\textrm{Pl}}} V$ and $\lambda >0$ is a non-zero constant, the dynamical equations become self-similar given the definitions in Eqs.~\eqref{eqqq:eqqq4.3} and \eqref{eq:def-u-wgb-model}. This class of potentials gives rise to scaling solutions and they are ubiquitous in quintessence models (see e.g. Ref.~\cite{vandeBruck:2017voa}, where the exponential potential is the large field limit of the one proposed therein). Then, taking the derivatives of $x$, $y$, $z$ and $u$, and using Eq.~\eqref{eq:sf-eq}, the scalar field equation, the continuity equation \eqref{eq:3.26eq}, and Eq.~\eqref{eqqq:eqqqq4.4}, we arrive at \begin{align}&\label{eq:dx-eq-dimensionless} x' = -\frac{1}{1+ux} \left[(3x+u)(1+ux)-\sqrt{\frac{3}{2}}\lambda y^2 +(x+u)\frac{\dot W}{W^2} \right],\\
\label{eq:dy-eq-dimensionless}&y' =-\frac{y}{1+ux} \left(\frac{\dot W}{W^2} + \sqrt{\frac{3}{2}}\lambda x\right),\\
\label{eq:dz-eq-dimensionless}&z' =-z\left[\frac{\dot W}{W^2}\frac{1}{1+ux} +\frac{3}{2} \left(1+w_{\mathrm{M}}\right) \right],\\
&\label{eq:eq-u-no-assump-xi-coupling}u' = \frac{2}{1+ux} \left(u\frac{\dot W}{W^2} +6\xi_{,\phi\phi} W^2 x\right).\end{align} $\dot W/W^2$ can be expressed in terms of the dimensionless variables using Eq.~\eqref{eq:eq-dotW-matter}. Plugging Eq.~\eqref{eq:eq-Phi} into Eq.~\eqref{eq:eq-dotW-matter} we find \begin{equation}\label{eqq:4.22} -\frac{\dot W}{W^2} = x\left(3x+u\right) +\frac{3}{2} z^2 \left(1+w_{\textrm{M}}\right).\end{equation} Primes denote derivatives with respect to the number of elapsing e-folds, which we choose to be $\textrm{d}N \equiv \textrm{d} \ln a = H\textrm{d}t$.\

$\dot W/W^2$ can be related to the Hubble flow parameter $\epsilon_H$, which is defined as \cite{Baumann:2009ds} \begin{equation}\label{eq:def-Hubble-flow-param-1H} \epsilon_H \equiv -\frac{\dot H}{H^2}~.\end{equation} $\epsilon_H$ parametrises the acceleration of spatial slices. Using Eq.~\eqref{eq:H-func-of-W} we find\begin{equation}\label{eq:rel-epsilonH-dotWW2} \epsilon_H = -\frac{1}{1+ux} \left[\frac{\dot W}{W^2} +\frac{(ux)^{\bullet}}{H} \right].\end{equation}

Notice that no assumption regarding $\xi(\phi)$ has been made so far. All the previous equations are valid for any coupling function, including the exponential one \begin{equation}\label{eqqq:eqqqq4.11} \xi(\phi) = \xi_0 e^{\kappa \phi/M_{\textrm{Pl}}},\end{equation} such that $\xi_{,\phi\phi} = \frac{\kappa}{M_{\textrm{Pl}}}\xi_{,\phi}$, where $\kappa>0$ is a non-zero constant.\footnote{The strength of the exponential $\kappa$ in Eq.~\eqref{eqqq:eqqqq4.11} should not be confused with the scalar $\kappa$ defined in Eq.~\eqref{eq:eqqq-def-kappa-scalar}, discussed in Sec.~\ref{WGM} and App.~\ref{app:I}.} Assuming such a coupling function, Eq.~\eqref{eq:eq-u-no-assump-xi-coupling} is a self-similar equation too and the explicit dependence on $W$ drops out of the equation\footnote{For a linear coupling function $\xi(\phi) \propto \phi$, the explicit dependence on the Weyl parameter $W$ also disappears from Eq.~\eqref{eq:eq-u-no-assump-xi-coupling}. However, as we will argue later on (see footnote~\ref{footnote:7-seven}), the linear coupling case can be discarded on the basis of our knowledge of the past history of the Universe.} \begin{equation}\label{eq:du-eq-dimensionless}u' = \frac{2u}{1+ux} \left(\frac{\dot W}{W^2} +\sqrt{\frac{3}{2}}\kappa x\right).\end{equation}

\subsection{Fixed Points} 
\label{sec:fixed-points-section}
Fixed points of the dynamical system of equations \eqref{eq:dx-eq-dimensionless}, \eqref{eq:dy-eq-dimensionless} and \eqref{eq:du-eq-dimensionless} are denoted by $x_c$, $y_c$ and $u_c$, respectively, and correspond to regions of the phase space where $x' = y' = u' = 0$. Among those, we can list the following ones: \begin{itemize} \item \textbf{M}: $(x_c,y_c,u_c) = (0,0,0)$ and $\left.\epsilon_H\right|_c = \frac{3}{2}\left(1+w_{\textrm{M}}\right)$;
\item \textbf{K}$\pm$: $(x_c,y_c,u_c) = (\pm 1, 0, 0)$ and $\left.\epsilon_H\right|_c = 3$;
\item \textbf{I}: $(x_c,y_c,u_c) = \left(\frac{\lambda}{\sqrt{6}},\sqrt{1-\frac{\lambda^2}{6}},0\right)$ and $\left.\epsilon_H\right|_c = \frac{\lambda^2}{2}$;
\item \textbf{ScI}: $(x_c,y_c,u_c) = \left(\sqrt{\frac{3}{2}\frac{\left(1+w_{\textrm{M}}\right)^2}{\lambda^2}},\sqrt{\frac{3}{2}\frac{1-w^2_{\textrm{M}}}{\lambda^2}},0\right)$ and $\left.\epsilon_H\right|_c = \frac{3}{2}\left(1+w_{\textrm{M}}\right)$; \end{itemize} which are fixed points in GR too because $W_c = H_c$ at those fixed points (see Eq.~\eqref{eqqq:eqqqq4.4}). $z_c$ is determined from the constraint equation \eqref{eq:constraint-eq-v1-xyz}, so we do not really need to impose $z' = 0$; it follows from the constraint equation.\footnote{Notice from Eq.~\eqref{eq:dz-eq-dimensionless} that, in order for $z' = 0$, either $z_c = 0$ or $\left.\epsilon_H\right|_c = \frac{3}{2} \left(1+w_{\textrm{M}}\right)$ (see Eq.~\eqref{eq:rel-epsilonH-dotWW2}). Both are satisfied (although not simultaneously) by the fixed points above and by the ones we discuss later. Besides this, we remark that, unlike $z$, $u' = 0$ is imposed as $x'=y' = 0$ are satisfied for $u = u(N)$ and $x_c = \pm 1$ and $y_c = 0$, for example. As was argued already, this is due to the fact that the modifications of gravity are included in all the definitions of the dimensionless variables, Eqs.~\eqref{eqqq:eqqq4.3} and \eqref{eq:def-u-wgb-model}, whilst in our previous work, Ref.~\cite{TerenteDiaz:2023iqk}, these were only represented by $u$, which is defined in Eq.~(17) of that paper.}\

\textbf{M} represents the regime of the background fluid domination. \textbf{K}$\pm$ stands for the period of kination (dominance of the kinetic energy of the scalar field $\dot \phi^2/2$) and \textbf{I} corresponds to a regime of power law inflation (when $\lambda <\sqrt{2}$, so that $\left.\epsilon_H\right|_{\textrm{\textbf{I}}}<1$) where matter fields do not contribute ($z_{\textrm{\textbf{I}}}=0$; see Eq.~\eqref{eq:constraint-eq-v1-xyz}). Notice that this latter fixed point exists if $\lambda<\sqrt{6}$, so $y_{\textrm{\textbf{I}}}$ is real (remember that $\lambda >0$). Lastly, we have the scaling regime (\textbf{ScI}). In this case, the EoS parameter of the scalar field, \begin{equation} w_{\phi} \equiv \frac{\frac{1}{2} \dot \phi^2-V(\phi)}{\frac{1}{2} \dot \phi^2+V(\phi)}~,\end{equation} becomes equal to that of the background fluid, $w_{\textrm{M}}$. This fixed point yields \begin{equation}\label{eq:Omegahat_phi_ScI} x^2_{\textrm{\textbf{ScI}}} + y^2_{\textrm{\textbf{ScI}}} = \frac{3}{\lambda^2} \left(1+w_{\textrm{M}}\right),\end{equation} and, by virtue of Eq.~\eqref{eq:constraint-eq-v1-xyz}, we obtain the following existence condition: \begin{equation}\label{des:condition-existence-fixed-point-scI} \lambda > \sqrt{3(1+w_{\textrm{M}})}\geq \sqrt{3}~.\end{equation} Therefore, we observe that \textbf{I} and \textbf{ScI} coexist whenever $\sqrt{6}>\lambda > \sqrt{3}$. If \textbf{I} corresponds to power law inflation, then both fixed points cannot coexist because $\lambda <\sqrt{2}$ is required, as was pointed out already.\ 

In addition to the fixed points listed above, we find five other points: \begin{itemize}\item \textbf{dS}: $(x_c,y_c,u_c) = \left(0,1,\sqrt{\frac{3}{2}}\lambda\right)$ and $\left.\epsilon_H\right|_c = 0$;
\item \textbf{ScII}$\pm$: $(x_c,y_c,u_c) = \left(\frac{\kappa}{\sqrt{6}}\frac{1+3w_{\textrm{M}}}{3(1-w^2_{\textrm{M}})}\Delta_{\pm}(\kappa,w_{\textrm{M}}),0,\frac{\kappa}{\sqrt{6}}\frac{1}{1+w_{\textrm{M}}}\Delta_{\pm}(\kappa,w_{\textrm{M}})\right)$, where \begin{equation}\label{eq:def-Deltapm}\Delta_{\pm}(\kappa,w_{\textrm{M}}) \equiv 1\pm \sqrt{1-\frac{18}{\kappa^2} \frac{(1+w_{\textrm{M}})(1-w^2_{\textrm{M}})}{1+3w_{\textrm{M}}}}~,\end{equation} and $\left.\epsilon_H\right|_c = \frac{3}{2}\left(1+w_{\textrm{M}}\right)$;
\item \textbf{\^{K}}$\pm$: $(x_c,y_c,u_c) = \left(\pm 1, 0, \sqrt{\frac{3}{2}}\kappa \mp 3\right)$ and $\left.\epsilon_H\right|_c = \frac{\mp \frac{3\kappa}{2\sqrt{6}}}{1\mp \frac{3\kappa}{2\sqrt{6}}}$.\end{itemize} The first fixed point corresponds to the de Sitter expansion, where the total energy density consists only of the CC. This fixed point exists in GBDE models even for $\lambda \neq 0$. It is also found in the SGB model (see Refs.~\cite{TerenteDiaz:2023iqk,Koivisto:2006ai}) because $x_{\textrm{\textbf{dS}}} = 0$ and $W_{\textrm{\textbf{dS}}} = H_{\textrm{\textbf{dS}}}$ (see Eq.~\eqref{eqqq:eqqqq4.4}).\ 

Besides the de Sitter fixed point, we have a second scaling regime which consists of two fixed points (\textbf{ScII}$\pm$), depending on the root sign in Eq.~\eqref{eq:def-Deltapm}. These fixed points exist if \begin{equation}\label{bounds-on-kappa-scII} \kappa^2 \geq 18\frac{(1+w_{\textrm{M}})(1-w^2_{\textrm{M}})}{1+3w_{\textrm{M}}}~,\end{equation} so that $x_{\textrm{\textbf{ScII}$\pm$}}$ and $u_{\textrm{\textbf{ScII}$\pm$}}$ are real. Then $1\leq \Delta_{+}<2$ and $0<\Delta_{-}\leq 1$. Also, Eq.~\eqref{eq:constraint-eq-v1-xyz} demands \begin{equation}\label{eqqq:eq4.34}  x^2_{\textrm{\textbf{ScII}$\pm$}} = \frac{\kappa^2}{54}\frac{(1+3w_{\textrm{M}})^2}{(1-w^2_{\textrm{M}})^2} \Delta_{\pm}^2<1~.\end{equation} Combining it with Eq.~\eqref{bounds-on-kappa-scII} we obtain \begin{equation}\label{eqqq:eq4.35}\Delta_{\pm}^2 <\frac{3(1-w_{\textrm{M}})}{1+3w_{\textrm{M}}}~.\end{equation} Given that $1\leq \Delta_{+}^2 < 4$ and $0<\Delta_{-}^2 \leq 1$, we have $0 \leq w_{\textrm{M}} < \frac{1}{3}$ for $\Delta_{+}$ and $0 \leq w_{\textrm{M}} < 1$ for $\Delta_{-}$ (we remind the reader that the whole range of values of $w_{\textrm{M}}$ is assumed to be $0\leq w_{\textrm{M}} <1$; see below Eq.~\eqref{eq:3.26eq}). Therefore, the scaling fixed point \textbf{ScII}$+$ in particular does not exist if the background fluid under consideration corresponds to radiation for example, where $w_{\textrm{M}} = w_r = 1/3$. The scaling regime is still represented by \textbf{ScII}$-$ in that case though.\ 

Finally, the fourth and fifth fixed points listed above have vanishing potential energy and the energy density of the background fluid is zero. We denote them by `\textbf{\^{K}}$\pm$' and call the corresponding regime `pseudo-kination' since $3W^2_{\textrm{\textbf{\^{K}}}\pm}M^2_{\textrm{Pl}} = \dot \phi_{\textrm{\textbf{\^{K}}}\pm}^2/2$, although $W_{\textrm{\textbf{\^{K}}}\pm}\neq H_{\textrm{\textbf{\^{K}}}\pm}$. Irrespective of the sign, if $\kappa \gg 1$, then $\left.\epsilon_H\right|_{\textrm{\textbf{\^{K}}}\pm} \rightarrow 1$. This value of the Hubble flow parameter, for large $\kappa$, leads into a universe that does not accelerate.\

For $x_c = +1$ (i.e. \textbf{\^{K}}$+$) and $\kappa < 2\sqrt{\frac{2}{3}}$, $\left.\epsilon_H\right|_{\textrm{\textbf{\^{K}}}+}<0$ (remember that $\kappa$ is defined to be positive; see below Eq.~\eqref{eqqq:eqqqq4.11}). This implies that the Universe accelerates faster than during de Sitter, where it accelerates exponentially.\footnote{In order to see why it accelerates faster than exponential, we solve $\epsilon_H = -\alpha$, where $\alpha>0$ is constant ($\epsilon_H$ is constant at the fixed points \textbf{\^{K}}$\pm$, indeed) \begin{equation}H(t) = \frac{H_0}{1-\alpha H_0 t}~,\end{equation} where $H_0 = H(t=0)$. Solving for $a(t)$ we obtain \begin{equation}a(t) = \frac{a_0}{(1-\alpha H_0t)^{1/\alpha}}~,\end{equation} such that $a_0 = a(t=0)$. Assuming an expanding universe, $H_0 >0$ and $t<t_s \equiv 1/(\alpha H_0)$. Then, $a(t)$ diverges (approaches $a(t_s)$) faster than if $a(t) \propto e^{H_0t}$.}\ 

Lastly, choosing $\lambda = \kappa$ allows for all $x_c$, $y_c$ and $u_c$ being non-zero at once as can be seen from Eqs.~\eqref{eq:dy-eq-dimensionless} and \eqref{eq:du-eq-dimensionless}. The same occurs in the SGB model (see Ref.~\cite{TerenteDiaz:2023iqk}). The corresponding fixed `curves' are \begin{itemize} \item \textbf{ScIII}: $(x,y,u) = \left(\frac{\frac{3}{2}(1+w_{\textrm{M}})}{\sqrt{\frac{3}{2}}\lambda -\frac{3}{2} u_c(1+w_{\textrm{M}})},\sqrt{\frac{\sqrt{\frac{3}{2}}\lambda u_c +\frac{3}{2}(1+w_{\textrm{M}})\left(\frac{3}{2}(1-w_{\textrm{M}})-\sqrt{\frac{3}{2}}\lambda u_c +\frac{1}{2}(1+3w_{\textrm{M}})u_c^2\right)}{\left(\sqrt{\frac{3}{2}}\lambda -\frac{3}{2} u_c(1+w_{\textrm{M}})\right)^2}},u_c\right)$ and $\epsilon_H|_c = \frac{3}{2}(1+w_{\textrm{M}})$;
\item $\widehat{\textrm{\textbf{IV}}}$: $(x,y,u) = \left(x_c,\sqrt{1-x_c^2},-3x_c + \sqrt{\frac{3}{2}}\lambda\right)$ and $\epsilon_H|_c = \frac{\sqrt{\frac{3}{2}}\lambda x_c}{1-3x_c^2+\sqrt{\frac{3}{2}}\lambda x_c}$.
\end{itemize} We then have a third scaling solution, \textbf{ScIII}, and a regime of zero energy density of the background fluid, corresponding to the fixed `curve' $\widehat{\textrm{\textbf{IV}}}$. This curve is denoted that way following the convention in Ref.~\cite{TerenteDiaz:2023iqk} (see Table~I in that paper), although neither of those two fixed `curves' are found in the SGB model strictly because $W_c \neq H_c$, so carets are used on the latter to differentiate it from the fixed curve in the SGB model (similarly to what was done with the pseudo-kination fixed points we just discussed).\ 

Those two regimes (especially the scaling one) could be of interest, although in this work we shall consider $\kappa \neq \lambda$. We leave the stability analysis of these two fixed curves and possible solutions for future works.       

\subsection{Stability}
\label{subsubsec-stability}
To perform the stability analysis of the fixed points shown above, we perturb Eqs.~\eqref{eq:dx-eq-dimensionless}, \eqref{eq:dy-eq-dimensionless} and \eqref{eq:du-eq-dimensionless} linearly such that \begin{align}\nonumber&\left(1+u_cx_c\right) \delta x' = -\left[3-9x_c^2 -2u_cx_c -\frac{3}{2} (1+w_{\textrm{M}}) \left(1-3x^2_c-y_c^2-2u_cx_c\right) \right]\delta x+\\
\label{eq:dx-eq-dimensionless-pert}&+y_c\left[\sqrt{6} \lambda-3(1+w_{\textrm{M}})(x_c+u_c)\right]\delta y - \left[1-x_c^2 -\frac{3}{2} \left(1+w_{\textrm{M}}\right) \left(1-x_c^2-y_c^2 \right) \right] \delta u~,\end{align} \begin{align}\nonumber&(1+u_cx_c) \delta y' = -y_c \left[\sqrt{\frac{3}{2}}\lambda -u_c -3x_c(1-w_{\textrm{M}}) \right] \delta x-\left[\left(\sqrt{\frac{3}{2}}\lambda -3x_c -u_c \right)x_c-\right.\\
\label{eq:dy-eq-dimensionless-pert}&\left.-\frac{3}{2} (1+w_{\textrm{M}}) \left(1-x_c^2 -3y_c^2 \right)\right]\delta y +y_c x_c \delta u~,\end{align} \begin{align}\nonumber&\left(1+u_c x_c \right) \delta u' = 2u_c \left[\sqrt{\frac{3}{2}} \kappa - u_c -3x_c \left(1-w_{\textrm{M}} \right) \right] \delta x +6u_c y_c \left(1+w_{\textrm{M}} \right) \delta y +\\
\label{eq:du-eq-dimensionless-pert}&+2 \left[\left(\sqrt{\frac{3}{2}} \kappa-3x_c -2u_c \right) x_c -\frac{3}{2} \left(1+w_{\textrm{M}} \right) \left(1-x_c^2 - y_c^2 \right) \right] \delta u~,\end{align} where the constraint equation \eqref{eq:constraint-eq-v1-xyz} was used to replace $z^2$ in Eq.~\eqref{eqq:4.22}.\

We begin our stability analysis with the fixed points \textbf{M} and \textbf{K}$\pm$, given the relevance of the latter in particular in early stages in models of quintessential inflation. In the case of \textbf{M}, which corresponds to the stage of the background fluid domination, as was commented already, the eigenvalues, which we denote by `$m$', are the following ones: \begin{align}&m_1 = -\frac{3}{2}\left(1-w_{\textrm{M}}\right),\\
&m_2 = \frac{3}{2} \left(1+w_{\textrm{M}}\right),\\
&m_3 = -3\left(1+w_{\textrm{M}}\right).\end{align} One of the eigenvalues, $m_2$, which corresponds to the eigenvector $\mathbf{v} = (0,1,0)$,\footnote{We order eigenvector components as $\mathbf{v} = (v_x,v_y,v_u)$.} is non-negative, whereas $m_1, m_3<0$. In view of this, \textbf{M} is a saddle point and is unstable in the $y$ direction.\ 

Regarding \textbf{K}$\pm$, the eigenvalues are given by \begin{align}&m_1 = 3\left(1-w_{\textrm{M}}\right),\\
&m_2 = 3\mp \sqrt{\frac{3}{2}}\lambda~,\\
&m_3 = -6\pm \sqrt{6} \kappa~,\end{align} where the upper sign corresponds to the fixed point \textbf{K}$+$. $m_1$, corresponding to the eigenvector $\mathbf{v} = (1,0,0)$, is always positive, irrespective of $\lambda$ and $\kappa$. Thus, the kination period is unstable in the $x$ direction. The stability of these fixed points, \textbf{M} and \textbf{K}$\pm$, that is shown here, resembles the one in the SGB model \cite{Koivisto:2006ai,TerenteDiaz:2023iqk}.\

We now analyse the stability of the fixed points \textbf{ScI} and \textbf{dS}, which stand out amongst the rest of the fixed points given our interest in quintessence cosmology. For \textbf{ScI}, the eigenvalues can be easily calculated to be \begin{align}\label{eq:eq4.51}&m_1 = 3(1+w_{\textrm{M}}) \left(\frac{\kappa}{\lambda} -1 \right),\\
\label{eqqq:eqqq4.56}&m_{2,3} =-\frac{3}{4}(1-w_{\textrm{M}}) \left\{1\mp \sqrt{1-8\frac{1+w_{\textrm{M}}}{1-w_{\textrm{M}}}\left[1-\frac{3}{\lambda^2}(1+w_{\textrm{M}}) \right]}\right\}.\end{align} Since $\lambda > \sqrt{3(1+w_{\textrm{M}})}$ (see Eq.~\eqref{eq:Omegahat_phi_ScI}), $m_{2,3}$ are complex eigenvalues whose real part is always negative. On the other hand, if $\kappa > \lambda$, $m_{1}$ will be positive, and the fixed point \textbf{ScI} will be a saddle point. This same feature is shared with the SGB model \cite{TerenteDiaz:2023iqk,Koivisto:2006ai}, and it is an important one because the scaling regime, in which the expansion of the Universe does not accelerate in contrast to observations \cite{Weinberg:2013agg}, is brought to an end.\footnote{\label{footnote:7-seven}For a linear coupling function $\xi(\phi)\propto \phi$, we have that $\kappa = 0$ because $\xi_{,\phi\phi} = 0$. This implies that $m_1<0$ in Eq.~\eqref{eq:eq4.51}, irrespective of $\lambda$. Then, the scaling fixed point is stable. Moreover, we see that \textbf{dS} is unstable (see Eq.~\eqref{eq:eigenvalues-m23-dS}, where $m_2>0$). The stability of the scaling fixed point \textbf{ScI} contradicts observations indicating that the Universe undergoes a phase of current accelerating expansion, which cannot correspond to the scaling regime with $\epsilon_H = \frac{3}{2}(1+w_{\textrm{M}})$ as commented already. Therefore, the linear coupling can be discarded. Also, $\kappa = 0$ implies that either $u_c = 0$ or $\left.\epsilon_H\right|_c = 0$ (see Eqs.~\eqref{eq:du-eq-dimensionless} and \eqref{eq:rel-epsilonH-dotWW2}). This implies that no other scaling regime with $u_c \neq 0$ exists. The same conclusions can be drawn in the SGB model (see Ref.~\cite{TerenteDiaz:2023iqk}).}\ 

The Universe can enter the de Sitter phase depending on the stability of this fixed point. The eigenvalues are \begin{align}\label{eqqq:eqqq4.59}&m_1 = -3\left(1+w_{\textrm{M}}\right),\\
\label{eq:eigenvalues-m23-dS}&m_{2,3} = -\frac{3}{2} \left[1\mp \sqrt{1-\frac{4}{3}\lambda^2 \left(\frac{\kappa}{\lambda}-1\right)}\right].\end{align} As long as $\kappa > \lambda$, all the eigenvalues will have negative real parts, and \textbf{dS} will be a stable fixed point (an attractor). The condition $\kappa >\lambda$ was sufficient in order for \textbf{ScI} to be unstable (see Eq.~\eqref{eq:eq4.51}), and therefore, the Universe may transition from \textbf{ScI} to \textbf{dS}. This reproduces the evolution of the Universe as we know it: from radiation/matter domination to an eventual exponentially expanding universe. The fact that \textbf{dS} may be a stable fixed point is a well-known result in the SGB model as well \cite{Koivisto:2006ai,TerenteDiaz:2023iqk}.\

For completeness, we include the stability analysis of \textbf{I} before getting to the other two fixed points with $u_c \neq 0$. In this case, the eigenvalues read \begin{align}&m_1 =  \lambda^2 \left(\frac{\kappa}{\lambda}-1\right),\\
&m_2 = \frac{1}{2}\left(\lambda^2 -6\right),\\
&m_3 = \lambda^2 -3\left(1+w_{\textrm{M}}\right).\end{align} Since $\lambda^2 <6$ in order for \textbf{I} to exist (so that $y_{\textrm{\textbf{I}}}$ is real, as was commented earlier), we have that $m_2<0$. Given the range of values of $w_{\textrm{M}}$, $m_3$ may be non-negative ($\lambda^2$ would have to be smaller than $3(1+w_{\textrm{M}})$ otherwise). Considering power law inflation though, such that $\lambda^2 < 2$, we see that $m_3$ is definitely negative. Regardless of $w_{\textrm{M}}$ and $\lambda$, $m_1$ will be positive if $\kappa >\lambda$. Then, \textbf{I} will be unstable if $\kappa >\lambda$.\  

The fixed point \textbf{\^{K}}$+$ could lead to a phase of faster than exponential acceleration, $\left.\epsilon_H\right|_{\textrm{\textbf{\^{K}}}+} < 0$, if $\kappa< 2\sqrt{\frac{2}{3}}$ (remember that we are assuming that $\kappa$ does not vanish; see Eq.~\eqref{eqqq:eqqqq4.11}). We study the stability of the two fixed points \textbf{\^{K}}$\pm$. The eigenvalues are \begin{align}\label{eq:m1-Khatpm}&m_1 = -\frac{2}{2\mp \sqrt{\frac{3}{2}}\kappa} \left[\pm \sqrt{\frac{3}{2}}\kappa +\frac{3}{2} \left(1+w_{\textrm{M}}\right) \left(2\mp \sqrt{\frac{3}{2}}\kappa \right) \right],\\
&m_2 = \frac{\mp \sqrt{\frac{3}{2}}\lambda}{2\mp \sqrt{\frac{3}{2}}\kappa} \left(\frac{\kappa}{\lambda} -1 \right),\\
\label{eqqq:QQQ:QQQ3.45}&m_3 = -\frac{2}{2\mp \sqrt{\frac{3}{2}}\kappa} \left(3\mp \sqrt{\frac{3}{2}}\kappa\right),\end{align} the upper sign corresponding to \textbf{\^{K}}$+$. In the case of \textbf{\^{K}}$-$, the fixed point will be unstable whenever $\kappa > \lambda$ (because $m_2>0$), regardless of the magnitude of $\kappa$ (given that $m_1$ and $m_3$ are always negative for \textbf{\^{K}}$-$). For \textbf{\^{K}}$+$ and $\kappa <2\sqrt{\frac{2}{3}}$, $m_1$ and $m_3$ are both negative too. If we assume that $\kappa>\lambda$, then $m_2$, corresponding to the eigenvector $\mathbf{v} = (0,1,0)$, will be negative and the phase of super-acceleration will be \emph{stable}.\footnote{`Super-acceleration' in relation to $\epsilon_H < 0$ is used, for example, in Refs.~\cite{Jarv:2009zf,MohseniSadjadi:2020qnm}.} If not, it will be unstable in the $y$ direction. Irrespective of the sign of $x_c$, if $\kappa \gg 1$, but $\lambda >\kappa$, the asymptotic phase of zero acceleration, corresponding to $\left.\epsilon_H\right|_{\textrm{\textbf{\^{K}}}\pm} = 1$, will be stable as well.\

Notice that, for that super-accelerated phase (see Eq.~\eqref{eqqq:eqqqq4.4}), \begin{equation}\left.\frac{H}{W} \right|_{\textrm{\textbf{\^{K}}}+}= 1+u_{\textrm{\textbf{\^{K}}}+}x_{\textrm{\textbf{\^{K}}}+}=1+\left(\sqrt{\frac{3}{2}}\kappa - 3\right)\end{equation} is negative (remember that $\kappa <2\sqrt{\frac{2}{3}}$). For the rest of the fixed points treated above, the ratio $H/W$ is $1$ (positive in the case of \textbf{ScII}$\pm$ as one can check from the values of $x$ and $u$ at those fixed points; see before Eq.~\eqref{eq:def-Deltapm}) because $u_c = 0$. The same is true for \textbf{\^{K}}$-$ given that \begin{equation}\left.\frac{H}{W} \right|_{\textrm{\textbf{\^{K}}}-}= 1+u_{\textrm{\textbf{\^{K}}}-}x_{\textrm{\textbf{\^{K}}}-}=1-\left(\sqrt{\frac{3}{2}}\kappa + 3\right)\end{equation} is always negative for $\kappa >0$.\

Finally, for the second scaling solution, \textbf{ScII}$\pm$, the eigenvalues are \begin{align}&m_1 = \frac{3}{2} \left(1+w_{\textrm{M}}\right) \left(1 - \frac{\lambda}{\kappa} \right),\\
\label{eqq:eqqq4.71}&m_{2,3} = -\frac{3}{4} \left(1-w_{\textrm{M}}\right) \left\{1\mp \sqrt{1+8\frac{1+w_{\textrm{M}}}{1-w_{\textrm{M}}}\left[1-\frac{1-3w_{\textrm{M}}}{3(1-w_{\textrm{M}})}\Delta_{\pm} -\frac{12}{\kappa^2} \frac{\left(1+w_{\textrm{M}}\right)^2}{1+3w_{\textrm{M}}}\right]}\right\}.\end{align} We see that $m_1$, which corresponds to the eigenvector $\mathbf{v}=(0,1,0)$, is positive (and hence neither of the scaling fixed points \textbf{ScII}$\pm$ is stable) whenever $\kappa> \lambda$ (same condition as that of the former scaling solution; see Eq.~\eqref{eq:eq4.51}). The instability is in the $y$ direction. On the other hand, $m_{2,3}$ do not depend on the magnitude of $\lambda$.\ 

From the existence condition of the fixed points (see Eq.~\eqref{bounds-on-kappa-scII}) we find \begin{equation} \frac{1-3w_{\textrm{M}}}{3(1-w_{\textrm{M}})}\Delta_{\pm} +\frac{12}{\kappa^2} \frac{(1+w_{\textrm{M}})^2}{1+3w_{\textrm{M}}}\leq \frac{(1-3w_{\textrm{M}})\Delta_{\pm} +2}{3(1-w_{\textrm{M}})}~.\end{equation} Also, since $1\leq  \Delta_{+}< 2$ and $0\leq w_{\textrm{M}} < \frac{1}{3}$ for the fixed point \textbf{ScII}$+$ (see Eq.~\eqref{eqqq:eq4.35}), then \begin{equation}  \frac{1-3w_{\textrm{M}}}{3(1-w_{\textrm{M}})}\Delta_{+} +\frac{12}{\kappa^2} \frac{(1+w_{\textrm{M}})^2}{1+3w_{\textrm{M}}}<\frac{2}{3} \frac{2-3w_{\textrm{M}}}{1-w_{\textrm{M}}} \leq \frac{4}{3}~.\end{equation} In the case of \textbf{ScII}$-$, we assume that $0\leq w_{\textrm{M}} < \frac{1}{3}$ as well, although $w_{\textrm{M}}$ was constrained to be smaller than $1$. Since $0<\Delta_{-}\leq 1$, we have \begin{equation}\label{eqqq:eqqq4.74}\frac{1-3w_{\textrm{M}}}{3(1-w_{\textrm{M}})}\Delta_{-} +\frac{12}{\kappa^2} \frac{(1+w_{\textrm{M}})^2}{1+3w_{\textrm{M}}} \leq 1~.\end{equation} Thus only in the case of \textbf{ScII}$-$ will $m_{2,3}$ always be real (again, assuming $0\leq w_{\textrm{M}} < \frac{1}{3}$). For complex $m_{2,3}$, the real part of both eigenvalues is negative and \textbf{ScII}$+$ is stable if $\lambda >\kappa$. For the rest of the possibilities, that depends on the value of $w_{\textrm{M}}$ under consideration.
\begin{table}
\begin{centering}
\begin{tabular}{|l|c|l|l|}
\cline{3-4} \cline{4-4} 
\multicolumn{1}{l}{\multirow{1}{*}{}} & \multirow{1}{*}{} & Fixed point: $(x_c,y_c,u_c)$ & Stability\tabularnewline
\hline 
\multicolumn{2}{|l|}{\textbf{M}} & $(0,0,0)$  & Unstable 
\tabularnewline
\hline 
\multicolumn{2}{|l|}{\textbf{K}$\pm$} & $(\pm1,0,0)$  & Unstable 
\tabularnewline
\hline 
\multicolumn{2}{|l|}{\textbf{I}} & $\left(\frac{\lambda}{\sqrt{6}},\sqrt{1-\frac{\lambda^2}{6}},0\right)$  & Unstable whenever $\kappa > \lambda$ 
\tabularnewline
\hline 
\multicolumn{2}{|l|}{\textbf{ScI}} & $\left(\sqrt{\frac{3}{2} \frac{(1+w_{\textrm{M}})^2}{\lambda^2}},\sqrt{\frac{3}{2}\frac{1-w^2_{\textrm{M}}}{\lambda^2}},0\right)$  & Unstable whenever $\kappa > \lambda$ 
\tabularnewline
\hline 
\multicolumn{2}{|l|}{\textbf{dS}} & $\left(0,1,\sqrt{\frac{3}{2}}\lambda\right)$  & Stable unless $\kappa <  \lambda$
\tabularnewline
\hline 
\multicolumn{2}{|l|}{\textbf{ScII}$\pm$} & $\left(\frac{\kappa}{\sqrt{6}}\frac{1+3w_{\textrm{M}}}{3(1-w^2_{\textrm{M}})}\Delta_{\pm},0,\frac{\kappa}{\sqrt{6}}\frac{1}{1+w_{\textrm{M}}}\Delta_{\pm}\right)$  & Unstable whenever $\kappa > \lambda$ 
\tabularnewline
\hline 
\multirow{2}{*}{\textbf{\^{K}}$\pm$} & \textbf{\^{K}}$+$ & $\left(+1,0,\sqrt{\frac{3}{2}}\kappa- 3\right)$  & \makecell{Unstable whenever $\kappa < \lambda$ \\ and $\kappa <2\sqrt{\frac{2}{3}}$} \tabularnewline
\cline{2-4} \cline{3-4} \cline{4-4} 
 & \textbf{\^{K}}$-$ & $\left(-1,0,\sqrt{\frac{3}{2}}\kappa+ 3\right)$  & \multirow{1}{*}{Unstable whenever $\kappa > \lambda$ }\tabularnewline
\hline 
\end{tabular}
\par\end{centering}
\caption[Summary of the fixed points of the Weyl-Gauss-Bonnet model and their stability according to $\kappa$ and $\lambda$ for an exponential coupling function]{\label{tab:fixed-points-wgb-stability}Summary of the fixed points considered in this work and their stability according to $\kappa$ and $\lambda$ for an exponential coupling function (see Eq.~\eqref{eqqq:eqqqq4.11} and remember that $\kappa \neq \lambda$ is assumed). We determine the stability of each fixed point assuming an EoS parameter of the background fluid in the range $0\leq w_{\textrm{M}} <1$. $\Delta_{\pm}(\kappa,w_{\textrm{M}})$ is defined in Eq.~\eqref{eq:def-Deltapm}.}
\end{table}

A summary of all the fixed points we studied in this section, and their stability, is provided in Table~\ref{tab:fixed-points-wgb-stability}.

\subsection{Numerical Simulations}
\label{sec:numerical-simulations-sub}
In view of the fixed points shown in Sec.~\ref{sec:fixed-points-section} and their stability, we are interested in the trajectories that begin close to kination, pass near the scaling regime \textbf{ScI} and approach the de Sitter fixed point, depicting the evolution of the Universe as suggested by the observations. To that end, Eqs.~\eqref{eq:dx-eq-dimensionless}, \eqref{eq:dy-eq-dimensionless} and \eqref{eq:du-eq-dimensionless} are solved numerically, taking the constraint equation \eqref{eq:constraint-eq-v1-xyz} into account. We select those trajectories in the phase space that satisfy $\Omega_{\textrm{M}} = 0.3147\pm 0.0074$ and $w_{\textrm{f}} = -0.957\pm 0.080$ \cite{Planck:2018vyg} at $N=0$, which corresponds to the present time. To define $w_{\textrm{f}}$, we use the relation \cite{TerenteDiaz:2023iqk}\begin{equation}\label{eqqq:eqqq4.84} \epsilon_H = \frac{3}{2} \frac{\rho_{\phi}(1+w_{\textrm{f}}) + \rho_{\textrm{M}}(1+w_{\textrm{M}})}{\rho_{\phi} + \rho_{\textrm{M}}}~.\end{equation} $w_{\textrm{f}}$ can be interpreted as the EoS parameter of the DE fluid causing the accelerated expansion of the Universe \cite{Weinberg:2013agg}. Then \begin{equation}\label{eqqq:eqqq4.85}w_{\textrm{f}} = -1+\frac{1}{x^2+y^2} \left[\frac{2}{3} \epsilon_H -\left(1-x^2-y^2\right) \left(1+w_{\textrm{M}}\right)\right],\end{equation} where $\epsilon_H$ is given in Eq.~\eqref{eq:rel-epsilonH-dotWW2}. The background fluid EoS parameter is chosen to be that of pressureless matter, $w_{\textrm{M}} = w_{m} = 0$ (therefore $\Omega_{\textrm{M}} = \Omega_m$).\ 

As initial values, we set $u_0 = 10^{-23}$ and $y_0 = 10^{-3}$. Such a small initial value of $u$ is imposed to ensure a negligible initial contribution from the GB coupling, given that the trajectories in phase space begin near the fixed points \textbf{K}$\pm$ and \textbf{M}, in accordance with the evolution prescribed by the quintessence models. The initial value of $x$ is chosen differently within the range $(-1,1)$. All the trajectories obtained from the simulations end up at the de Sitter fixed point $y_{\textrm{\textbf{dS}}} = 1$ (these trajectories correspond to the different values assigned to $x_0$). Since $\delta u \propto \exp\left[3\left(\frac{\kappa}{\lambda} -1 \right)N\right]$ during the scaling regime (see Eq.~\eqref{eq:eq4.51}, where $w_{\textrm{M}} = w_m = 0$), $\kappa/\lambda$ can be chosen to be of order $\mathcal{O}(1)$, so it takes longer for the Universe to abandon the scaling regime.\

\begin{figure}
\begin{minipage}{.5\linewidth}
\centering
\subfloat[\label{fig:xy_plane_phase_1wgb}Phase portrait for $\lambda = 4.47\sqrt{\frac{2}{3}}$ and $\kappa/\lambda = 1.4$.]{\includegraphics[scale=.3]{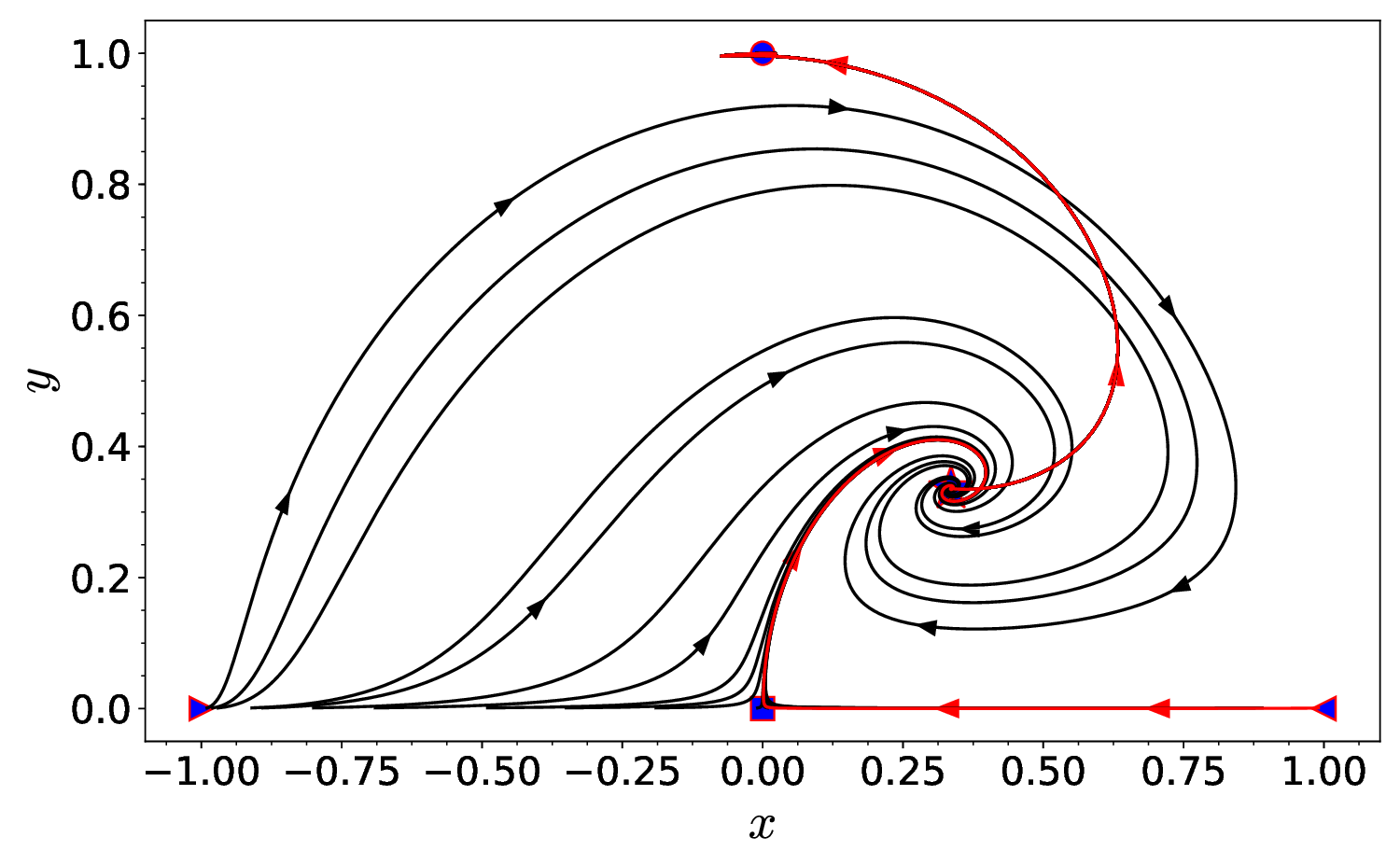}}
\end{minipage}%
\begin{minipage}{.5\linewidth}
\centering
\subfloat[\label{fig:Omegas_wf_1wgb}$\Omega$'s and $w_{\textrm{f}}$ drawn from the red trajectory.]{\includegraphics[scale=.3]{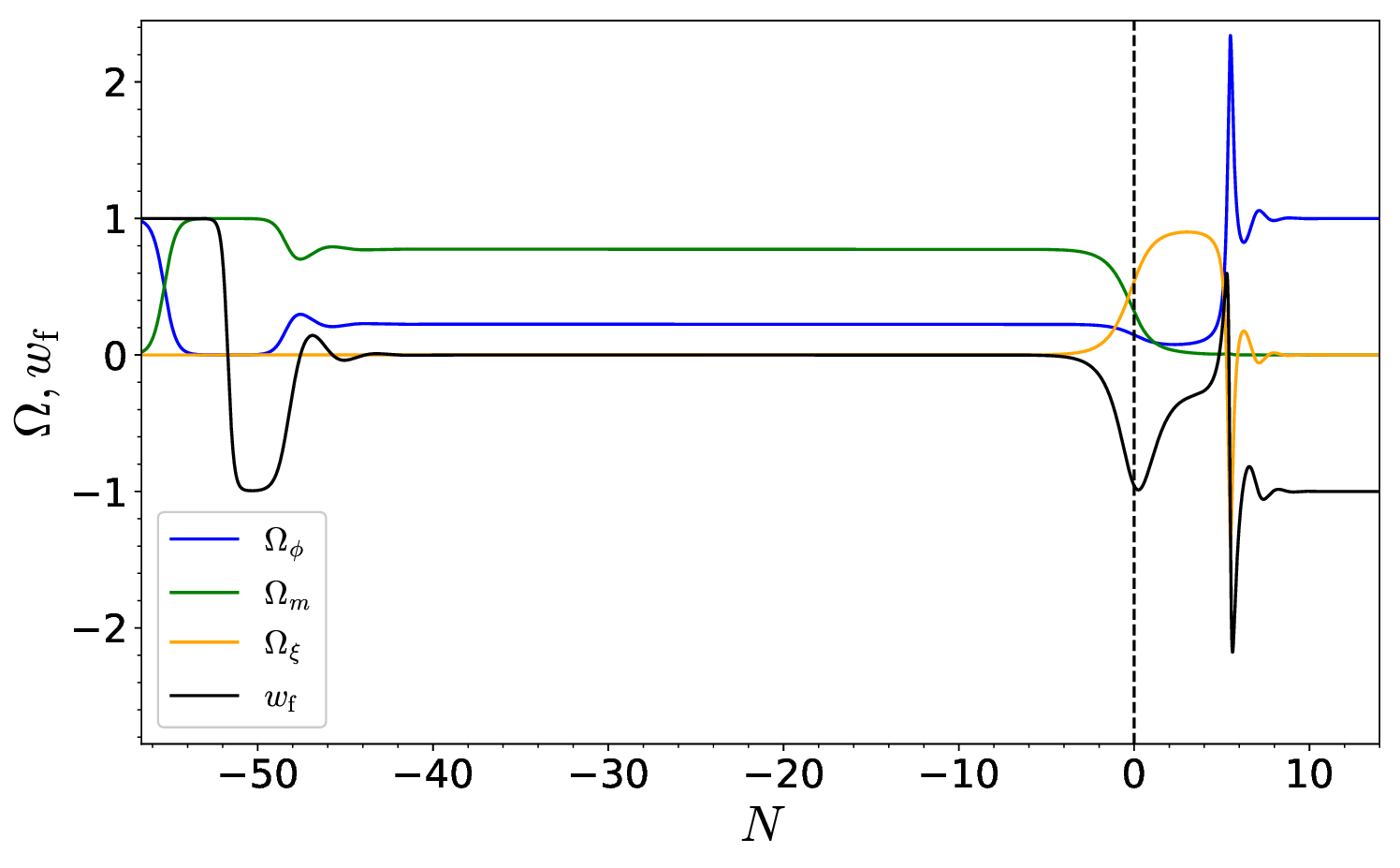}}
\end{minipage}\par\medskip
\begin{minipage}{.5\linewidth}
\centering
\subfloat[\label{fig:xu_plane_phase_2wgb}Phase portrait for $\lambda = 0.72\sqrt{3}$ and $\kappa/\lambda = 4.41$.]{\includegraphics[scale=.3]{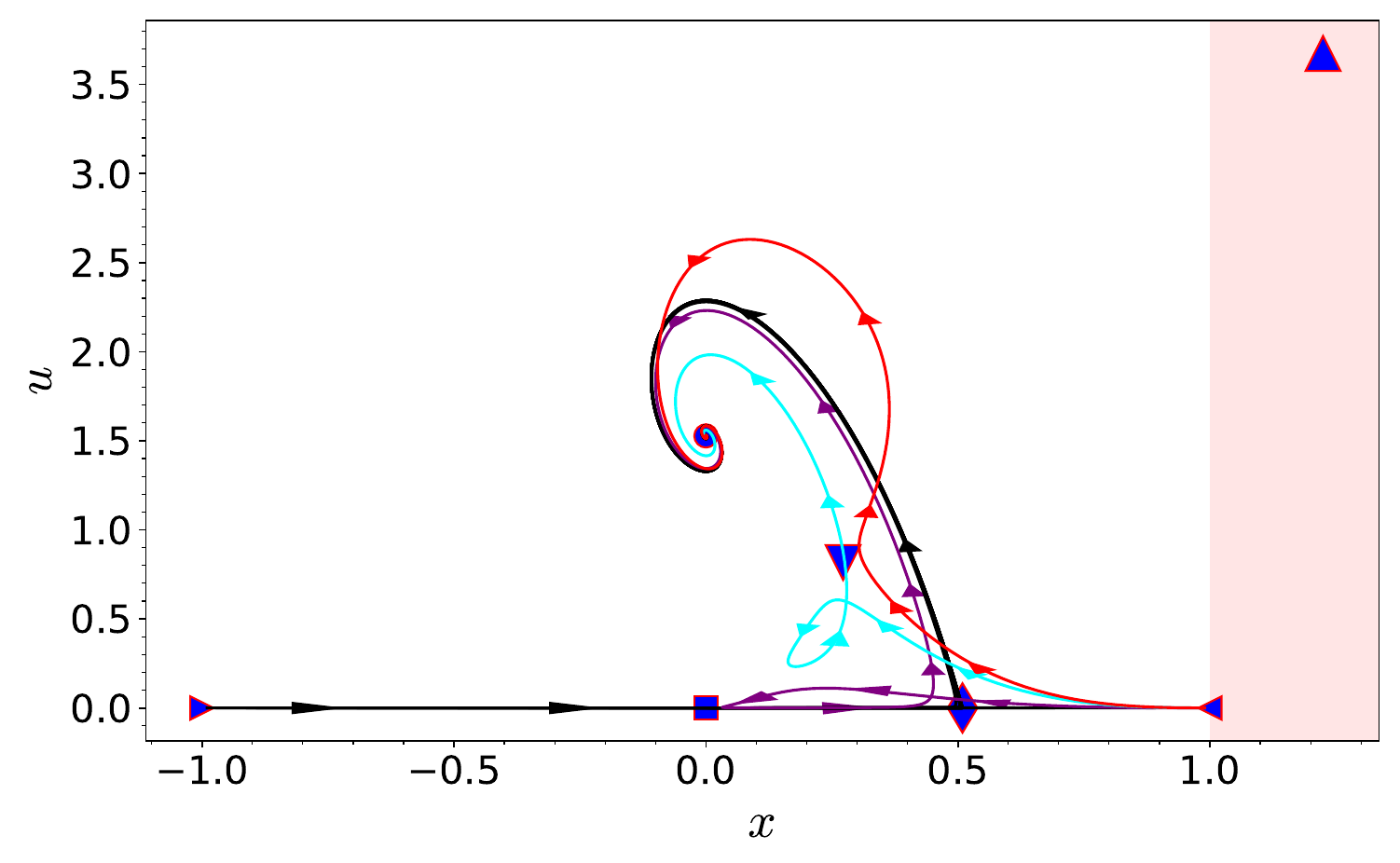}}
\end{minipage}%
\begin{minipage}{.5\linewidth}
\centering
\subfloat[\label{fig:Omegas_wf_2wgb}$\Omega$'s and $w_{\textrm{f}}$ drawn from the red trajectory.]{\includegraphics[scale=.3]{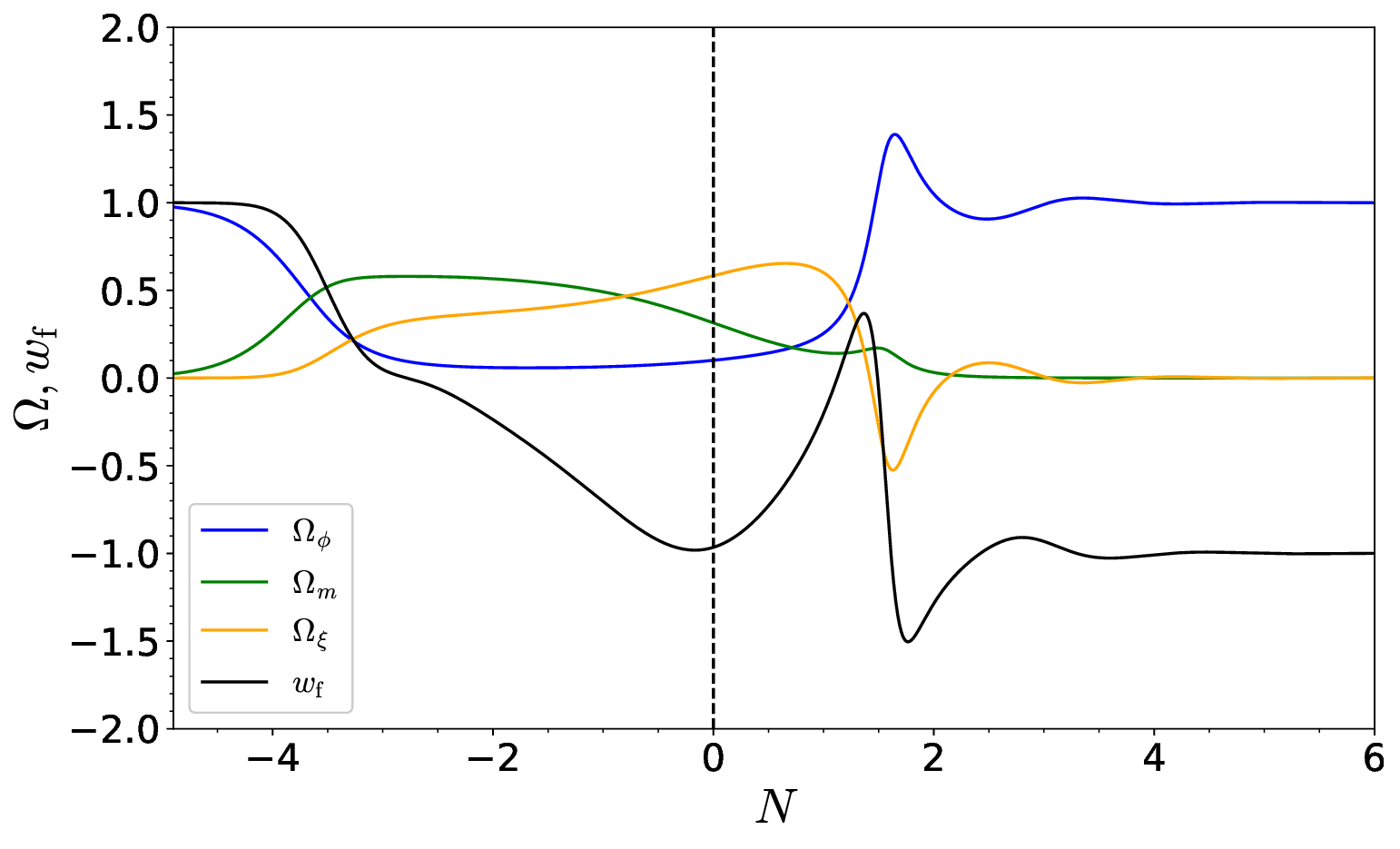}}
\end{minipage}\par\medskip
\begin{minipage}{.5\linewidth}
\centering
\subfloat[\label{fig:xyu_3d_phase_1wgb}$3$D phase portrait for $\lambda = 4.47\sqrt{\frac{2}{3}}$ and $\kappa/\lambda = 1.4$.]{\includegraphics[scale=.37]{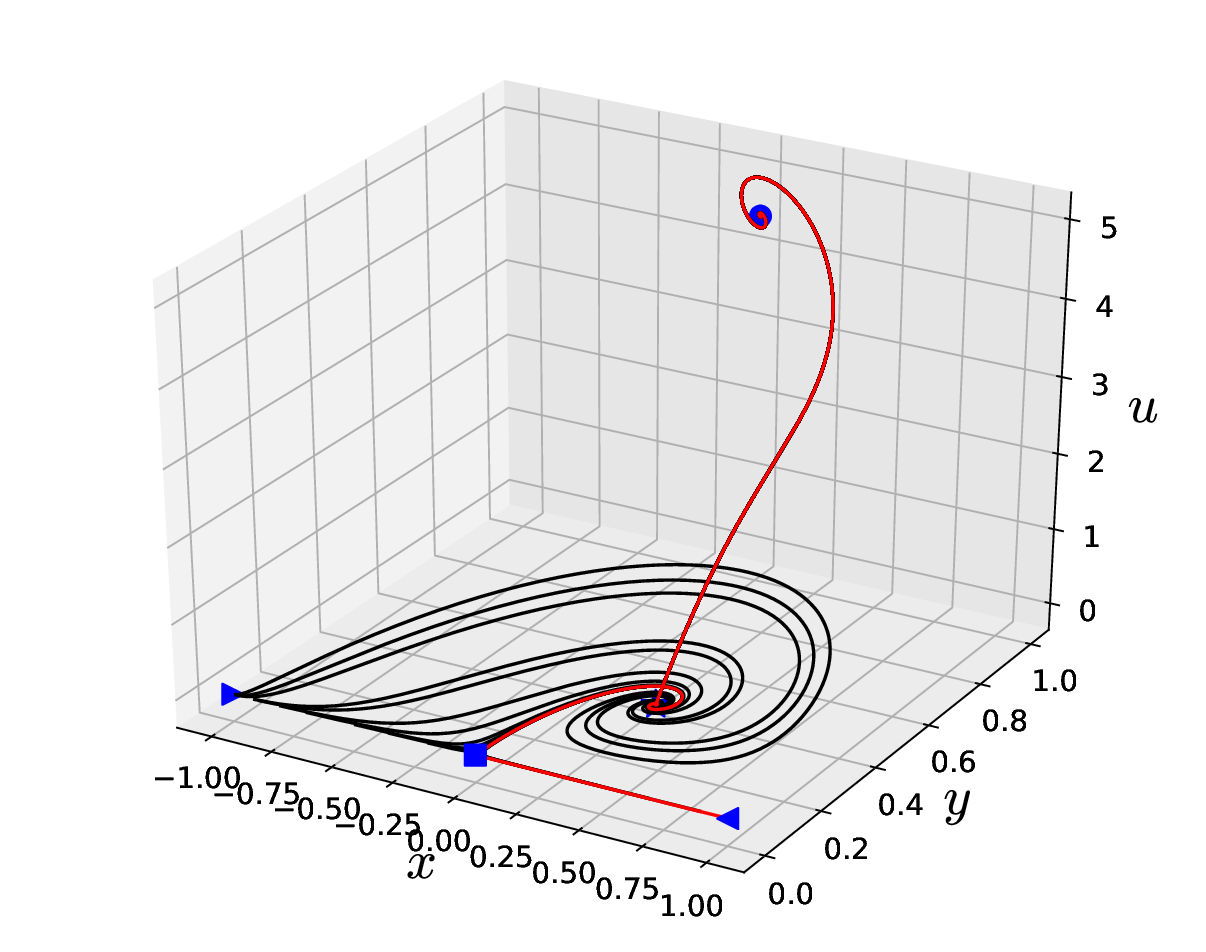}}
\end{minipage}%
\begin{minipage}{.5\linewidth}
\centering
\subfloat[\label{fig:xyu_3d_phase_2wgb}$3$D phase portrait for $\lambda = 0.72\sqrt{3}$ and $\kappa/\lambda = 4.41$.]{\includegraphics[scale=.37]{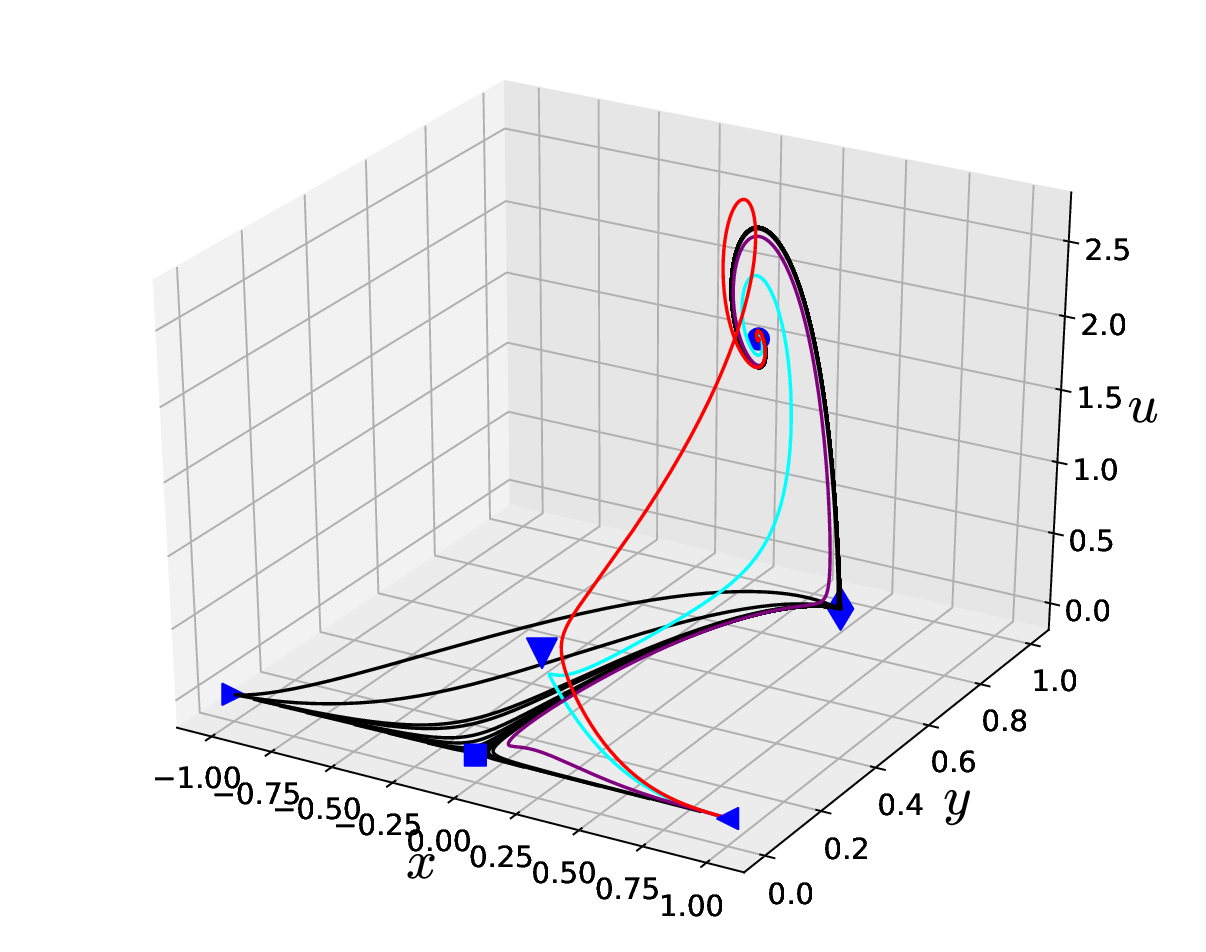}}
\end{minipage}\par\medskip
\caption[Phase portraits and density parameters and $w_{\textrm{f}}$ for some values of the parameters $\lambda$ and $\kappa$.]{\label{fig:phase_portrait_Omegas_wf_12wgb}Phase portraits and density parameters and $w_{\textrm{f}}$ for some values of the parameters $\lambda$ and $\kappa$. In Fig.~\ref{fig:xy_plane_phase_1wgb}, all the trajectories begin either close to \textbf{K}$\pm$ or near \textbf{M}. They all converge at \textbf{ScI} and transition to \textbf{dS}. In Fig.~\ref{fig:xu_plane_phase_2wgb}, the red and cyan trajectories in particular pass near the scaling regime \textbf{ScII}$-$ (triangle pointing downwards) before reaching \textbf{dS}. The rest converge at \textbf{I} instead (the diamond in between \textbf{K}$+$ and \textbf{M} in the plot). The stripe indicates the forbidden region explicitly due to the constraint equation \eqref{eq:constraint-eq-v1-xyz} and, the triangle pointing upwards, the scaling fixed point \textbf{ScII}$+$. Same phase portraits in $3$D are shown in Figs.~\ref{fig:xyu_3d_phase_1wgb} and \ref{fig:xyu_3d_phase_2wgb}.} 
\end{figure}

We see in Figs.~\ref{fig:xy_plane_phase_1wgb} and \ref{fig:Omegas_wf_1wgb} the results of some simulations for $\kappa/\lambda = 1.4$ (so the ratio $\kappa/\lambda$ is of order $\mathcal{O}(1)$, as argued above) depicting the evolution of interest as explained at the beginning of this section. The $3$D phase portrait is shown in Fig.~\ref{fig:xyu_3d_phase_1wgb}. In Figs.~\ref{fig:xy_plane_phase_1wgb} and \ref{fig:xyu_3d_phase_1wgb}, the \textbf{M} fixed point has been represented by a square. The kination fixed points \textbf{K}$\pm$ correspond to the left and right leaning triangles, depending on the sign. The de Sitter fixed point is given by the circle instead, and the scaling regime corresponds to a star onto which the trajectories converge. It can be noticed that, before the de Sitter phase is reached, the density parameter of the scalar field (see Eq.~\eqref{eqqq:eqqqq4.55}) becomes larger than $1$ given that $\Omega_{\xi}$ (see Eq.~\eqref{eq:constraint-eq-Friedmannv2}) turns negative. We obtained this in many other simulations, with different values of $\lambda$ and $\kappa$. Likewise, the growth of $w_{\textrm{f}}$ reaches the value of the EoS parameter of pressureless matter and surpasses it for a brief moment of time. There is an equally short period where it takes on phantom values before settling down at $w_{\textrm{\textbf{dS}}} = -1$, which is the expected value during de Sitter. Note that slightly phantom DE is preferred by the observational data \cite{Planck:2018vyg}.\ 

We observe another feature shared with other simulations: the density parameter $\Omega_{\xi}$ is always the dominant contribution to DE at the present time. This occurs unless the scalar field energy density takes over during the scaling regime, which is discarded given that the field is supposed to be light and may alter the successful predictions of Big Bang Nucleosynthesis (BBN).\ 

The Universe may enter the second scaling regime \textbf{ScII}$\pm$ instead, where $y_{\textrm{\textbf{ScII}$\pm$}} = 0$, starting from kination. If $\Omega_{\textrm{M}}$ is larger than the sum of the rest of the density parameters, $\Omega_{\phi} +\Omega_{\xi}$, during the scaling regime, then $\left.\Omega_{\textrm{M}}\right|_{\textrm{\textbf{ScII}$\pm$}}>0.5$.\footnote{One might argue that this is not necessarily true if $\Omega_{\xi}$ is negative. However, this is discarded in this situation given that $\Omega_{\xi} <0$ implies $-2<ux<0$ (see Eq.~\eqref{eq:constraint-eq-Friedmannv2}). At \textbf{ScII}$\pm$, we obtain \begin{equation} u_{\textrm{\textbf{ScII}$\pm$}}x_{\textrm{\textbf{ScII}$\pm$}} = \frac{1+3w_{\textrm{M}}}{3(1-w_{\textrm{M}})} u_{\textrm{\textbf{ScII}$\pm$}}^2~,\end{equation} and because $0\leq w_{\textrm{M}}<1$, $\left.\Omega_{\xi}\right|_{\textrm{\textbf{ScII}$\pm$}}$ is never negative.} For pressureless matter ($w_{\textrm{M}} = w_m = 0$), $\left.\Omega_{m}\right|_{\textrm{\textbf{ScII}$\pm$}}$ reads \begin{equation}\label{eqqq:eqqq4.90} \left.\Omega_{m}\right|_{\textrm{\textbf{ScII}$\pm$}}=\frac{54\left(54-\kappa^2 \Delta_{\pm}^2\right)}{\left(54+3\kappa^2 \Delta_{\pm}^2\right)^2}~.\end{equation} Then $\left.\Omega_{m}\right|_{\textrm{\textbf{ScII}$\pm$}}>0.5$ implies $\kappa \Delta_{\pm} < \sqrt{6}$. The fixed point \textbf{ScII}$+$ can be discarded because $\kappa > 3\sqrt{2}$ in order for the fixed point to exist, and $1\leq \Delta_{+}$ (see Eq.~\eqref{bounds-on-kappa-scII} and discussion below), which means that $\kappa\leq \kappa \Delta_{+}<\sqrt{6}$. Regarding the fixed point \textbf{ScII}$-$, we have (see Eq.~\eqref{eq:def-Deltapm}) \begin{equation}\kappa \Delta_{-} = \kappa -\sqrt{\kappa^2 -18} <\sqrt{6}~,\end{equation} and hence $\kappa > 2\sqrt{6}$ for $\Omega_{m}$ to be larger than $\Omega_{\phi}+\Omega_{\xi}$ during the scaling regime, in agreement with the BBN predictions.

If the fixed points \textbf{ScII}$\pm$ are unstable only in the $y$ direction (see discussion below Eq.~\eqref{eqq:eqqq4.71}), then $m_{2,3}$ should be complex, given that the real part in that case is always negative (see Eq.~\eqref{eqq:eqqq4.71}). Unfortunately, this cannot happen in the case of \textbf{ScII}$-$ (see Eq.~\eqref{eqqq:eqqq4.74} and discussion below), and only at this fixed point can $\Omega_{m}$ dominate over $\Omega_{\phi} + \Omega_{\xi}$.\

An example of trajectories going through the second scaling solution (corresponding to the fixed point \textbf{ScII}$-$, which is being represented by a triangle pointing downwards) are the red and cyan ones in Fig.~\ref{fig:xu_plane_phase_2wgb}, where $\kappa/\lambda = 4.41$ and $u_0 = 10^{-5}$. The purple line no longer passes near the scaling regime, but approaches the \textbf{I} fixed point, represented by the diamond in the plot. The corresponding $3$D phase portrait is shown in Fig.~\ref{fig:xyu_3d_phase_2wgb}. In Fig.~\ref{fig:xu_plane_phase_2wgb} we show the density parameters and $w_{\textrm{f}}$ associated with the red trajectory, which are compatible with the constraints on $\Omega_{m}$ and $w_{\textrm{f}}$. Unfortunately, the scaling regime does not last long enough for this particular case, although we see that $\Omega_{m}$ takes on the value that we expect in view of Eq.~\eqref{eqqq:eqqq4.90} and the corresponding value of $\kappa$. Also, in line with what we explained earlier, the fixed point \textbf{ScII}$+$ falls within the forbidden region because $\kappa >3\sqrt{2}$. Such a fixed point is represented by a triangle pointing upwards. The forbidden region corresponds to the pale red coloured stripe, which excludes values of $x>1$ that violate the constraint equation \eqref{eq:constraint-eq-v1-xyz}.   

\subsection{The Speed of Gravitational Waves}
In this last section we examine the evolution of the $\alpha_T$ parameter. We first write it in terms of the dimensionless variables defined in Eqs.~\eqref{eqqq:eqqq4.3} and \eqref{eq:def-u-wgb-model}, and of the Hubble flow parameter $\epsilon_H$ (see Eqs.~\eqref{eq:alphaT-wgb-eqv2} and \eqref{eq:rel-epsilonH-dotWW2}) \begin{equation}\label{eqqq:eqqq.4.77} \alpha_T = -\frac{2}{1-2ux} \left\{(1+2ux)(ux)' -ux \left[1-ux-\epsilon_H \left(1+ux\right) \right] \right\}.\end{equation} It can be readily seen that \begin{equation} \left.\alpha_T\right|_{\textrm{\textbf{M}}} = \left.\alpha_T\right|_{\textrm{\textbf{K}}\pm} = \left.\alpha_T\right|_{\textrm{\textbf{I}}} = \left.\alpha_T \right|_{\textrm{\textbf{ScI}}}=\left.\alpha_T \right|_{\textrm{\textbf{dS}}}=0~.\end{equation} For the rest of the fixed points, the $\alpha_T$ parameter is given by \begin{align}\label{eq:eqqq4.79}&\left.\alpha_T\right|_{\textrm{\textbf{ScII}$\pm$}} = -\frac{18(1+w_{\textrm{M}})(1-w^2_{\textrm{M}}) -2\kappa^2 (1+3w_{\textrm{M}}) \Delta_{\pm}}{27(1+w_{\textrm{M}})(1-w_{\textrm{M}}^2)-2\kappa^2 (1+3w_{\textrm{M}})\Delta_{\pm}}\left[2-\frac{\kappa^2}{18} \frac{(1+3w_{\textrm{M}})(5+3w_{\textrm{M}})}{(1+w_{\textrm{M}})(1-w^2_{\textrm{M}})}\Delta_{\pm}\right],\\
&\left.\alpha_T \right|_{\textrm{\textbf{\^{K}}}\pm} = -\frac{(6\mp \sqrt{6} \kappa)(4\mp \sqrt{6} \kappa)}{7\mp \sqrt{6} \kappa}~.\end{align} $\left.\alpha_T \right|_{\textrm{\textbf{\^{K}}}+}$ in particular is zero if $\kappa = 3\sqrt{\frac{2}{3}}$, which does not satisfy the bound $\kappa <2\sqrt{\frac{2}{3}}$ and falls out of the super-accelerated phase region. The other possibility is $\kappa = 2\sqrt{\frac{2}{3}}$, but in that case $u_c x_c = -1$ and $H_c = 0$ (see Eq.~\eqref{eqqq:eqqqq4.4}). Since $\kappa >0$, $\left.\alpha_T \right|_{\textrm{\textbf{\^{K}}}-}$ never vanishes exactly.\

Regarding $\left.\alpha_T\right|_{\textrm{\textbf{ScII}$\pm$}}$, in the case of the fixed point \textbf{ScII}$+$, the $\alpha_T$ parameter cannot be zero due to the existence conditions of the fixed point derived below Eq.~\eqref{eq:def-Deltapm} and the range of values of $w_{\textrm{M}}$ under consideration. One might have $\left.\alpha_T\right|_{\textrm{\textbf{ScII}$-$}}=0$ though. However, given $w_{\textrm{M}} = w_m = 0$, $\left.\alpha_T \right|_{\textrm{\textbf{ScII}$-$}}=0$ if $\kappa^2\Delta_{-}(\kappa,0) = 36/5$ or $\kappa^2 \Delta_{-}(\kappa,0) = 9$ (see Eq.~\eqref{eq:eqqq4.79}). We do not obtain solutions for $\kappa$ in those cases. Thus, $\alpha_T$ never vanishes at \textbf{ScII}$-$ when $w_{\textrm{M}} = 0$.\

Although the value of $|\alpha_T|$ at the fixed points is useful to have an indication of its evolution, the current state of the Universe cannot be described by any of those fixed points. Moreover, $|\alpha_T|$ does not have to be exactly zero, instead it must satisfy the bound in Eq.~\eqref{bounds-on-cgw2}. Then, we show the evolution of $\alpha_T$ connecting the different fixed points and make sure it does not surpass the upper bound from the observational constraint. 
\begin{figure}
\begin{centering}
\includegraphics[scale=0.59]{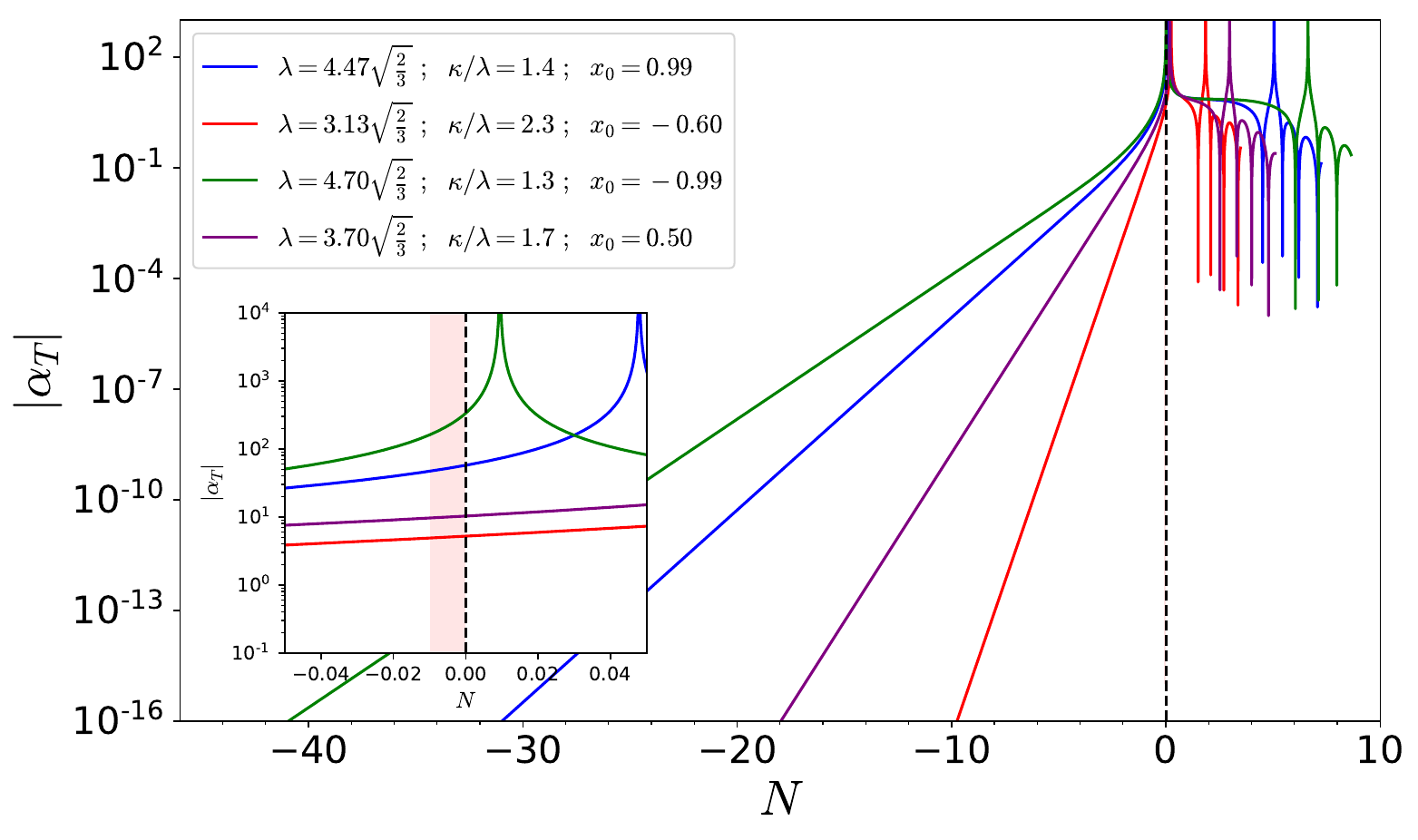}
\par\end{centering}
\caption[$|\alpha_T|$ predicted by the WGB model for some values of the parameters $\lambda$ and $\kappa$ and different initial conditions of $x$]{\label{fig:alphaT-4values-comp-obs1} $|\alpha_T|$ predicted by the WGB model for some values of the parameters $\lambda$ and $\kappa$ and different initial conditions of $x$. The stripe in the smallest graph on the left indicates the excluded region starting at redshift $z=0.00980\pm 0.00079$ (see Ref.~\cite{Hjorth:2017yza}). We see that the bound $|\alpha_T| < 10^{-15}$ is grossly violated for the cases under consideration. These cases have been selected given that they did satisfy the constraints on $\Omega_{\textrm{M}}$ and $w_{\textrm{f}}$.}
\end{figure}

That bound is applicable in the relatively recent history of the Universe. To be precise, $|\alpha_T|$ is constrained from redshift\footnote{If one consults Ref.~\cite{LIGOScientific:2018hze} the redshift value, $z=0.0099$, is the geocentric redshift of the host galaxy. The cosmological redshift by contrast, which is the one shown here and what we need in this case, is given in Ref.~\cite{Hjorth:2017yza}.} $z=0.00980\pm 0.00079$ until today ($z=0$).\footnote{\label{footnote:7th-footnote}Do not confuse `$z$' here denoting the redshift with the dimensionless variable defined in Eq.~\eqref{eqqq:eqqq4.3} and related with the energy density of the background fluid.} Selecting models which reproduce trajectories that go through the scaling solution \textbf{ScI}, before entering de Sitter, with different values of the parameters $\kappa$ and $\lambda$, and different initial condition $x_0$, we plot the evolution of the $|\alpha_T|$ parameter over the number of e-folds $N$ (see Fig.~\ref{fig:alphaT-4values-comp-obs1}). In that plot, we zoom in on the excluded region, marked by a vertical, pale red coloured stripe, and we observe that the bound is violated by many orders of magnitude, similarly to the SGB model \cite{TerenteDiaz:2023iqk}. The cases shown in the plot were chosen given that they did satisfy the constraints on $\Omega_{m}$ and $w_{\textrm{f}}$ at the present time.\

If we examine Eq.~\eqref{eqqq:eqqq.4.77}, we can understand why the bound is so strongly violated. We notice that for $ux$ of order $\mathcal{O}(0.1)$-$\mathcal{O}(1)$, then $\Omega_{\xi}$ is of order $\mathcal{O}(0.1)$ (see Eq.~\eqref{eq:constraint-eq-Friedmannv2}). We observed numerically that $\Omega_{\xi}$ is always the density parameter that dominates over the rest by the time $\Omega_{m}$ and $w_{\textrm{f}}$ satisfy the constraints. In addition to this, we remark that it grows at that time. Then $ux \rightarrow 1/2$ (i.e. $\Omega_{\xi} \rightarrow 0.6$) while the numerator in Eq.~\eqref{eqqq:eqqq.4.77} remains of order $\mathcal{O}(0.1)$-$\mathcal{O}(1)$. This causes $|\alpha_T|$ to increase near the present time, although it does not diverge because $(ux)'$ increases eventually as well, even though $\Omega_{\xi}$ remains constant. This can be seen by simply taking the derivative of Eq.~\eqref{eq:constraint-eq-Friedmannv2} \begin{equation} \Omega_{\xi}' = \frac{2(1-\Omega_{\xi})}{1+ux} (ux)'.\end{equation} When $\Omega_{\xi} \rightarrow 1$, $\Omega_{\xi}' \rightarrow 0$.\ 

The fact that $|\alpha_T|$ does not remain below the observational bound $10^{-15}$ while we approach the present time indicates that the exponential coupling function in the WGB model must be discarded. Similar results are obtained in the SGB model \cite{TerenteDiaz:2023iqk}. Another feature of Fig.~\ref{fig:alphaT-4values-comp-obs1} to point out\footnote{We would like to thank the referee for bringing this aspect to our attention.} is the appearance of gradient instabilities \cite{Joyce:2014kja}, where $\alpha_T <-1$. This issue is related to the attractive behaviour in the $u$ direction of the phase space of the pseudo-kination fixed point \textbf{\^{K}}$+$ when $\kappa >3\sqrt{\frac{2}{3}}$ (see Eq.~\eqref{eqqq:QQQ:QQQ3.45}), and the fact that $\lambda >3\sqrt{\frac{2}{3}}$ in order for the \textbf{I} fixed point not to exist and replace \textbf{ScI} (see below Eq.~\eqref{des:condition-existence-fixed-point-scI}). Since $\kappa>\lambda$ in order for trajectories to leave the scaling fixed point \textbf{ScI} and approach de Sitter (and hence $\kappa > 3\sqrt{\frac{2}{3}}$), the pseudo-kination fixed point will divert them causing $\alpha_T<-1$ in Eq.~\eqref{eqqq:eqqq.4.77} (because $u_{\textrm{ \textbf{\^{K}}$+$}}x_{\textrm{ \textbf{\^{K}}$+$}} > 3(\sqrt{\frac{3}{2}}-1) >\frac{1}{2}$ and hence $ux$ must approach $1/2$).  

\subsection{The Case of $\alpha_T = 0$}
In this section, instead of specifying the functional form of $\xi(\phi)$ and consider examples other than the exponential case, we impose the condition $\alpha_T = 0$. In the SGB model, such a condition translates into Eq.~\eqref{eq:cgweqc-metric}. In the WGB model, however, $\alpha_T = 0$ leads to the following equation (see Eq.~\eqref{eq:alphaT-wgb-eqv2}): \begin{equation}\label{eqqq:eqq4.89} \ddot \xi = \dot \xi W\left(1-2\frac{\dot \xi W}{M^2_{\textrm{Pl}}}\right).\end{equation} Using Eq.~\eqref{eq:ratio-W-H}, Eq.~\eqref{eqqq:eqq4.89} becomes \begin{equation}\label{eqqq:eq4.90} \ddot \xi = -\frac{\dot\xi  H}{2\mu} \left(1+2\mu - \sqrt{1+8\mu} \right).\end{equation} 

Now, the equation of $\dot H$ in the SGB model when $\ddot \xi = H \dot \xi$ reads (see Eq.~\eqref{eq:dot-H-eq-GB-metric}) \begin{equation}\label{eqqq:eq4.87} -2\dot H \left(M^2_{\textrm{Pl}}-4\dot \xi H \right) = \dot \phi^2 + \rho_{\textrm{M}}\left(1+w_{\textrm{M}}\right).\end{equation} This implies that in de Sitter, $\dot \phi = 0$ (so $\dot H = 0$). However, this value of $\dot \phi$ is problematic when we integrate Eq.~\eqref{eq:cgweqc-metric}. Indeed, the integral on the LHS of the equation \begin{equation} \int_{0}^{\dot{\xi}_1} \textrm{d}\ln(|\dot \xi|) = \int_{a_{\textrm{\textbf{dS}}}}^{a_1} \textrm{d}\ln a~,\end{equation} does not converge. In the WGB model however, since we have $\Phi$ in addition to $H$, there is a chance that $\dot H = 0$ but still $\dot \phi \neq 0$. To verify this, we must write an equation analogous to Eq.~\eqref{eqqq:eq4.87}. Bearing Eq.~\eqref{eqqq:eq4.90} in mind, we write Eq.~\eqref{eq:ij-00-v2} as \begin{equation}\label{eq:H-dot-eq-WGB-model-v22} -\frac{2M^2_{\textrm{Pl}}\dot H}{\sqrt{1+8\mu}} = \dot \phi^2 +\rho_{\textrm{M}}(1+w_{\textrm{M}}) -4\ddot \xi H^2 \frac{(1-\sqrt{1+8\mu})^2}{16\mu^2 \sqrt{1+8\mu}} -4\dot \xi H^3 \frac{(1-\sqrt{1+8\mu})^3}{64\mu^3}~,\end{equation} where Eq.~\eqref{eq:ratio-W-H} was used. Plugging Eq.~\eqref{eqqq:eq4.90} into Eq.~\eqref{eq:H-dot-eq-WGB-model-v22}, we obtain \begin{equation} -\frac{2M^2_{\textrm{Pl}}\dot H}{\sqrt{1+8\mu}} = \dot \phi^2 + \rho_{\textrm{M}} \left(1+w_{\textrm{M}}\right) +3H^2 M^2_{\textrm{Pl}}\frac{\left(1-\sqrt{1+8\mu}\right)^2}{16\mu^2 \sqrt{1+8\mu}}\left(1+4\mu-\sqrt{1+8\mu}\right).\end{equation} We see that the last term vanishes when $|\mu| \ll 1$, recovering Eq~\eqref{eqqq:eq4.87} for $|\dot \xi H| \ll M_{\textrm{Pl}}^2$ ($\mu$ was defined in Eq.~\eqref{eq:def-mu-dimensionless}). Unfortunately, each term on the RHS is non-negative meaning that $\dot H = 0$ implies $\dot \phi = 0$ and hence $\dot \xi = 0$, same as in the SGB model.\ 

Knowing this, we proceed to analyse the stability of the de Sitter and scaling fixed points, \textbf{dS} and \textbf{ScI} respectively, when the new condition $\alpha_T = 0$ is imposed, in line with was done in Ref.~\cite{TerenteDiaz:2023iqk}. Firstly, we write Eq.~\eqref{eqqq:eqq4.89} as \begin{equation}\label{eqqq:eqqq4.98}(ux)' = \frac{ux}{1+ux} \left[1-\left(3x+2u\right)x-\frac{3}{2}\left(1+w_{\textrm{M}}\right)\left(1-x^2-y^2\right) \right],\end{equation} where we used Eq.~\eqref{eqq:4.22}, the constraint equation \eqref{eq:constraint-eq-v1-xyz} and Eqs.~\eqref{eq:def-u-wgb-model} and \eqref{eq:eq-u-no-assump-xi-coupling}, given that we do not assume a coupling function form in particular. This is equivalent to $\alpha_T = 0$ in Eq.~\eqref{eqqq:eqqq.4.77}. We however make it clear that we use Eq.~\eqref{eq:eq-u-no-assump-xi-coupling} to write it that way.\ 

We perturb Eq.~\eqref{eqqq:eqqq4.98} around the de Sitter fixed point (which is still a fixed point of the system despite Eq.~\eqref{eqqq:eqqq4.98}), that can only correspond to $x = 0$ and $z = 0$ as we just argued \begin{equation}\label{eq:deltax-dS-constraint-cgw} \delta x' = \delta x~.\end{equation} This equation implies that $\delta x \propto e^{N}$. The rest of the equations read\footnote{Notice that, although we have not assumed a specific coupling function $\xi(\phi)$ in this section, $u_{\textrm{\textbf{dS}}} = \sqrt{\frac{3}{2}} \lambda$, as in the case of the exponential function. This can be seen from Eq.~\eqref{eq:dx-eq-dimensionless} with $x_{\textrm{\textbf{dS}}} = 0$ and $y_{\textrm{\textbf{dS}}} = 1$.} \begin{align}\label{eq:deltax-prime-dS}&\delta x' = -3\delta x-\sqrt{\frac{3}{2}}\lambda\left(1 +3w_{\textrm{M}}\right)\delta y -\delta u~,\\
&\delta y' = -3(1+w_{\textrm{M}}) \delta y~.\end{align} The second equation gives $\delta y \propto e^{-3(1+w_{\textrm{M}})N}$. Using Eq.~\eqref{eq:deltax-dS-constraint-cgw}, Eq.~\eqref{eq:deltax-prime-dS} gives $\delta u$. Since $\delta y$ dies out but $\delta x$ grows, $\delta u$ will increase too. Then we see that the de Sitter fixed point \textbf{dS} is unstable, as in the SGB model \cite{TerenteDiaz:2023iqk}.\

Regarding the scaling regime \textbf{ScI}, since $u_{\textrm{\textbf{ScI}}} = 0$, this is still a fixed point despite Eq.~\eqref{eqqq:eqqq4.98}. Linearising Eq.~\eqref{eqqq:eqqq4.98}, we obtain \begin{equation} \delta u' = -\frac{1}{2}\left(1+3w_{\textrm{M}}\right)\delta u~.\end{equation} Hence $\delta u \propto e^{-\frac{1}{2}(1+3w_{\textrm{M}})N}$, and given the eigenvalues in Eq.~\eqref{eqqq:eqqq4.56} (which remain the same despite $\alpha_T = 0$), we have that \textbf{ScI} becomes stable, a result similar to the one obtained in the SGB model in Ref.~\cite{TerenteDiaz:2023iqk} and in quintessence models in GR. When it comes to the second scaling regime, where $y_{\textrm{\textbf{ScII}$\pm$}} = 0$, we can see that they are not even fixed points of the new system of equations that includes Eq.~\eqref{eqqq:eqqq4.98}.\

Consequently, the WGB model with $\alpha_T =0$ cannot reproduce a long scaling regime with small $u$ followed by a period of accelerated expansion, and hence can be discarded.    

\section{Summary and Conclusions}
Gauss-Bonnet (GB) gravity with $\xi(\phi)\mathcal{G}$ (the scalar-Gauss-Bonnet (SGB) model), where $\mathcal{G}$ is the GB term, has become a popular dark energy (DE) model and it is known to predict the propagation speed of gravitational waves (GWs) $c_{\textrm{GW}} \neq 1$ (in units where $c=1$). For simple coupling functions $\xi(\phi)$ which reproduce the past evolution of the Universe as we know it, the SGB model does not satisfy the observational constraints on $|\alpha_T| \equiv |c_{\textrm{GW}}^2-1|$ from GRB170817A \cite{LIGOScientific:2017zic} when applied to the late Universe in particular \cite{TerenteDiaz:2023iqk}. The same occurs with Horndeski's theory \cite{Horndeski:1974wa}. The model is actually a subclass of that theory and one can recover the SGB action for a non-trivial choice of the functions of the theory \cite{Kobayashi:2011nu}.\ 

In spite of those well-known results, it was shown recently that Horndeski's theory where $G(\phi,X)R$ with $X\equiv -\frac{1}{2} g^{\mu\nu} \partial_{\mu} \phi \partial_{\nu} \phi$ does predict GWs propagating at the speed of light in vacuum when formulated in the Palatini formalism \cite{Kubota:2020ehu}. All the previous claims of $c_{\textrm{GW}} \neq 1$ were made assuming the metric case instead, where the connection is the Levi-Civita (LC) one, which fully depends on the metric and its first derivative. In the Palatini formalism however, the connection is independent of the metric and obeys some field equations, meaning that the geometry differs from the pseudo-Riemannian one. As a consequence of the absence of second order derivatives of the metric in the action, no `counter terms' need to be added to avoid the Ostrogradski ghost \cite{deRham:2016wji}, those being responsible for $c_{\textrm{GW}} \neq 1$. Given the unsuccessful predictions of the SGB model of the metric formalism, in the present work we considered the SGB action but in Weyl geometry where the connection (the Weyl connection) has zero torsion but it is not metric compatible. We named this model {\em Weyl-Gauss-Bonnet} (WGB). A similar action was analysed in Ref.~\cite{BeltranJimenez:2014iie} but ours considers the field-dependent coupling to the GB term and restricts to four dimensions. Thus, the goal in our work was to investigate whether the SGB gravity, as a DE model, when formulated in a different geometry, fulfilled the constraint \eqref{bounds-on-cgw2} and yet displayed a period of matter domination followed by an accelerated expanding phase of the Universe in accordance with other constraints on the density and equation of state (EoS) parameters.\

We assumed a spatially-flat FLRW metric and a background Weyl vector with temporal component $\Phi(t)$ and no spatial components breaking spatial isotropy. The resulting action once the LC and non-Riemannian parts were written separately was given in Eq.~\eqref{eq:full-actionv2}. The $\alpha_T$ parameter was calculated in Sec.~\ref{sec:speed-GWs}. In Sec.~\ref{sec:dynamical-systems-analysis-full}, we performed a dynamical systems analysis assuming an exponential potential, which is common in quintessence scenarios. The dimensionless variables in Eqs.~\eqref{eqqq:eqqq4.3} and \eqref{eq:def-u-wgb-model} were formulated in terms of the newly defined Weyl parameter $W \equiv H - \Phi$, thereby including the modifications to Einstein's gravity in contrast to the metric formalism where these are accounted for by $u$ alone. Fixed points of the dynamical system were found in Sec.~\ref{sec:fixed-points-section} under the assumption that $\xi(\phi)$ is an exponential function (see Eq.~\eqref{eqqq:eqqqq4.11}). This choice makes all the equations self-similar so the explicit dependence on $W$ disappears. Another possibility was the linear GB coupling (because $\xi_{,\phi\phi} = 0$; see Eq.~\eqref{eq:eq-u-no-assump-xi-coupling}), but it was argued that this did not lead to the expected evolution of the Universe, as was noticed in the metric formalism as well \cite{TerenteDiaz:2023iqk}. A stability analysis of the fixed points was carried out in Sec.~\ref{subsubsec-stability}.\

Among the fixed points calculated in this work, the scaling \textbf{ScI} (where $u=0$) and de Sitter \textbf{dS} ones had been found in the SGB model of the metric formalism too \cite{TerenteDiaz:2023iqk}, and were shown to satisfy the same stability conditions in both the WGB and SGB models. While the scaling fixed point is unstable if $\kappa>\lambda$, \textbf{dS} becomes an attractor and the Universe is predicted to abandon a period of matter domination and enter a regime of accelerated expansion, where $\kappa$ and $\lambda$ are the strength of the exponentials involved in the GB coupling and the scalar potential, as shown in Eqs.~\eqref{eqqq:eqqqq4.11} and \eqref{eq:Vphi} 
respectively. If $\kappa<\lambda$ however, the roles are switched over and de Sitter becomes unstable (a saddle point) whereas \textbf{ScI} is stable. On top of these and other known fixed points, we found a second scaling regime consisting of two fixed points \textbf{ScII}$\pm$ (where $y=0$) and a regime of `pseudo-kination' where the Weyl parameter, $W_c$, satisfied $3W_c^2 M^2_{\textrm{Pl}} = \dot \phi^2_c/2$ but $W_c\neq H_c$. This regime is associated with two fixed points \textbf{\^{K}}$\pm$ as well. In the case of \textbf{ScII}$+$, it was shown not to exist if the EoS parameter of the background fluid fell within the range $\frac{1}{3} \leq w_{\textrm{M}}<1$, thereby excluding background radiation but including pressureless matter $w_{\textrm{M}} = 0$. Regarding the pseudo-kination regime, we found that \textbf{\^{K}}$+$ led to a universe that accelerated faster than during de Sitter, where it accelerates exponentially, if $\kappa <2\sqrt{\frac{2}{3}}$. Also, regardless of the sign of $x_{\textrm{\textbf{\^{K}}$\pm$}}$, if $\kappa \gg 1$, then the Universe was predicted not to accelerate. Moreover, both situations, the super-accelerated and uniform expansion phases, were shown to be stable if $\kappa>\lambda$ and $\kappa <\lambda$, respectively. The scaling regime \textbf{ScII}$\pm$ was determined to be unstable whenever $\kappa >\lambda$, same as in the case of \textbf{ScI}. The instability in the former case was found to be in the $y$ direction of the phase space. We found also that \textbf{ScII}$+$ is always stable only if $\lambda >\kappa$ because two of the three eigenvalues, which do not depend on $\lambda$, are complex with negative real parts. The summary of the fixed points and their stability was provided in Table~\ref{tab:fixed-points-wgb-stability}.\

In Sec.~\ref{sec:numerical-simulations-sub} we performed some numerical simulations assuming pressureless background matter $w_{\textrm{M}} = w_m = 0$ and checked trajectories crossing both scaling regimes. When it comes to those approaching \textbf{ScI}, we plotted the 2D phase portrait in Fig.~\ref{fig:xy_plane_phase_1wgb} and the 3D one in Fig.~\ref{fig:xyu_3d_phase_1wgb} for some values of $\lambda$ and $\kappa$. These were chosen among other simulations given that they satisfied the constraints on $\Omega_{m}$ and $w_{\textrm{f}}$ \cite{Planck:2018vyg} at present time ($N=0$), where $w_{\rm f}$ is the EoS parameter of the DE fluid responsible for the current accelerated expansion of the Universe. The evolution of the density parameters and $w_{\textrm{f}}$ for one of the trajectories in the phase portrait was displayed in Fig.~\ref{fig:Omegas_wf_1wgb}. The system's behaviour is reproduced as expected: starting from a period of kination, the Universe enters a long period of background fluid domination which gives way to the present epoch of DE domination. An eventual period of de Sitter is predicted after a transient epoch where the effective density parameter associated with the GB coupling $\xi(\phi)$, $\Omega_{\xi}$, dominates and $w_{\textrm{f}}$ takes on phantom values (i.e. $w_{\textrm{f}}<-1$). We pointed out that a common feature of all the simulations was that, during DE domination, $\Omega_{\xi}$ was always dominant over the density parameter of the scalar field $\Omega_{\phi}$.\ 

A similar but much shorter period of matter domination was reproduced in the case of the second scaling solution \textbf{ScII}$-$ (see Fig.~\ref{fig:Omegas_wf_2wgb}). The corresponding 2D phase diagram was shown in Fig.~\ref{fig:xu_plane_phase_2wgb} and the 3D phase space in Fig.~\ref{fig:xyu_3d_phase_2wgb}. In order for $\Omega_m$ to dominate over $\Omega_{\phi}+\Omega_{\xi}$ during the scaling regime (so BBN predictions are not affected), it was demonstrated that $\kappa>2\sqrt{6}$. As to \textbf{ScII}$+$, the bound translated into $\kappa <\sqrt{6}$ and was at odds with the existence condition of that fixed point. Indeed, we saw that \textbf{ScII}$+$ was located in the excluded region of the phase portrait \ref{fig:xu_plane_phase_2wgb} (the coloured stripe).\

The $\alpha_T$ parameter was written in terms of the dimensionless variables, defined in Eqs.~\eqref{eqqq:eqqq4.3} and \eqref{eq:def-u-wgb-model}, their derivatives, and the Hubble flow parameter $\epsilon_H$ in Eq.~\eqref{eqqq:eqqq.4.77}. We noticed that it is always non-vanishing for \textbf{\^{K}}$-$, and for \textbf{\^{K}}$+$ only if one considers the super-accelerated regime. Also, $\alpha_T$ never vanishes at \textbf{ScII}$-$ if $w_{\textrm{M}} = 0$. However, it was argued that the value of $|\alpha_T|$ at the fixed points was not enough because the current Universe had to be somewhere in between the scaling and de Sitter fixed points in the phase space. We plotted in Fig.~\ref{fig:alphaT-4values-comp-obs1} the evolution of $|\alpha_T|$ for different values of the constants $\lambda$ and $\kappa$ and initial conditions $x_0$. Unfortunately, the constraint in Eq.~\eqref{bounds-on-cgw2} was found to be grossly violated at present time, when $\Omega_{m}$ and $w_{\textrm{f}}$ satisfy the aforementioned constraints. It is to that moment that Eq.~\eqref{bounds-on-cgw2} applies approximately given that the binary neutron star system's merger occurred in the relatively recent past history of the Universe. This was indicated in an inset plot, where the constrained region corresponds to the coloured narrow stripe. Furthermore, the model suffers from the so-called `gradient instabilities' \cite{Joyce:2014kja}, where $\alpha_T <-1$, when the trajectories connect the scaling and de Sitter fixed points. It was argued that the pseudo-kination \textbf{\^K}$+$ is responsible for this given that it is stable in the $u$ direction of the phase space when $\kappa >3\sqrt{\frac{2}{3}}$, and that condition must be fulfilled in order for the scaling fixed point to exist and be a saddle, and for the trajectories to approach the de Sitter one.\

Despite the negative results, similar to the ones we obtained in \cite{TerenteDiaz:2023iqk} for the SGB model in the metric formalism, the analysis performed in this work did not consider coupling functions other than the exponential and linear cases (so the equations are self-similar), which might lead to different conclusions. We then regarded the case of $\alpha_T = 0$, without making any assumption on the coupling function, and analysed the stability of the scaling and de Sitter fixed points in particular given their importance in late time cosmology. We found that the latter was only possible if $\dot \phi = 0$, same as in the SGB model, despite the presence of the new variable $\Phi(t)$ in the equations. The subsequent analysis of the stability indicated that, whereas the de Sitter regime became unstable, the scaling one was stable, contradicting the known evolution of the Universe. Consequently, when $\alpha_T = 0$ for a non-constant coupling function $\xi(\phi)$, we encountered the same stability of \textbf{ScI} and \textbf{dS} as that in Ref.~\cite{TerenteDiaz:2023iqk}.\

Assuming that the Weyl vector has vanishing spatial components, one can apply the very same results of this work to a connection compatible with the metric but with non-zero torsion. We showed this in App.~\ref{sec:projective_trans_and_torsion}. In App.~\ref{app:I} we introduced such a connection and the geometric structures used in this manuscript briefly. The background equations used in Sec.~\ref{sec:theoretical_framework_ref} and those including the homogeneous spatial components of the Weyl vector were detailed in App.~\ref{app:II}.\

As mentioned above, we have regarded only exponential and linear coupling functions in this work, given that the equations were self-similar and the dynamical systems analysis could be easily done. On top of this, the case of $\kappa = \lambda$, which was analytically considered in Ref.~\cite{Nojiri:2005vv} in the metric case, was not tackled here (because, again, it was not amenable to the presented methods of analysis). It would be interesting to apply an analysis similar to that of the referenced paper, although two fixed `curves' were calculated in Sec.~\ref{sec:fixed-points-section} already, one of them corresponding to a third scaling solution of cosmological interest. Also, despite that the homogeneous spatial components of the Weyl vector were fully neglected under the assumption of spatial isotropy, one might regard the presence of such components so that they are relevant in the dynamical analysis but subdominant enough not to generate a large scale anisotropy. During inflation, this and even the fully dominant presence of the vector field were shown to be consistent with observations in Ref.~\cite{Dimopoulos:2006ms}. We leave these questions for future publications.\ 

With this work, we hope to have brought attention to the fact that SGB gravity could be rescued in the Palatini formalism with the aim of addressing the late time evolution of the Universe, where the stringent observational bounds on the speed of propagation of GWs are an unavoidable test for scalar-tensor theories at large.

\acknowledgments

We are grateful to T.~Koivisto for useful comments while this work was being completed. K.D. is supported (in part) by the STFC consolidated grant: ST/X000621/1. M.K. is supported by the Mar\'ia Zambrano grant, provided by the Ministry of Universities from the Next Generation funds of the European Union. This work is also partially supported by the MICINN (Spain) projects PID2019-107394GB-I00/AEI/10.13039/501100011033 and PID2022-139841NB-I00 (AEI/FEDER, UE). A.R. is supported by the Estonian Research Council grants PRG1055, RVTT3, RVTT7 and the CoE program TK202 ``Fundamental Universe''.\

For the purpose of open access, the authors have applied a Creative Commons Attribution (CC BY) license to any Author Accepted Manuscript version arising.        

\appendix
\section{\label{app:I}Palatini Formalism and Curvature Tensors}
In this section of the \textbf{Appendix}, we turn our attention to various geometric structures in the Palatini formalism. The Riemann tensor is defined only with respect to the connection and its first derivatives as \cite{Carroll:1997ar} \begin{equation}\label{eq:Riemann-def} \tensor{R}{^{\alpha}_{\beta\mu\nu}} \equiv \partial_{\mu} \Gamma^{\alpha}_{\nu\beta} -\partial_{\nu} \Gamma^{\alpha}_{\mu\beta} +\Gamma^{\alpha}_{\mu\lambda} \Gamma^{\lambda}_{\nu\beta} -\Gamma^{\alpha}_{\nu\lambda} \Gamma^{\lambda}_{\mu\beta}~.\end{equation} In the metric formalism, the above expression can be written in terms of the second derivatives of the metric by virtue of Eq.~\eqref{eq:LC-conn}. In the Palatini formalism, on the other hand, the Riemann tensor does not depend on the derivatives of the metric tensor. However, in that case, we can write Eq.~\eqref{eq:Riemann-def} as the sum of the LC part, and the distortion tensor (defined in Eq.~\eqref{eq:conn-generic}) and its derivatives \begin{equation}\label{eqqq:eqqq2.13} \tensor{R}{^{\alpha}_{\beta\mu\nu}} = \tensor{\mathring{R}}{^{\alpha}_{\beta\mu\nu}}+\tensor{\kappa}{^{\alpha}_{\beta\mu\nu}}~,\end{equation} where we introduced a tensor $\tensor{\kappa}{^{\alpha}_{\beta\mu\nu}}$, given by \begin{equation}\label{eqqq:eqqq2.14} \tensor{\kappa}{^{\alpha}_{\beta\mu\nu}} \equiv \mathring{\nabla}_{\mu} \tensor{\kappa}{_{\nu\beta}^{\alpha}} -\mathring{\nabla}_{\nu} \tensor{\kappa}{_{\mu\beta}^{\alpha}} +\tensor{\kappa}{_{\mu\lambda}^{\alpha}} \tensor{\kappa}{_{\nu\beta}^{\lambda}} -\tensor{\kappa}{_{\nu\lambda}^{\alpha}} \tensor{\kappa}{_{\mu\beta}^{\lambda}}.\end{equation} 

An important second-rank tensor that is related to the Riemann one is the Ricci tensor, which is defined as \begin{equation}R_{\mu\nu} \equiv \tensor{R}{^{\sigma}_{\mu\sigma\nu}} = \mathring{R}_{\mu\nu} + \kappa_{\mu\nu}~,\end{equation} where $\kappa_{\mu\nu} \equiv \tensor{\kappa}{^{\sigma}_{\mu\sigma\nu}}$. While $\mathring{R}_{\mu\nu}$ is symmetric, $R_{\mu\nu}$ does not possess any symmetry \emph{a priori}.\ 

Another related tensor, which in the metric formalism coincides with the Ricci tensor, is the co-Ricci tensor \begin{equation}\tilde{R}_{\mu\nu} \equiv g^{\alpha\beta} R_{\mu\alpha\nu\beta} = \mathring{R}_{\mu\nu} + \tilde{\kappa}_{\mu\nu}~,
\label{eq:coRicci}
\end{equation} where $\tilde{\kappa}_{\mu\nu} \equiv g^{\alpha\beta} \kappa_{\mu\alpha\nu\beta}$.\

The last second-rank tensor that can be obtained from the Riemann tensor is the `homothetic curvature tensor', $\bar{R}_{\mu\nu}$. In the Palatini formalism, the Riemann tensor is not generically antisymmetric in its first two indices. The only symmetry it has is the antisymmetry in the last two indices, which can be inferred from Eq.~\eqref{eq:Riemann-def}. Therefore, $\bar{R}_{\mu\nu}$ is defined as \begin{equation} \bar{R}_{\mu\nu} \equiv \tensor{R}{^{\sigma}_{\sigma\mu\nu}} = \bar{\kappa}_{\mu\nu}~,\label{eq:homothetic}\end{equation} where \begin{equation}\label{eqqq:eqqq2.18} \bar{\kappa}_{\mu\nu} \equiv \tensor{\kappa}{^{\sigma}_{\sigma\mu\nu}} = \mathring{\nabla}_{\mu} \tensor{\kappa}{_{\nu\sigma}^{\sigma}} -\mathring{\nabla}_{\nu} \tensor{\kappa}{_{\mu\sigma}^{\sigma}}.\end{equation} This tensor is fully antisymmetric. Consequently, the Ricci scalar \begin{equation} R \equiv g^{\mu\nu} R_{\mu\nu} = \tensor{\tilde{R}}{^{\sigma}_{\sigma}}=\mathring{R} + \kappa~,\end{equation} is the only non-vanishing scalar one can extract from the Riemann tensor. In the above expression, the scalar $\kappa$ is defined as \begin{equation}\label{eq:eqqq-def-kappa-scalar} \kappa \equiv g^{\mu\nu} \kappa_{\mu\nu} = \tensor{\tilde{\kappa}}{^{\sigma}_{\sigma}}~.\end{equation}

The Weyl connection considered in Eq.~\eqref{weyl-connection-consideration} is a concrete example of a more general one with distortion tensor \begin{equation}\label{eq:examples-distortion} \tensor{\kappa}{_{\mu\nu}^{\alpha}} = c_1\delta_{\mu}^{\alpha} A_{\nu} + c_2\delta_{\nu}^{\alpha} A_{\mu} -c_3 g_{\mu\nu} A^{\alpha},\end{equation} where $c_1$, $c_2$ and $c_3$ are constants. The torsion and non-metricity tensors read (see Eqs.~\eqref{eq:T-rel-kappa} and \eqref{eq:Q-rel-kappa}, respectively) \begin{align}\label{eq:the-T}&\tensor{T}{_{\mu\nu}^{\alpha}} = \left(c_1-c_2\right) \left(\delta_{\mu}^{\alpha} A_{\nu} - \delta_{\nu}^{\alpha} A_{\mu} \right),\\
\label{eq:the-Q}&Q_{\alpha\mu\nu} =-2c_2g_{\mu\nu} A_{\alpha} -\left(c_1-c_3\right)\left(g_{\mu\alpha} A_{\nu} + g_{\nu\alpha} A_{\mu} \right).\end{align} We see that for $(c_1,c_2,c_3) = (1,1,1)$, the torsion tensor vanishes although the non-metricity tensor is not zero, leading to Eq.~\eqref{eqqq:__QQ2.10}. Besides, the choice $(c_1,c_2,c_3) = (1,0,1)$ leads to a connection compatible with the metric (i.e. with vanishing non-metricity tensor), but with non-zero torsion. A connection with non-zero torsion and non-metricity tensors corresponds to $(c_1,c_2,c_3) = (0,1,0)$. This latter connection is well-known for being the most general solution to the connection field equations derived from the Einstein-Hilbert action in the Palatini formalism \cite{Dadhich:2012htv,Bernal:2016lhq}.\

In the case of the Weyl connection, the contributions from the distortion tensor to the Ricci, co-Ricci and homothetic curvature tensors read \begin{align}&\kappa_{\mu\nu} = \mathring{\nabla}_{\mu} A_{\nu} -3\mathring{\nabla}_{\nu} A_{\mu} +2A_{\mu} A_{\nu} -g_{\mu\nu} \left(\mathring{\nabla}_{\sigma} A^{\sigma} +2A_{\sigma} A^{\sigma} \right),\\
&\tilde{\kappa}_{\mu\nu} = -\mathring{\nabla}_{\mu} A_{\nu} -\mathring{\nabla}_{\nu} A_{\mu} +2A_{\mu} A_{\nu} -g_{\mu\nu} \left(\mathring{\nabla}_{\sigma} A^{\sigma} +2A_{\sigma} A^{\sigma} \right),\\
&\bar{\kappa}_{\mu\nu} = 4\left(\mathring{\nabla}_{\mu} A_{\nu} - \mathring{\nabla}_{\nu} A_{\mu} \right),\end{align} respectively. One can easily show that the relation among the three tensors above can be written as \begin{equation} \kappa_{\mu\nu} = \tilde{\kappa}_{\mu\nu} + \frac{1}{2} \bar{\kappa}_{\mu\nu}~.\end{equation} 

Notice that $\tilde{\kappa}_{\mu\nu}$ (and thus $\tilde{R}_{\mu\nu}$) is symmetric in this particular case. Moreover, $\tilde{\kappa}_{\mu\nu}$ is the symmetric part of $\kappa_{\mu\nu}$; i.e. $\kappa_{(\mu\nu)} = \tilde{\kappa}_{\mu\nu}$. The antisymmetric part, on the other hand, is given by $\kappa_{[\mu\nu]} = \frac{1}{2}\bar{\kappa}_{\mu\nu}$.\ 

The scalar $\kappa$, defined in Eq.~\eqref{eq:eqqq-def-kappa-scalar}, becomes \begin{equation}\kappa = -6\left(\mathring{\nabla}_{\sigma} A^{\sigma} + A_{\sigma} A^{\sigma} \right).\end{equation} Regarding $\tensor{\kappa}{^{\alpha}_{\beta\mu\nu}}$, we have the following symmetry relations: \begin{align}&\kappa_{\alpha\beta\mu\nu} =  \bar{\kappa}_{\alpha\beta\mu\nu}+\frac{1}{4} g_{\alpha\beta}\bar{\kappa}_{\mu\nu}~,\\
\label{eq:kappa-secsymm}&\kappa_{\alpha\beta\mu\nu} - \kappa_{\mu\nu\alpha\beta} = \frac{1}{4}\left(g_{\alpha\beta} \bar{\kappa}_{\mu\nu}-g_{\mu\nu} \bar{\kappa}_{\alpha\beta} - g_{\beta\nu} \bar{\kappa}_{\mu\alpha}+g_{\mu\alpha} \bar{\kappa}_{\beta\nu} -g_{\mu\beta} \bar{\kappa}_{\alpha\nu}+g_{\alpha\nu} \bar{\kappa}_{\mu\beta} \right),\end{align} where $\bar{\kappa}_{\alpha\beta\mu\nu} \equiv \kappa_{[\alpha\beta]\mu\nu}$.\

Among the various options discussed below Eq.~\eqref{eq:the-Q}, the second connection with torsion is addressed in App.~\ref{sec:projective_trans_and_torsion}.

\section{\label{app:II}Homogeneous Spatial Vector}
For completeness, we derive the homogeneous equations when the Weyl vector has non-zero homogeneous spatial components. This induces background anisotropy, as expected. To that end, we vary the action with respect to the inverse metric first.\ 

The variation of the scalar-Gauss-Bonnet part of the action in Eq.~\eqref{eq:full-actionv2} is \begin{align}\nonumber&\delta S_{\textrm{SGB}} = \frac{M^2_{\textrm{Pl}}}{2}\int \textrm{d}^4x \sqrt{-g} \left\{G_{\mu\nu} -\frac{4}{M^2_{\textrm{Pl}}}\left(\tensor{R}{_{\mu}^{\alpha}_{\nu}^{\beta}}+2\tensor{S}{^{\alpha}_{\mu}} \delta_{\nu}^{\beta} -g_{\mu\nu} G^{\alpha\beta} - R_{\mu\nu} g^{\alpha\beta} \right) \nabla_{\alpha} \partial_{\beta} \xi -\right.\\
&\left.-\frac{1}{M^2_{\textrm{Pl}}}\left[\partial_{\mu} \phi \partial_{\nu}\phi -\frac{1}{2} g_{\mu\nu} \left(g^{\alpha\beta} \partial_{\alpha} \phi \partial_{\beta} \phi +2V\right) \right] \right\} \delta g^{\mu\nu},\end{align} where $S_{\mu\nu}$ is the traceless tensor defined as \begin{equation} S_{\mu\nu} \equiv R_{\mu\nu} - \frac{1}{4} g_{\mu\nu} R~.\end{equation} We remind the reader we dropped the rings (overcircles) here because, as was argued below Eq.~\eqref{eq:full-actionv2}, every geometric quantity is defined with respect to the LC connection once the LC and non-Riemannian parts (the latter depending on the Weyl vector and its derivatives) have been written separately. Then \begin{align}\nonumber&\delta S_{\textrm{SGB}} = \frac{M^2_{\textrm{Pl}}}{2} \int \textrm{d}^4x \sqrt{-g}\left\{\left[3H^2 -12\frac{\dot \xi}{M^2_{\textrm{Pl}}} H^3 -\frac{1}{M^2_{\textrm{Pl}}} \rho_{\phi} \right]\delta g^{00} -\left[2\dot H + 3H^2 -4\frac{\ddot \xi}{M^2_{\textrm{Pl}}} H^2 -\right.\right.\\
\label{eq:deltaSsgb-var}&\left.\left.-8\frac{\dot \xi}{M^2_{\textrm{Pl}}} H^3 -8\frac{\dot \xi}{M^2_{\textrm{Pl}}} H \dot H +\frac{1}{M^2_{\textrm{Pl}}} P_{\phi} \right] g_{ij}\delta g^{ij} \right\},\end{align} where the spatially-flat FLRW metric was assumed, and $\rho_{\phi}$ and $P_{\phi}$ are defined as the energy density and pressure of the canonical scalar field $\phi$ \begin{align}&\rho_{\phi} \equiv \frac{1}{2} \dot \phi^2 + V(\phi)~,\\
&P_{\phi} \equiv \frac{1}{2} \dot \phi^2- V(\phi)~,\end{align} respectively.\

Regarding the action of the vector field $A_{\mu}$, which we shall call `$S_A$', such that $S_{\textrm{WGB}} = S_{\textrm{SGB}}+S_A+S_{\textrm{M}}$ (see Eq.~\eqref{eqqq:eqqqq2.35} where $S_{\textrm{M}}$ was introduced) and \begin{align}\nonumber&S_A = - 4 \int \textrm{d}^4x \sqrt{-g} \left[\frac{3}{4}M^2_{\textrm{Pl}} A_{\sigma} A^{\sigma} -\left(G^{\mu\nu} - \nabla^{\mu} A^{\nu} \right)A_{\mu} \partial_{\nu} \xi -\left(\nabla_{\sigma} A^{\sigma} + A_{\sigma} A^{\sigma} \right) \partial_{\rho} \xi A^{\rho} -\right.\\
\label{eq:SA-full-action}&\left.-\frac{1}{2} \xi \Upsilon F^{\mu\nu} F_{\mu\nu} \right],\end{align} where we defined \begin{align}F_{\mu\nu} \equiv \nabla_{\mu} A_{\nu} - \nabla_{\nu} A_{\mu} = \partial_{\mu} A_{\nu} - \partial_{\nu} A_{\mu}~,\end{align} 
i.e. $F_{\mu\nu} = \frac{1}{4} \bar{\kappa}_{\mu\nu}$, we have \begin{align}\nonumber&\delta S_A = -4\int \textrm{d}^4x \sqrt{-g} \left\{\left[\frac{3}{8} M^2_{\textrm{Pl}}\Phi^2 -\frac{1}{2} \dot \xi W a^{-2} A_i A^{i} -\frac{9}{2} \dot \xi H \Phi W -\frac{3}{2} \dot \xi \Phi^3-\frac{1}{2}\xi  \Upsilon a^{-2} \dot{A_i} \dot{A^{i}}+\right.\right.\\
\nonumber&\left.\left.+\frac{3}{8}M^2_{\textrm{Pl}} a^{-2} A_i A^{i} \right] \delta g^{00} -\frac{3}{2}M^2_{\textrm{Pl}}\left[\Phi - 2\frac{\dot \xi}{M^2_{\textrm{Pl}}} \left(W^2-\frac{1}{3} a^{-2} A_j A^{j} \right)\right] A_i \delta g^{0i} +\left[\frac{1}{2}\left(\frac{3}{2}M^2_{\textrm{Pl}} +\right.\right.\right.\\
\nonumber&\left.\left.\left.+\ddot \xi +3 H \dot \xi -2\dot \xi \Phi \right)A_i A_j +\dot \xi \dot{A_i} A_j +\xi \Upsilon \dot{A_i} \dot{A_j} +\left(\ddot \xi H \Phi -\frac{1}{2} \ddot \xi \Phi^2 +\frac{3}{8}M^2_{\textrm{Pl}} \Phi^2 -\right.\right.\right.\\
\nonumber&\left.\left.\left.-\frac{3}{8}M^2_{\textrm{Pl}} a^{-2} A_kA^{k} +\dot \xi H \dot \Phi +\dot \xi \dot H \Phi -\dot \xi \Phi \dot \Phi -\frac{1}{2} \dot \xi \Phi^3 +\frac{3}{2} \dot \xi H^2 \Phi -\frac{1}{2} \dot \xi W a^{-2} A_k A^{k} -\right.\right.\right.\\
\label{eq:deltaSA-var}&\left.\left.\left.-\frac{1}{2} \xi \Upsilon a^{-2} \dot{A_k} \dot{A^{k}} \right)g_{ij}\right] \delta g^{ij}\right\},\end{align} after some tedious calculations. $A_i(t)$ are the homogeneous spatial components of the Weyl vector; i.e. \begin{equation} A_{\mu}(t) \equiv \left(-\Phi(t),A_i(t)\right).\end{equation} Spatial indices are raised and lowered with respect to $\delta_{ij}$.\ 

The off-diagonal part in Eq.~\eqref{eq:deltaSA-var} vanishes because of the $\Phi$ equation which, in the presence of the homogeneous spatial component, is \begin{equation} \Phi = 2\frac{\dot \xi}{M^2_{\textrm{Pl}}}\left(W^2 -\frac{1}{3} a^{-2} A_j A^{j} \right).\end{equation} The equation of $A_i$, which can be obtained from Eq.~\eqref{eq:A-full}, reads \begin{equation} \Upsilon\left[\xi \ddot{A_i} + \left(\dot \xi + \xi H\right) \dot{A_i} \right] = \left(\frac{3}{4} M^2_{\mathrm{Pl}}+\dot \xi W\right) A_i~.\end{equation} 

The $M^2_{\textrm{Pl}} G_{ij} = T^{\textrm{eff}}_{ij}$ metric field equations, $T^{\textrm{eff}}_{\mu\nu}$ being the effective energy-momentum tensor, can be read off from Eqs.~\eqref{eq:deltaSsgb-var} and \eqref{eq:deltaSA-var} (including the matter fields) \begin{align}\nonumber&-M^2_{\textrm{Pl}}\left(2\dot H + 3H^2 \right)g_{ij} = -\left(4\ddot \xi H^2 + 8\dot \xi H^3 + 8\dot \xi H \dot H -P_{\phi}-P_{\textrm{M}} \right) g_{ij}+4\left(\frac{3}{2}M^2_{\textrm{Pl}} + \ddot \xi +3H \dot \xi -\right.\\
\nonumber&\left.-2\Phi \dot \xi \right) A_i A_j +8\dot \xi \dot{A}_{\left(i\right.}A_{\left.j\right)}+8\xi \Upsilon \dot{A_i} \dot{A_j}+8\left(\ddot \xi H \Phi -\frac{1}{2} \ddot \xi \Phi^2 +\frac{3}{8}M^2_{\textrm{Pl}} \Phi^2 -\frac{3}{8}M^2_{\textrm{Pl}} a^{-2} A_kA^{k} +\dot \xi H \dot \Phi +\right.\\
&\left.+\dot \xi \dot H \Phi -\dot \xi \Phi \dot \Phi -\frac{1}{2} \dot \xi \Phi^3 +\frac{3}{2} \dot \xi H^2 \Phi -\frac{1}{2} \dot \xi W a^{-2} A_k A^{k}-\frac{1}{2} \xi \Upsilon a^{-2} \dot{A_k} \dot{A^{k}}\right)g_{ij}~.\end{align} Assuming that the homogeneous spatial vector lies along the $z$ direction without the loss of generality, $\mathbf{A}(t) = (0,0,A_z(t))$, we see that $T^{\textrm{eff}}_{xx} = T^{\textrm{eff}}_{yy} \neq T^{\textrm{eff}}_{zz}$, which means spatial anisotropy (and the off-diagonal components vanish).\ 

Finally, $M^2_{\textrm{Pl}} G_{00} = T^{\textrm{eff}}_{00}$ reads \begin{equation} 3H^2 M^2_{\textrm{Pl}} = \rho_{\phi} +\rho_{\textrm{M}} +3\left(\Phi^2+a^{-2} A_i A^{i} \right)M^2_{\textrm{Pl}} -4\xi \Upsilon a^{-2} \dot{A_i} \dot{A^{i}} +12\dot \xi W\left(W^2-\frac{1}{3} a^{-2} A_i A^{i} \right).\end{equation}   

\section{\label{sec:projective_trans_and_torsion}Projective Transformations and Torsion}
In this section of the \textbf{Appendix}, the homogeneous equations \eqref{eq:eq-W2-matter}-\eqref{eqqq:eqqqq3.18} and the tensor perturbation equation \eqref{eqqqq:eqqqqq3.23} are shown to be valid for a connection that is metric compatible but has non-zero torsion. For this purpose, we review the so-called `projective transformations', which are given by (see Ref.~\cite{Schouten:1954book}) \begin{align}&\Gamma^{\alpha}_{\mu\nu} \rightarrow \Gamma^{\alpha}_{\mu\nu} + B_{\mu} \delta_{\nu}^{\alpha}~,\end{align} where $B_{\mu}$ is an arbitrary vector. This amounts to \begin{align}&\tensor{\kappa}{_{\mu\nu}^{\alpha}} \rightarrow \tensor{\kappa}{_{\mu\nu}^{\alpha}} + B_{\mu} \delta_{\nu}^{\alpha}~.\end{align} The $\kappa$ tensors defined in App.~\ref{app:I} transform as \begin{align}&\tensor{\kappa}{^{\alpha}_{\beta\mu\nu}} \rightarrow \tensor{\kappa}{^{\alpha}_{\beta\mu\nu}}+\delta_{\beta}^{\alpha} \mathring{F}_{\mu\nu}~,\\
&\kappa_{\mu\nu} \rightarrow \kappa_{\mu\nu} +\mathring{F}_{\mu\nu}~,\\
&\tilde{\kappa}_{\mu\nu} \rightarrow \tilde{\kappa}_{\mu\nu} -\mathring{F}_{\mu\nu}~,\\
&\bar{\kappa}_{\mu\nu} \rightarrow \bar{\kappa}_{\mu\nu} +4\mathring{F}_{\mu\nu}~,\end{align} where $\mathring{F}_{\mu\nu} \equiv \mathring{\nabla}_{\mu} B_{\nu} - \mathring{\nabla}_{\nu} B_{\mu}$. 
The scalar $\kappa$ is invariant under these transformations because the tensor $\mathring{F}_{\mu\nu}$ is antisymmetric. Notice also that the tensor $\bar{\kappa}_{\alpha\beta\mu\nu}$, which was defined as the antisymmetrisation of the first two indices of $\kappa_{\alpha\beta\mu\nu}$ below Eq.~\eqref{eq:kappa-secsymm}, is invariant as well.\

Now, let us apply this transformation to the Weyl connection. The $\mathcal{G}$ in Eq.~\eqref{eqqq:eqqqq2.32} becomes \begin{align}&\mathcal{G} \rightarrow \mathcal{G}-4\mathring{F}^{\mu\nu} \mathring{F}_{\mu\nu} - 4\Upsilon\left(\frac{1}{2} \bar{\kappa}^{\mu\nu} + \mathring{F}^{\mu\nu} \right)\mathring{F}_{\mu\nu}~.\end{align} Under this transformation, the distortion tensor of the Weyl connection is given by \begin{align} \tensor{\kappa}{_{\mu\nu}^{\alpha}} \rightarrow \delta_{\mu}^{\alpha} A_{\nu} + \delta_{\nu}^{\alpha} \left(A_{\mu} + B_{\mu} \right) - g_{\mu\nu} A^{\alpha}.\end{align} If $B_{\mu} = - A_{\mu}$, we obtain the connection compatible with the metric but with non-zero torsion which was the second of the three cases extracted from Eq.~\eqref{eq:examples-distortion} with $(c_1,c_2,c_3) = (1,0,1)$. Furthermore, if the projective transformation is applied to the LC connection, the resulting connection corresponds to the third example, which had non-vanishing torsion and non-metricity tensors, $(c_1,c_2,c_3) = (0,1,0)$. For $B_{\mu} = -A_{\mu}$, we have \begin{align}&\mathring{F}_{\mu\nu} =  -\frac{1}{4} \bar{\kappa}_{\mu\nu}~,\end{align} hence the $\mathcal{G}$ with torsion and vanishing non-metricity is \begin{align}\mathcal{G} = \mathcal{G}^{\textrm{Weyl}}-\frac{1-\Upsilon}{4}\bar{\kappa}^{\mu\nu} \bar{\kappa}_{\mu\nu}~.\end{align} We see that $\mathcal{G}$ and $\mathcal{G}^{\textrm{Weyl}}$ only differ by the term $\bar{\kappa}^{\mu\nu}\bar{\kappa}_{\mu\nu}$. Actually, we can see that the new $\mathcal{G}$ does not depend on $\Upsilon$ \begin{align}&\mathcal{G} = \mathring{\mathcal{G}} + 2\mathring{R} \kappa - 8\mathring{R}^{\mu\nu} \tilde{\kappa}_{\mu\nu} + 2\mathring{R}^{\alpha\beta\mu\nu} \bar{\kappa}_{\alpha\beta\mu\nu}+\kappa^2 -4\tilde{\kappa}^{\mu\nu} \tilde{\kappa}_{\mu\nu} + \bar{\kappa}^{\alpha\beta\mu\nu} \bar{\kappa}_{\alpha\beta\mu\nu}-\frac{1}{4} \bar{\kappa}^{\mu\nu} \bar{\kappa}_{\mu\nu}~.\end{align} So the calculations are the same as in the case of the Weyl connection if we simply set $\Upsilon = 1$ (see Eq.~\eqref{eqqq:eqqqq2.32}).\

If $A_i(t) = 0$, the $\Upsilon$ term does not contribute to the homogeneous and tensor perturbations equations as it multiplies $\bar{\kappa}_{\mu\nu}$ (which vanishes in that case; see Eq.~\eqref{eq:A-full}). Consequently, for a background Weyl vector that has solely a non-zero temporal component (so it does not break spatial isotropy), the results derived in this work are the same as those obtained assuming the $(c_1,c_2,c_3) = (1,0,1)$ connection with non-zero torsion and metric compatible. 






\begin{thebibliography}{99}
\bibitem{SupernovaSearchTeam:1998fmf}
A.~G.~Riess \textit{et al.} [Supernova Search Team], \emph{Observational evidence from supernovae for an accelerating universe and a cosmological constant}, \href{https://doi.org/10.1086/300499}{\emph{Astron. J.} {116 1009-1038} (1998)} [\href{https://arxiv.org/abs/astro-ph/9805201}{\texttt{astro-ph/9805201}}].

\bibitem{SupernovaCosmologyProject:1998vns}
S.~Perlmutter \textit{et al.} [Supernova Cosmology Project], \emph{Measurements of $\Omega$ and $\Lambda$ from $42$ high redshift supernovae}, \href{https://doi.org/10.1086/307221}{\emph{Astrophys. J.} {517 565-586} (1999)} [\href{https://arxiv.org/abs/astro-ph/9812133}{\texttt{astro-ph/9812133}}].

\bibitem{Planck:2018vyg}
N.~Aghanim \textit{et al.} [Planck Collaboration], \emph{Planck 2018 results. VI. Cosmological parameters}, \href{https://doi.org/10.1051/0004-6361/201833910}{\emph{Astron.~Astrophys.~}{641 A6} (2020)} [Erratum: \href{https://doi.org/10.1051/0004-6361/201833910e}{Astron.~Astrophys. 652, C4 (2021)}] [\href{https://arxiv.org/abs/1807.06209}{\texttt{1807.06209}}].

\bibitem{Carroll:2000fy}
S.~M.~Carroll, \emph{The Cosmological constant}, \href{https://doi.org/10.12942/lrr-2001-1}{\emph{Living Rev. Rel.}{~4, 1} (2001)} [\href{https://arxiv.org/abs/astro-ph/0004075}{\texttt{astro-ph/0004075}}].

\bibitem{Martin:2012bt}
J.~Martin, \emph{Everything You Always Wanted To Know About The Cosmological Constant Problem (But Were Afraid To Ask)}, \href{https://doi.org/10.1016/j.crhy.2012.04.008}{\emph{Comptes Rendus Physique}{~13, 566-665} (2012)} [\href{https://arxiv.org/abs/1205.3365}{\texttt{1205.3365}}].

\bibitem{Joyce:2014kja}
A.~Joyce, B.~Jain, J.~Khoury and M.~Trodden, \emph{Beyond the Cosmological Standard Model}, \href{https://doi.org/10.1016/j.physrep.2014.12.002}{\emph{Phys. Rept.}{~568, 1-98} (2015)} [\href{https://arxiv.org/abs/1407.0059}{\texttt{1407.0059}}].

\bibitem{Weinberg:1988cp}
S.~Weinberg, \emph{The Cosmological Constant Problem}, \href{https://doi.org/10.1103/RevModPhys.61.1}{\emph{Rev. Mod. Phys.}{~61, 1-23} (1989)}.

\bibitem{Weinberg:2000yb}
S.~Weinberg, \emph{The Cosmological constant problems}, \href{https://inspirehep.net/conferences/972507}{\emph{Contribution to 4th International Symposium on Sources and Detection of Dark Matter in the Universe}{~(DM 2000)} (2000)} [\href{https://arxiv.org/abs/astro-ph/0005265}{\texttt{astro-ph/0005265}}].

\bibitem{Caldwell:1997ii}
R.~R.~Caldwell, R.~Dave and P.~J.~Steinhardt, \emph{Cosmological imprint of an energy component with general equation of state}, \href{https://doi.org/10.1103/PhysRevLett.80.1582}{\emph{Phys. Rev. Lett.}{~80, 1582-1585} (1998)} [\href{https://arxiv.org/abs/astro-ph/9708069}{\texttt{astro-ph/9708069}}].

\bibitem{Zlatev:1998tr}
I.~Zlatev, Li-Min~Wang and P.~J.~Steinhardt, \emph{Quintessence, cosmic coincidence, and the cosmological constant}, \href{https://doi.org/10.1103/PhysRevLett.82.896}{\emph{Phys. Rev. Lett.}{~82, 896-899} (1999)} [\href{https://arxiv.org/abs/astro-ph/9807002}{\texttt{astro-ph/9807002}}].

\bibitem{Copeland:2006wr}
E.~J.~Copeland, M.~Sami and S.~Tsujikawa, \emph{Dynamics of dark energy}, \href{https://doi.org/10.1142/S021827180600942X}{\emph{Int. J. Mod. Phys. D}{~15, 1753-1936} (2006)} [\href{https://arxiv.org/abs/hep-th/0603057}{\texttt{hep-th/0603057}}].

\bibitem{Tsujikawa:2010sc}
S.~Tsujikawa, \emph{Dark energy: investigation and modeling}, \href{https://doi.org/10.1007/978-90-481-8685-3_8}{\emph{ASSL}{~Vol. 370, 331-402, Springer, Dordrecht} (2011)} [\href{https://arxiv.org/abs/1004.1493}{\texttt{1004.1493}}].

\bibitem{Tsujikawa:2013fta}
S.~Tsujikawa, \emph{Quintessence: A Review}, \href{https://doi.org/10.1088/0264-9381/30/21/214003}{\emph{Class. Quant. Grav.}{~30, 214003} (2013)} [\href{https://arxiv.org/abs/1304.1961}{\texttt{1304.1961}}].

\bibitem{Peebles:1998qn}
P.~J.~E.~Peebles and A.~Vilenkin, \emph{Quintessential inflation}, \href{https://doi.org/10.1103/PhysRevD.59.063505}{\emph{Phys. Rev. D}{~59, 063505} (1999)} [\href{https://arxiv.org/abs/astro-ph/9810509}{\texttt{astro-ph/9810509}}].

\bibitem{Konstantinos:2001MQI}
K.~Dimopoulos and J.~W.~F.~Valle, \emph{Modeling Quintessential Inflation}, \href{https://doi.org/10.1016/S0927-6505\%2802\%2900115-9}{\emph{Astropart. Phys.} {18 287-306} (2002)} [\href{https://arxiv.org/abs/astro-ph/0111417}{\texttt{astro-ph/0111417}}].

\bibitem{deHaro:2016ftq}
J.~de Haro, \emph{On the viability of quintessential inflation models from observational data}, \href{https://doi.org/10.1007/s10714-016-2173-8}{\emph{Gen. Rel. Grav.} {49, 1, 6} (2017)} [\href{https://arxiv.org/abs/1602.07138}{\texttt{1602.07138}}].

\bibitem{Dimopoulos:2021xld}
K.~Dimopoulos, \emph{Jointly modelling Cosmic Inflation and Dark Energy}, \href{https://doi.org/10.1088/1742-6596/2105/1/012001}{\emph{J. Phys. Conf. Ser.}{~2105, 1, 012001} (2021)} [\href{https://arxiv.org/abs/2106.14966}{\texttt{2106.14966}}].

\bibitem{Starobinsky:1980te}
A.~A.~Starobinsky, \emph{A New Type of Isotropic Cosmological Models Without Singularity}, \href{https://doi.org/10.1016/0370-2693(80)90670-X}{\emph{Phys. Lett. B~}{91, 99-102} (1980)}.

\bibitem{Guth:1980zm}
A.~H.~Guth, \emph{The Inflationary Universe: A Possible Solution to the Horizon and Flatness Problems}, \href{https://doi.org/10.1103/PhysRevD.23.347}{\emph{Phys. Rev. D} {23 347-356} (1981)}.

\bibitem{Linde:1981mu}
A.~D.~Linde, \emph{A New Inflationary Universe Scenario: A Possible Solution of the Horizon, Flatness, Homogeneity, Isotropy and Primordial Monopole Problems}, \href{https://doi.org/10.1016/0370-2693(82)91219-9}{\emph{Phys. Lett. B} {108 389-393} (1982)}.

\bibitem{Capozziello:2002rd}
S.~Capozziello, \emph{Curvature quintessence}, \href{https://doi.org/10.1142/S0218271802002025}{\emph{Int. J. Mod. Phys. D~}{11, 483-492} (2002)} [\href{https://arxiv.org/abs/gr-qc/0201033}{\texttt{gr-qc/0201033}}].

\bibitem{Carroll:2003wy}
S.~M.~Carroll, V.~Duvvuri, M.~Trodden and M.~S.~Turner, \emph{Is cosmic speed~-~up due to new gravitational physics?}, \href{https://doi.org/10.1103/PhysRevD.70.043528}{\emph{Phys. Rev. D~}{70, 043528} (2004)} [\href{https://arxiv.org/abs/astro-ph/0306438}{\texttt{astro-ph/0306438}}].

\bibitem{Koivisto:2006xf}
T.~Koivisto and D.~F.~Mota, \emph{Cosmology and Astrophysical Constraints of Gauss-Bonnet Dark Energy}, \href{https://doi.org/10.1016/j.physletb.2006.11.048}{\emph{Phys. Lett. B}{~644, 104-108} (2007)} [\href{https://arxiv.org/abs/astro-ph/0606078}{\ttfamily astro-ph/0606078}].

\bibitem{Nojiri:2017ncd}
S.~Nojiri, S.~D.~Odintsov and V.~K.~Oikonomou, \emph{Modified Gravity Theories on a Nutshell: Inflation, Bounce and Late-time Evolution}, \href{https://doi.org/10.1016/j.physrep.2017.06.001}{\emph{Phys. Rept.}{~692, 1-104} (2017)} [\href{https://arxiv.org/abs/1705.11098}{\texttt{1705.11098}}].

\bibitem{vandeBruck:2017voa}
C.~van de Bruck, K.~Dimopoulos, C.~Longden and C.~Owen, \emph{Gauss-Bonnet-coupled Quintessential Inflation}, \href{https://arxiv.org/abs/1707.06839}{\texttt{1707.06839}}.

\bibitem{deHaro:2021swo}
J.~de Haro and L.~A.~Saló, \emph{A Review of Quintessential Inflation}, \href{https://doi.org/10.3390/galaxies9040073}{\emph{Galaxies~}{9, 4, 73} (2021)}[\href{https://arxiv.org/abs/2108.11144}{\texttt{2108.11144}}].

\bibitem{LIGOScientific:2016aoc}
B.~P.~Abbott \textit{et al.} [LIGO Scientific, Virgo \textit{et al.} Collaboration], \emph{Observation of Gravitational Waves from a Binary Black Hole Merger}, \href{https://doi.org/10.1103/PhysRevLett.116.061102}{\emph{Phys. Rev. Lett.~}{116, 6, 061102} (2016)} [\href{https://arxiv.org/abs/1602.03837}{\texttt{1602.03837}}].

\bibitem{LIGOScientific:2016sjg}
B.~P.~Abbott \textit{et al.} [LIGO Scientific, Virgo \textit{et al.} Collaboration], \emph{GW151226: Observation of Gravitational Waves from a 22-Solar-Mass Binary Black Hole Coalescence}, \href{https://doi.org/10.1103/PhysRevLett.116.241103}{\emph{Phys. Rev. Lett.~}{116, 24, 241103} (2016)} [\href{https://arxiv.org/abs/1606.04855}{\texttt{1606.04855}}].

\bibitem{LIGOScientific:2017vwq}
B.~P.~Abbott \textit{et al.} [LIGO Scientific, Virgo \textit{et al.} Collaboration], \emph{GW170817: Observation of Gravitational Waves from a Binary Neutron Star Inspiral}, \href{https://doi.org/10.1103/PhysRevLett.119.161101}{\emph{Phys. Rev. Lett.~}{119, 16, 161101} (2017)} [\href{https://arxiv.org/abs/1710.05832}{\texttt{1710.05832}}].

\bibitem{Sakstein:2017xjx}
J.~Sakstein and B.~Jain, \emph{Implications of the Neutron Star Merger GW170817 for Cosmological Scalar-Tensor Theories}, \href{https://doi.org/10.1103/PhysRevLett.119.251303}{\emph{Phys. Rev. Lett.~}{119, 25, 251303} (2017)} [\href{https://arxiv.org/abs/1710.05893}{\texttt{1710.05893}}].

\bibitem{Ezquiaga:2017ekz}
J.~M.~Ezquiaga and M.~Zumalac\'arregui, \emph{Dark Energy After GW170817: Dead Ends and the Road Ahead}, \href{https://doi.org/10.1103/PhysRevLett.119.251304}{\emph{Phys. Rev. Lett.~}{119, 25, 251304} (2017)} [\href{https://arxiv.org/abs/1710.05901}{\texttt{1710.05901}}].

\bibitem{Baker:2017hug}
T.~Baker, E.~Bellini, P.~G.~Ferreira, M.~Lagos, J.~Noller and I.~Sawicki, \emph{Strong Constraints on Cosmological Gravity from GW170817 and GRB 170817A}, \href{https://doi.org/10.1103/PhysRevLett.119.251301}{\emph{Phys. Rev. Lett.~}{119, 25, 251301} (2017)} [\href{https://arxiv.org/abs/1710.06394}{\texttt{1710.06394}}].

\bibitem{Arai:2017hxj}
S.~Arai and A.~Nishizawa, \emph{Generalized framework for testing gravity with gravitational-wave propagation. II. Constraints on Horndeski theory}, \href{https://doi.org/10.1103/PhysRevD.97.104038}{\emph{Phys. Rev. D~}{97, 10, 104038} (2018)} [\href{https://arxiv.org/abs/1711.03776}{\texttt{1711.03776}}].

\bibitem{Creminelli:2017sry}
P.~Creminelli and F.~Vernizzi, \emph{Dark Energy after GW170817 and GRB170817A}, \href{https://doi.org/10.1103/PhysRevLett.119.251302}{\emph{Phys. Rev. Lett.~}{119, 25, 251302} (2017)} [\href{https://arxiv.org/abs/1710.05877}{\texttt{1710.05877}}].

\bibitem{Langlois:2017dyl}
D.~Langlois, R.~Saito, D.~Yamauchi and K.~Noui, \emph{Scalar-tensor theories and modified gravity in the wake of GW170817}, \href{https://doi.org/10.1103/PhysRevD.97.061501}{\emph{Phys. Rev. D~}{97, 6, 061501} (2018)} [\href{https://arxiv.org/abs/1711.07403}{\texttt{1711.07403}}].

\bibitem{LIGOScientific:2017ync}
B.~P.~Abbott \textit{et al.} [LIGO Scientific, Virgo \textit{et al.} Collaboration], \emph{Multi-messenger Observations of a Binary Neutron Star Merger}, \href{https://doi.org/10.3847/2041-8213/aa91c9}{\emph{Astrophys. J. Lett.~}{848, 2, L12} (2017)} [\href{https://arxiv.org/abs/1710.05833}{\texttt{1710.05833}}].

\bibitem{LIGOScientific:2017zic}
B.~P.~Abbott \textit{et al.} [LIGO Scientific, Virgo \textit{et al.} Collaboration], \emph{Gravitational Waves and Gamma-rays from a Binary Neutron Star Merger: GW170817 and GRB 170817A}, \href{https://doi.org/10.3847/2041-8213/aa920c}{\emph{Astrophys. J. Lett.~}{848, 2, L13} (2017)} [\href{https://arxiv.org/abs/1710.05834}{\texttt{1710.05834}}].

\bibitem{Horndeski:1974wa}
G.~W.~Horndeski, \emph{Second-order scalar-tensor field equations in a four-dimensional space}, \href{https://doi.org/10.1007/BF01807638}{\emph{Int. J. Theor. Phys.~}{10, 363-384} (1974)}.

\bibitem{deRham:2018red}
C.~de Rham and S.~Melville, \emph{Gravitational Rainbows: LIGO and Dark Energy at its Cutoff}, \href{https://doi.org/10.1103/PhysRevLett.121.221101}{\emph{Phys. Rev. Lett.~}{121, 22, 221101} (2018)} [\href{https://arxiv.org/abs/1806.09417}{\texttt{1806.09417}}].

\bibitem{Kobayashi:2011nu}
T.~Kobayashi, M.~Yamaguchi and J.~Yokoyama, \emph{Generalized G-inflation: Inflation with the most general second-order field equations}, \href{https://doi.org/10.1143/PTP.126.511}{\emph{Prog. Theor. Phys.~}{126, 511-529} (2011)} [\href{https://arxiv.org/abs/1105.5723}{\texttt{1105.5723}}].

\bibitem{Fernandes:2022zrq}
P.~G.~S.~Fernandes, P.~Carrilho, T.~Clifton and D.~J.~Mulryne, \emph{The 4D Einstein-Gauss-Bonnet theory of gravity: a review}, \href{https://doi.org/10.1088/1361-6382/ac500a}{\emph{Class. Quant. Grav.~}{39, 6, 063001} (2022)} [\href{https://arxiv.org/abs/2202.13908}{\texttt{2202.13908}}].



\bibitem{Odintsov:2019clh}
S.~D.~Odintsov and V.~K.~Oikonomou, \emph{Inflationary Phenomenology of Einstein Gauss-Bonnet Gravity Compatible with GW170817}, \href{https://doi.org/10.1016/j.physletb.2019.134874}{\emph{Phys. Lett. B~}{797, 134874} (2019)} [\href{https://arxiv.org/abs/1908.07555}{\texttt{1908.07555}}].

\bibitem{Odintsov:2020sqy}
S.~D.~Odintsov, V.~K.~Oikonomou and F.~P.~Fronimos, \emph{Rectifying Einstein-Gauss-Bonnet Inflation in View of GW170817}, \href{https://doi.org/10.1016/j.nuclphysb.2020.115135}{\emph{Nucl. Phys. B~}{958, 115135} (2020)} [\href{https://arxiv.org/abs/2003.13724}{\texttt{2003.13724}}].

\bibitem{Odintsov:2020zkl}
S.~D.~Odintsov and V.~K.~Oikonomou, \emph{Swampland implications of GW170817-compatible Einstein-Gauss-Bonnet gravity}, \href{https://doi.org/10.1016/j.physletb.2020.135437}{\emph{Phys. Lett. B~}{805, 135437} (2020)} [\href{https://arxiv.org/abs/2004.00479}{\texttt{2004.00479}}].

\bibitem{Hjorth:2017yza}
J.~Hjorth \textit{et al.}, \emph{The Distance to NGC 4993: The Host Galaxy of the Gravitational-wave Event GW170817}, \href{https://doi.org/10.3847/2041-8213/aa9110}{\emph{Astrophys. J. Lett.~}{848, 2, L31} (2017)} [\href{https://arxiv.org/abs/1710.05856}{\texttt{1710.05856}}].

\bibitem{TerenteDiaz:2023iqk}
J.~J.~Terente D\'iaz, K.~Dimopoulos, M.~Kar\v{c}iauskas and A.~Racioppi, \emph{Gauss-Bonnet Dark Energy and the speed of gravitational waves}, \href{https://doi.org/10.1088/1475-7516/2023/10/031}{\emph{JCAP~}{10, 031} (2023)} [\href{https://arxiv.org/abs/2307.06163}{\texttt{2307.06163}}].

\bibitem{Granda:2018tzi}
L.~N.~Granda and D.~F.~Jimenez, \emph{The speed of gravitational waves and power-law solutions in a scalar-tensor model}, \href{https://doi.org/10.1016/j.astropartphys.2018.08.001}{\emph{Astropart. Phys.~}{103, 115-121} (2018)} [\href{https://arxiv.org/abs/1802.03781}{\texttt{1802.03781}}].

\bibitem{Gong:2017kim}
Y.~Gong, E.~Papantonopoulos and Z.~Yi, \emph{Constraints on scalar-tensor theory of gravity by the recent observational results on gravitational waves}, \href{https://doi.org/10.1140/epjc/s10052-018-6227-9}{\emph{Eur. Phys. J. C~}{78, 9, 738} (2018)} [\href{https://arxiv.org/abs/1711.04102}{\texttt{1711.04102}}].

\bibitem{MohseniSadjadi:2023amn}
H.~Mohseni Sadjadi, \emph{Scalar-Gauss-Bonnet model, the coincidence problem and the gravitational wave speed}, \href{https://arxiv.org/abs/2309.07816}{\texttt{2309.07816}}.

\bibitem{MohseniSadjadi:2023cjd}
H.~Mohseni Sadjadi, \emph{Non-minimally coupled quintessence in the Gauss-Bonnet model, symmetry breaking, and cosmic acceleration}, \href{https://arxiv.org/abs/2310.03697}{\texttt{2310.03697}}.

\bibitem{Helpin:2019kcq}
T.~Helpin and M.~S.~Volkov, \emph{Varying the Horndeski Lagrangian within the Palatini approach}, \href{https://doi.org/10.1088/1475-7516/2020/01/044}{\emph{JCAP~}{01, 044} (2020)} [\href{https://arxiv.org/abs/1906.07607}{\texttt{1906.07607}}].

\bibitem{Helpin:2019vrv}
T.~Helpin and M.~S.~Volkov, \emph{A metric-affine version of the Horndeski theory}, \href{https://doi.org/10.1142/S0217751X20400102}{\emph{Int. J. Mod. Phys. A~}{35, 02n03, 2040010} (2020)} [\href{https://arxiv.org/abs/1911.12768}{\texttt{1911.12768}}].

\bibitem{Kubota:2020ehu}
M.~Kubota, Kin-Ya~Oda, K.~Shimada and M.~Yamaguchi, \emph{Cosmological Perturbations in Palatini Formalism}, \href{https://doi.org/10.1088/1475-7516/2021/03/006}{\emph{JCAP~}{03, 006} (2021)} [\href{https://arxiv.org/abs/2010.07867}{\texttt{2010.07867}}].

\bibitem{Dong:2021jtd}
Yu-Qi.~Dong and Yu-Xiao~Liu, \emph{Polarization modes of gravitational waves in Palatini-Horndeski theory}, \href{https://doi.org/10.1103/PhysRevD.105.064035}{\emph{Phys. Rev. D~}{105, 6, 6} (2022)} [\href{https://arxiv.org/abs/2111.07352}{\texttt{2111.07352}}].

\bibitem{Dong:2022cvf}
Yu-Qi~Dong, Yu-Qiang~Liu and Yu-Xiao~Liu, \emph{Constraining Palatini-Horndeski theory with gravitational waves after GW170817}, \href{https://doi.org/10.1140/epjc/s10052-023-11861-9}{\emph{Eur. Phys. J. C~}{83, 8, 702} (2023)} [\href{https://arxiv.org/abs/2211.12056}{\texttt{2211.12056}}].

\bibitem{Dimopoulos:2020pas}
K.~Dimopoulos and S.~S\'anchez L\'opez, \emph{Quintessential inflation in Palatini $f(R)$ gravity}, \href{https://doi.org/10.1103/PhysRevD.103.043533}{\emph{Phys. Rev. D~}{103, 4, 043533} (2021)} [\href{https://arxiv.org/abs/2012.06831}{\texttt{2012.06831}}].

\bibitem{Dimopoulos:2022rdp}
K.~Dimopoulos, A.~Karam, S.~S\'anchez L\'opez and E.~Tomberg, \emph{Palatini $R^2$ quintessential inflation}, \href{https://doi.org/10.1088/1475-7516/2022/10/076}{\emph{JCAP.~}{10, 076} (2022)} [\href{https://arxiv.org/abs/2206.14117}{\texttt{2206.14117}}].

\bibitem{Sharma:2021fou}
M.~Kumar Sharma and S.~Sur, \emph{Growth of matter perturbations in an interacting dark energy scenario emerging from metric-scalar-torsion couplings}, \href{https://arxiv.org/abs/2102.01525}{\texttt{2102.01525}}.

\bibitem{deRham:2016wji}
C.~de Rham and A.~Matas, \emph{Ostrogradsky in Theories with Multiple Fields}, \href{https://doi.org/10.1088/1475-7516/2016/06/041}{\emph{JCAP~}{06, 041} (2016)} [\href{https://arxiv.org/abs/1604.08638}{\texttt{1604.08638}}].

\bibitem{BeltranJimenez:2014iie}
J.~B.~Jimenez and T.~S.~Koivisto, \emph{Extended Gauss-Bonnet gravities in Weyl geometry}, \href{https://doi.org/10.1088/0264-9381/31/13/135002}{\emph{Class. Quant. Grav.~}{31, 135002} (2014)} [\href{https://arxiv.org/abs/1402.1846}{\texttt{1402.1846}}].

\bibitem{Copeland:1997et}
E.~J.~Copeland, A.~R.~Liddle and D.~Wands, \emph{Exponential potentials and cosmological scaling solutions}, \href{https://doi.org/10.1103/PhysRevD.57.4686}{\emph{Phys. Rev. D}{~57, 4686-4690} (1998)} [\href{https://arxiv.org/abs/gr-qc/9711068}{\texttt{gr-qc/9711068}}].

\bibitem{Ferreira:1997hj}
P.~G.~Ferreira and M.~Joyce, \emph{Cosmology with a primordial scaling field}, \href{https://doi.org/10.1103/PhysRevD.58.023503}{\emph{Phys. Rev. D}{~58, 023503} (1998)} [\href{https://arxiv.org/abs/astro-ph/9711102}{\texttt{astro-ph/9711102}}].

\bibitem{Barreiro:1999zs}
T.~Barreiro, E.~J.~Copeland and N.~J.~Nunes, \emph{Quintessence arising from exponential potentials}, \href{https://doi.org/10.1103/PhysRevD.61.127301}{\emph{Phys. Rev. D}{~61, 127301} (2000)} [\href{https://arxiv.org/abs/astro-ph/9910214}{\texttt{astro-ph/9910214}}].

\bibitem{Carroll:1997ar}
S.~M.~Carroll, \emph{Lecture Notes On General Relativity}, \href{https://arxiv.org/abs/gr-qc/9712019}{\texttt{gr-qc/9712019}}.



\bibitem{Bahamonde:2021gfp}
S.~Bahamonde, K.~F.~Dialektopoulos, C.~Escamilla-Rivera, G.~Farrugia, V.~Gakis, M.~Hendry, M.~Hohmann, J.~L.~Said, J.~Mifsud and E.~Di Valentino, \emph{Teleparallel gravity: from theory to cosmology}, \href{https://doi.org/10.1088/1361-6633/ac9cef}{\emph{Rept. Prog. Phys.~}{86, 2, 026901} (2023)} [\href{https://arxiv.org/abs/2106.13793}{\texttt{2106.13793}}].

\bibitem{Borunda:2008kf}
M.~Borunda, B.~Janssen and M.~Bastero-Gil, \emph{Palatini versus metric formulation in higher curvature gravity}, \href{https://doi.org/10.1088/1475-7516/2008/11/008}{\emph{JCAP~}{11, 008} (2008)} [\href{https://arxiv.org/abs/0804.4440}{\texttt{0804.4440}}].

\bibitem{Clifton:2011jh}
T.~Clifton, P.~G.~Ferreira, A.~Padilla and C.~Skordis, \emph{Modified Gravity and Cosmology}, \href{https://doi.org/10.1016/j.physrep.2012.01.001}{\emph{Phys. Rept.~}{513, 1-189} (2012)} [\href{https://arxiv.org/abs/1106.2476}{\texttt{1106.2476}}].

\bibitem{Baumann:2009ds}
D.~Baumann, \emph{Inflation}, \href{https://www.worldscientific.com/doi/abs/10.1142/9789814327183_0010}{\emph{Physics of the Large and the Small} {523-686} (2011)} [\href{https://arxiv.org/abs/0907.5424}{\texttt{0907.5424}}].

\bibitem{Koivisto:2006ai}
T.~Koivisto and D.~F.~Mota, \emph{Gauss-Bonnet Quintessence: Background Evolution, Large Scale Structure and Cosmological Constraints}, \href{https://doi.org/10.1103/PhysRevD.75.023518}{\emph{Phys. Rev. D}{~75, 023518} (2006)} [\href{https://arxiv.org/abs/hep-th/0609155}{\texttt{hep-th/0609155}}].


\bibitem{Weinberg:2013agg}
D.~H.~Weinberg, M.~J.~Mortonson, D.~J.~Eisenstein, C.~Hirata, A.~G.~Riess and E.~Rozo, \emph{Observational Probes of Cosmic Acceleration}, \href{https://doi.org/10.1016/j.physrep.2013.05.001}{\emph{Phys. Rept.~}{530, 87-255} (2013)} [\href{https://arxiv.org/abs/1201.2434}{\texttt{1201.2434}}].








\bibitem{Jarv:2009zf}
L.~Jarv, P.~Kuusk and M.~Saal, \emph{Remarks on (super-)accelerating cosmological models in general scalar-tensor theory}, \href{https://doi.org/10.3176/proc.2010.4.09}{\emph{Proc. Est. Acad. Sci. Phys. Math.~}{59, 306-312} (2010)} [\href{https://arxiv.org/abs/0903.4357}{\texttt{0903.4357}}].

\bibitem{MohseniSadjadi:2020qnm}
H.~Mohseni Sadjadi, \emph{On cosmic acceleration in four dimensional Einstein-Gauss-Bonnet gravity}, \href{https://doi.org/10.1016/j.dark.2020.100728}{\emph{Phys. Dark Univ.~}{30, 100728} (2020)} [\href{https://arxiv.org/abs/2005.10024}{\texttt{2005.10024}}].

\bibitem{LIGOScientific:2018hze}
B.~P.~Abbott \textit{et al.} [LIGO Scientific, Virgo \textit{et al.} Collaboration], \emph{Properties of the binary neutron star merger GW170817}, \href{https://doi.org/10.1103/PhysRevX.9.011001}{\emph{Phys. Rev. X~}{9, 1, 011001} (2019)} [\href{https://arxiv.org/abs/1805.11579}{\texttt{1805.11579}}].

\bibitem{Nojiri:2005vv}
S.~Nojiri, S.~D.~Odintsov and M.~Sasaki, \emph{Gauss-Bonnet dark energy}, \href{https://doi.org/10.1103/PhysRevD.71.123509}{\emph{Phys. Rev. D}{~71, 123509} (2005)} [\href{https://arxiv.org/abs/hep-th/0504052}{\texttt{hep-th/0504052}}].

\bibitem{Dimopoulos:2006ms}
K.~Dimopoulos, \emph{Can a vector field be responsible for the curvature perturbation in the Universe?}, \href{https://doi.org/10.1103/PhysRevD.74.083502}{\emph{Phys. Rev. D} {74, 083502} (2006)} [\href{https://arxiv.org/abs/hep-ph/0607229}{\texttt{hep-ph/0607229}}].

\bibitem{Dadhich:2012htv}
N.~Dadhich and J.~M.~Pons, \emph{On the equivalence of the Einstein-Hilbert and the Einstein-Palatini formulations of general relativity for an arbitrary connection}, \href{https://doi.org/10.1007/s10714-012-1393-9}{\emph{Gen. Rel. Grav.~}{44, 2337-2352} (2012)} [\href{https://arxiv.org/abs/1010.0869}{\texttt{1010.0869}}].

\bibitem{Bernal:2016lhq}
A.~N.~Bernal, B.~Janssen, A.~Jimenez-Cano, J.~A.~Orejuela, M.~Sanchez and P.~Sanchez-Moreno, \emph{On the (non-)uniqueness of the Levi-Civita solution in the Einstein-Hilbert-Palatini formalism}, \href{https://doi.org/10.1016/j.physletb.2017.03.001}{\emph{Phys. Lett. B~}{768, 280-287} (2017)} [\href{https://arxiv.org/abs/1606.08756}{\texttt{1606.08756}}].

\bibitem{Schouten:1954book}
J.~A.~Schouten, \emph{Ricci-Calculus: An Introduction to Tensor Analysis and Its Geometrical Applications}, \href{https://link.springer.com/book/10.1007/978-3-662-12927-2}{\emph{Springer-Verlag}{,~Berlin and Heidelberg} (1954)}.







\end{thebibliography}
\end{document}